\pdfoutput=1
\documentclass[preprint,superscriptaddress]{elsarticle}

\usepackage{lineno}

\modulolinenumbers[5]

\journal{Nuclear Instruments and Methods in Physics Research A, }
\usepackage{epsfig}
\usepackage{color}

\usepackage{url}
\usepackage{rotating}
\usepackage{amssymb}
\usepackage{dcolumn}% Align table columns on decimal point

\begin{document}

\begin{frontmatter}
%\begin{onecolumn}
\title{The NuMI Neutrino Beam}

% Group authors per affiliation:

%\author[address1]{X. XYZ1}

%\author[address2]{Y. XYZ2\corref{mycorrespondingauthor}}

%\address[address1]{Fermilab, Batavia, illinois}
%\address[address2]{University of, xxx, yyy}

%\end{onecolumn}

\address[Berkeley]{Lawrence Berkeley National Laboratory, Berkeley, California, 94720 USA}
\address[Cambridge]{Cavendish Laboratory, University of Cambridge, Madingley Road, Cambridge CB3 0HE, United Kingdom}
\address[Cincinnati]{Department of Physics, University of Cincinnati, Cincinnati, Ohio 45221, USA}
\address[FNAL]{Fermi National Accelerator Laboratory, Batavia, Illinois 60510, USA}
\address[RAL]{Rutherford Appleton Laboratory, Science and TechnologiesFacilities Council, Didcot, OX11 0QX, United Kingdom}
\address[UCL]{Department of Physics and Astronomy, University College London, Gower Street, London WC1E 6BT, United Kingdom}
\address[Caltech]{Lauritsen Laboratory, California Institute of Technology, Pasadena, California 91125, USA}
\address[Alabama]{Department of Physics and Astronomy, University of Alabama, Tuscaloosa, Alabama 35487, USA}
\address[ANL]{Argonne National Laboratory, Argonne, Illinois 60439, USA}
\address[Athens]{Department of Physics, University of Athens, GR-15771 Athens, Greece}
\address[NTUAthens]{Department of Physics, National Tech. University of Athens, GR-15780 Athens, Greece}
\address[Benedictine]{Physics Department, Benedictine University, Lisle, Illinois 60532, USA}
\address[BNL]{Brookhaven National Laboratory, Upton, New York 11973, USA}
\address[CdF]{APC -- Universit\'{e} Paris 7 Denis Diderot, 10, rue Alice Domon et L\'{e}onie Duquet, F-75205 Paris Cedex 13, France}
\address[Cleveland]{Cleveland Clinic, Cleveland, Ohio 44195, USA}
\address[Delhi]{Department of Physics \& Astrophysics, University of Delhi, Delhi 110007, India}
\address[GEHealth]{GE Healthcare, Florence South Carolina 29501, USA}
\address[Harvard]{Department of Physics, Harvard University, Cambridge, Massachusetts 02138, USA}
\address[HolyCross]{Holy Cross College, Notre Dame, Indiana 46556, USA}
\address[Houston]{Department of Physics, University of Houston, Houston, Texas 77204, USA}
\address[IIT]{Department of Physics, Illinois Institute of Technology, Chicago, Illinois 60616, USA}
\address[Iowa]{Department of Physics and Astronomy, Iowa State University, Ames, Iowa 50011 USA}
\address[Indiana]{Indiana University, Bloomington, Indiana 47405, USA}
\address[ITEP]{High Energy Experimental Physics Department, ITEP, B. Cheremushkinskaya, 25, 117218 Moscow, Russia}
\address[JMU]{Physics Department, James Madison University, Harrisonburg, Virginia 22807, USA}
\address[LASL]{Nuclear Nonproliferation Division, Threat Reduction Directorate, Los Alamos National Laboratory, Los Alamos, New Mexico 87545, USA}
\address[Lebedev]{Nuclear Physics Department, Lebedev Physical Institute, Leninsky Prospect 53, 119991 Moscow, Russia}
\address[LLL]{Lawrence Livermore National Laboratory, Livermore, California 94550, USA}
\address[LosAlamos]{Los Alamos National Laboratory, Los Alamos, New Mexico 87545, USA}
\address[Manchester]{School of Physics and Astronomy, University of Manchester, Oxford Road, Manchester M13 9PL, United Kingdom}
\address[MIT]{Lincoln Laboratory, Massachusetts Institute of Technology, Lexington, Massachusetts 02420, USA}
\address[Minnesota]{University of Minnesota, Minneapolis, Minnesota 55455, USA}
\address[Crookston]{Math, Science and Technology Department, University of Minnesota -- Crookston, Crookston, Minnesota 56716, USA}
\address[Duluth]{Department of Physics, University of Minnesota Duluth, Duluth, Minnesota 55812, USA}
\address[Ohio]{Center for Cosmology and Astro Particle Physics, Ohio State University, Columbus, Ohio 43210 USA}
\address[Otterbein]{Otterbein College, Westerville, Ohio 43081, USA}
\address[Oxford]{Subdepartment of Particle Physics, University of Oxford, Oxford OX1 3RH, United Kingdom}
\address[PennState]{Department of Physics, Pennsylvania State University, State College, Pennsylvania 16802, USA}
\address[PennU]{Department of Physics and Astronomy, University of Pennsylvania, Philadelphia, Pennsylvania 19104, USA}
\address[Pittsburgh]{Department of Physics and Astronomy, University of Pittsburgh, Pittsburgh, Pennsylvania 15260, USA}
\address[IHEP]{Institute for High Energy Physics, Protvino, Moscow Region RU-140284, Russia}
\address[Rochester]{Department of Physics and Astronomy, University of Rochester, New York 14627 USA}
\address[RoyalH]{Physics Department, Royal Holloway, University of London, Egham, Surrey, TW20 0EX, United Kingdom}
\address[Carolina]{Department of Physics and Astronomy, University of South Carolina, Columbia, South Carolina 29208, USA}
\address[SDakota]{South Dakota School of Mines and Technology, Rapid City, South Dakota 57701, USA}
\address[SLAC]{Stanford Linear Accelerator Center, Stanford, California 94309, USA}
\address[Stanford]{Department of Physics, Stanford University, Stanford, California 94305, USA}
\address[StJohnFisher]{Physics Department, St. John Fisher College, Rochester, New York 14618 USA}
\address[Sussex]{Department of Physics and Astronomy, University of Sussex, Falmer, Brighton BN1 9QH, United Kingdom}
\address[TexasAM]{Physics Department, Texas A\&M University, College Station, Texas 77843, USA}
\address[Texas]{Department of Physics, University of Texas at Austin, 1 University Station C1600, Austin, Texas 78712, USA}
\address[TechX]{Tech-X Corporation, Boulder, Colorado 80303, USA}
\address[Tufts]{Physics Department, Tufts University, Medford, Massachusetts 02155, USA}
\address[UNICAMP]{Universidade Estadual de Campinas, IFGW-UNICAMP, CP 6165, 13083-970, Campinas, SP, Brazil}
\address[UFG]{Instituto de F\'{i}sica, Universidade Federal de Goi\'{a}s, CP 131, 74001-970, Goi\^{a}nia, GO, Brazil}
\address[USP]{Instituto de F\'{i}sica, Universidade de S\~{a}o Paulo,  CP 66318, 05315-970, S\~{a}o Paulo, SP, Brazil}
\address[Warsaw]{Department of Physics, University of Warsaw, Pasteura 5, PL-02-093 Warsaw, Poland}
\address[Washington]{Physics Department, Western Washington University, Bellingham, Washington 98225, USA}
\address[WandM]{Department of Physics, College of William \& Mary, Williamsburg, Virginia 23187, USA}
\address[Wisconsin]{Physics Department, University of Wisconsin, Madison, Wisconsin 53706, USA}
\address[deceased]{Deceased.}

\author[FNAL]{P.~Adamson}
%\affiliation{\FNAL}
%\affiliation{\UCL}
%\affiliation{\Sussex}

\author[FNAL]{K.~Anderson}
%INCLUDED AS PER JIM HYLEN

%\author{C.~Andreopoulos}
%\affiliation{\RAL}
%\affiliation{\Athens}

\author[FNAL]{M.~Andrews}
%INCLUDED AS PER JIM HYLEN

\author[FNAL]{R.~Andrews}
%INCLUDED AS PER JIM HYLEN

\author[Iowa,ANL]{I.~Anghel}

%\author{K.~E.~Arms}
%\affiliation{\Minnesota}

%\author{R.~Armstrong}
%\affiliation{\Indiana}

\author[FNAL]{D.~Augustine}
%INCLUDED AS PER JIM HYLEN

\author[Cincinnati]{A.~Aurisano}

%\author{T.~H.~Fields}
%\affiliation{\ANL}

%\author{D.~J.~Auty}
%\affiliation{\Sussex}

\author[Stanford]{S.~Avvakumov}
% INCLUDED WITHOUT HAVING RECEIVED FEEDBACK FROM AUTHOR

\author[ANL]{D.~S.~Ayres}
% INCLUDED WITHOUT HAVING RECEIVED FEEDBACK FROM AUTHOR

%\author{C.~Backhouse}
%\affiliation{\Oxford}

\author[FNAL]{B.~Baller}

% ...

\date{\today}

%\maketitle

\begin{abstract}
This paper describes the hardware and operations of the Neutrinos at the Main Injector (NuMI) beam at Fermilab. It elaborates on the design considerations for the beam as a whole and for individual elements. The most important design details of individual components are described. Beam monitoring systems and procedures, including the tuning and alignment of the beam and NuMI long-term performance, are also discussed.
\end{abstract}


\end{frontmatter}

%\linenumbers

%\input{TitleFile}

%\tableofcontents 
%\newpage
%\newpage
\section{Introduction to The NuMI Beam}

The Neutrinos at the Main Injector (NuMI) neutrino beam \cite{numitdr, numitdh} was built at Fermilab to provide neutrinos for the MINOS \cite{PRD77} experiment, a long-baseline neutrino oscillation search, as well as for the COSMOS experiment \cite{cosmosref}, which was initially approved but subsequently withdrawn. Later, the NuMI beam was used for other experiments such as MINER$\nu$A \cite{MINERvA}, ArgoNeuT \cite{Argoneut}, and most recently the NO$\nu$A \cite{novaref} and the MINOS+ \cite{minosplusref} experiments. Neutrinos from NuMI have also been observed and studied by the MiniBooNE experiment at a large off-axis angle \cite{minibooneref}.

The NuMI beam facility produces neutrinos by steering a 120~GeV proton beam onto a narrow graphite target approximately 1~m in length through a collimating baffle. The produced hadrons are then focused in the forward direction and charge-sign-selected by two magnetic horns. Most of the hadrons subsequently decay into neutrinos (among other particles) in a long decay pipe. The resulting neutrino beam passes through dolomite rock and reaches the MINOS Near Detector (ND) 1.04~km downstream of the NuMI target. It then continues further north through the Earth's crust, encountering the MINOS Far Detector (FD) 734~km away in the Soudan Mine in Minnesota, and finally exiting the earth 12~km further north. This paper describes the NuMI beam hardware, operations and long-term performance. Its focus is the beam configuration in place during the data taking of the MINOS experiment, from May 20, 2005 until April 30, 2012. Developments after April 2012 are briefly discussed in Section \ref{sec:future}. The NuMI beam flux is described separately in a companion NuMI beam flux paper \cite{fluxpaper}.

\subsection{Historical Background} 

Starting gradually in the late 1980’s \cite{Davis, Haines, Kam0} and with ever-increasing frequency in the mid-1990’s \cite{Kam, Beier, Fukuda, IMB3, IMB, LSND, macro, Ambrosio, Soudan2a}, a plethora of experimental indications emerged suggesting that neutrinos change their flavor as they propagate through vacuum or matter. Thus a beam initially created as a pure muon-neutrino beam may develop an electron or tau neutrino flavor component as it propagates. If this phenomenon were indeed to be verified, it would profoundly affect our knowledge of neutrinos. Accordingly a number of experimental efforts were then being considered to investigate these issues under well-controlled experimental conditions which could be obtained with an accelerator-produced neutrino beam.

It was in the same time frame that the Main Injector accelerator was being constructed at Fermilab. This accelerator would use the Fermilab 8~GeV Booster as an injector and have the ability to accelerate protons up to 150~GeV\footnote{The Main Injector can operate at its design energy of 150~GeV but the optimal neutrino rate is obtained by running it at 120~GeV, taking advantage of a faster rep rate at this lower energy.}. The NuMI design specifications called for 3$\times10^{13}$ protons accelerated to 120~GeV every 1.87~s \cite{fnalnumitdh}. These parameters would allow creation of neutrino beams with much higher intensities than had been constructed previously and thus motivated several groups to seriously consider neutrino oscillation investigations.

Different groups proposed experiments to investigate different regions of the parameter space characterizing neutrino oscillations. The efforts that shaped the design of the NuMI beam involved studies of neutrinos over distances of the order of several hundred kilometers. There were three early proposals for such experiments based on a beam from the Main Injector; the one that had most influence on future developments proposed the use of the existing Soudan 2 Detector in the Soudan Underground Laboratory in Northern Minnesota \cite{P822ref, workshop1991}. This was followed by the more ambitious proposal \cite{MINOS} by the larger MINOS Collaboration to construct both a new beam at Fermilab and a new detector in the Soudan Laboratory. This proposal was approved and shaped the design of the NuMI beam.

\subsection{Design Considerations}
\label{sec:descons}

As a neutrino propagates, the flavor composition of its wave function oscillates; that is, the relative strengths of the component flavor amplitudes change as the particle travels through vacuum and/or matter. The probability that a neutrino with a flavor state $\alpha$ at birth will, after traveling a distance L, appear upon detection to have flavor $\beta$, is well-described by the two-flavor approximation: 
\begin{equation}
P_{\nu_{\alpha} \rightarrow \nu_{\beta}} = \sin^{2}2\theta \sin^{2}\biggl(\frac{1.27\Delta m^{2} L}{E}\biggr)
\label{equ:twoflavor}
\end{equation}
where the baseline $L$ is measured in km, the neutrino energy $E$ is measured in GeV, and $\Delta m^{2}$ is the mass difference squared between the two relevant mass states and is measured in eV$^{2}$/$c^{4}$. The first factor determines the amplitude of the oscillations, the second one their frequency. The effective mixing angle parameter  $\sin^{2}2\theta$ and the mass-squared difference  $\Delta m^{2}$ are functions of fundamental parameters characterizing the oscillations. The numerical factor 1.27 takes account of the units used in the above equation. 

For the oscillations to be significant $\Delta m^{2} L$ should be $\gtrsim E$ using the units above. Thus the optimum distance between the source and the detector depends upon the values of both $E$ and $\Delta m^{2}$. At the time of the NuMI beam design there were indications that $\Delta m^{2}$ lay somewhere in the range of $10^{-3} - 10^{-1}$eV$^{2}/c^{4}$ \cite{Beier, Fukuda}. A 120~GeV proton accelerator can efficiently produce a neutrino beam in the 1-10~GeV energy range.  Therefore a distance of a few hundred kilometers, such as the distance between Fermilab and Soudan, would sufficiently cover all possibilities. Furthermore once the distance was chosen, the only remaining free parameter in the design was the energy of the neutrinos. Flexibility in the energy of the beam was desirable to respond to new knowledge of $\Delta m^{2}$. 

A high-intensity beam was required to achieve a meaningful event rate at the MINOS detector several hundred kilometers away as the neutrino flux falls with the square of the distance from the decay point. Thus one of the primary design requirements for the beam line was the ability to withstand a high-power proton beam. The design beam power was somewhat optimistically set to 400~kW, a little above the maximum power that could be obtained from the Main Injector which was already pushing the beam technology at that time. A 400~kW beam power required 4$\times 10^{13}$ protons/cycle, whereas the Booster was capable of delivering 3$\times 10^{13}$ protons/cycle with 6 consecutive injections into the Main Injector. For a 2 s cycle time, this intensity corresponds to approximately 300~kW beam power. The use of ``slip-stacking'' techniques (discussed in more detail below) later allowed the number of protons/cycle to be increased, but at the cost of increasing the cycle time since more Booster batches need to be inserted. Thus the power does not increase as fast as protons/cycle.

Flexibility of the neutrino beam energy was achieved with a focusing system employing parabolic magnetic horns with adjustable separation between them and the target, which allowed tuning of the energy at which secondary particles were best focused \cite{Abramov:2001nr}. For low energies, since the divergence of the meson beam is greater and decay length shorter, wider downstream apertures and longer allowed decay paths are desirable. For higher energies the aperture area is less important but longer decay paths increase the flux. A relatively long (675~m) and not overly wide (2~m diameter) decay pipe geometry was chosen. In hindsight, the actual $\Delta m^{2}$ turned out to be relatively low and preferred a low-energy beam for oscillation measurements. Hence a beam designed today exclusively for MINOS would have a larger aperture and shorter decay flight paths. These considerations, together with the adopted values for the dimensions of the target chase and the decay pipe, are discussed more fully in Sections \ref{sec:hall} and \ref{sec:pipe}, respectively.

\section{NuMI Beam Components}
\begin{sidewaysfigure}
\begin{centering}
    \includegraphics[width=1.0\textwidth]{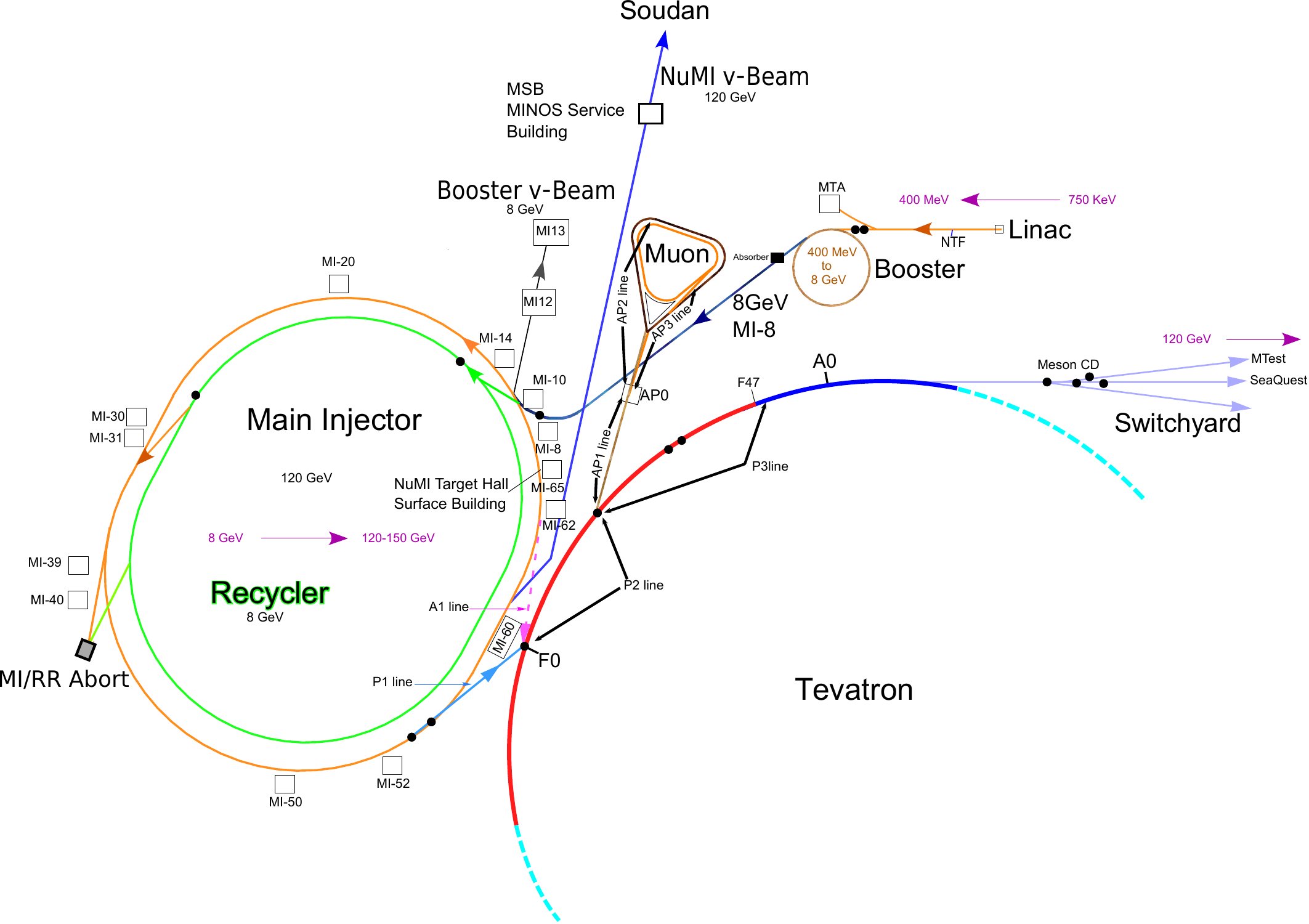}
    \caption{Fermilab Accelerator Complex. The proton accelerator cycle for the NuMI Beam starts with the Linac and is followed by the Booster and then the Main Injector. The Tevatron was operational during most of the MINOS run but was not used in neutrino production. The Recycler is used in the follow-up, post-MINOS experiments (see Sections \ref{sec:oper} and \ref{sec:future} for more operational details). A large number of beam lines shown were constructed for other experiments and are no longer in use or their function has changed. The AP1, AP2 and AP3 beam lines, AP0 target station and the ring named “Muon” formed the antiproton source which is no longer active and in the future some of these will be used for muon experiments. The P1 and A1 lines are proton and antiproton injection lines from the Main Injector to the Tevatron and are also no longer in use. The P2 and P3 lines use original Main Ring magnets and were part of the fixed target extraction complex. The squares labeled MI surrounding the Main Injector are various Main Injector service buildings.  }
\label{fig:complex}
\end{centering}
\end{sidewaysfigure}

The NuMI beam is presently the world's most powerful neutrino beam and is produced by the 120~GeV protons extracted from the Fermilab Main Injector. A layout of the accelerator complex at Fermilab is shown in Fig.~\ref{fig:complex}. Protons originate as H$^{-}$ ions in the Linac which accelerates them to 400~MeV. The ions are converted into protons in the Booster where they are accelerated to 8~GeV as 1.6~$\mu$s long batches with a 53~MHz bunch spacing. The Main Injector circumference is 7 times the circumference of the Booster. Thus the Main Injector can accommodate storage and acceleration of 6 Booster batches as one of these 7 “slots” is needed for the pulse kicker rise time. In the last few years a technique called “slip-stacking” was developed which allowed the combination of two batches into one after injection into the Main Injector. Use of this technique, together with other hardware improvements and better diagnostics, allowed a significant increase in  the proton intensity that was obtained \cite{Brown:2013idd}. 

During most of the MINOS run, one of the Main Injector slots was used by protons destined for the Antiproton Accumulator generating antiprotons for the Tevatron program, leaving five slots for MINOS and giving an 8~$\mu$s spill time. A spill is defined as one set of five or six proton beam batches accelerated together as described above. When the Accumulator was not operational all six slots were used for NuMI giving a 10~$\mu$s spill time. The cycle time for NuMI spills ranged from 2.1 to 2.4~s. The proton intensity during the MINOS run ranged from  2.2$\times 10^{13}$ protons on target (POT) in 2005 to approximately 3.6$\times 10^{13}$ POT in 2012. 

%\begin{figure*}

The protons destined for the NuMI beam line are extracted, bent downward to point at the MINOS Far Detector, and transported 350~m to the NuMI target. The global positioning system (GPS) was originally used to define the beam direction. The protons are incident on the graphite target and the produced hadrons are focused by two magnetic horns and then enter a 675~m long decay volume. The horns allow preferential selection of hadrons of one or the other charge sign. Pions and kaons constitute a major portion of the hadrons and predominantly decay via the modes $\pi^{+}\rightarrow \mu^{+} +\nu_{\mu}$ and $K^{+}\rightarrow \mu^{+} +\nu_{\mu}$ yielding a $\nu_{\mu}$ beam. There is also a few percent $\bar{\nu_{\mu}}$ component coming from negative hadrons and a small contamination of electron neutrinos ($\nu_{e}$) due to subdominant electronic decay mode of $K^{+}$ hadrons, decays of $K^{0}$ particles, and decays of tertiary muons \cite{fluxpaper}.

\begin{figure}
\begin{centering}
    \includegraphics[width=1.0\textwidth]{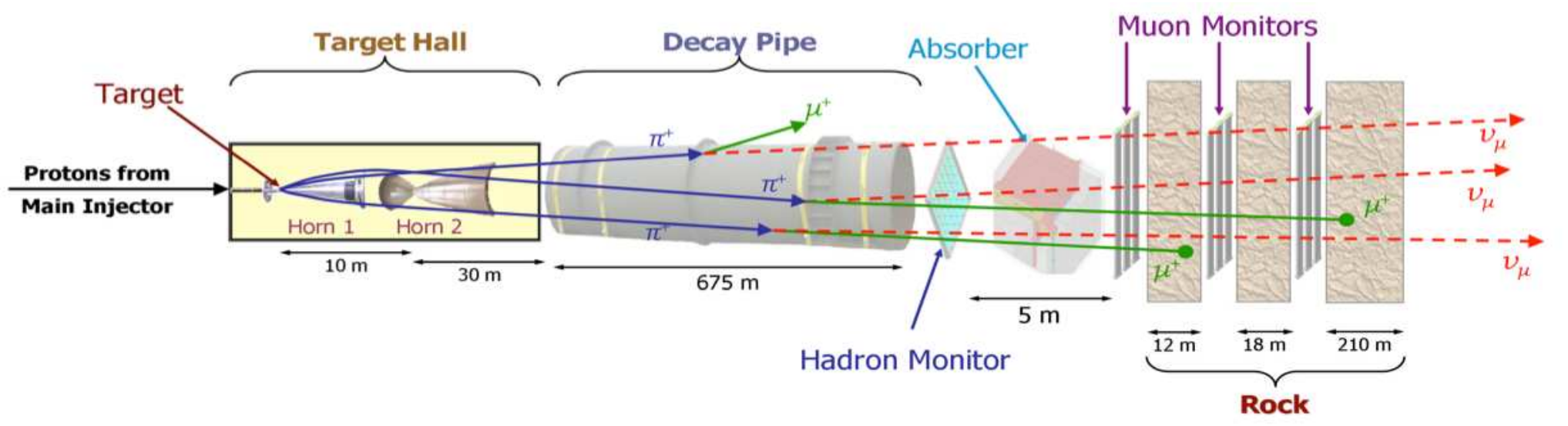}
    \caption{Schematic of the NuMI Beam. The individual components of the NuMI beam (not to scale) are shown together with the relevant dimensions. All the important elements are shown, including the target, the horns, the decay pipe, the hadron absorber, and the so-called muon shield which consists of the dolomite rock preceding the MINOS Near Detector. }
\label{fig:beamschematic}
\end{centering}
\end{figure}

A hadron monitor is located at the end of the decay volume just in front of the 5~m thick absorber to record the profile of the residual hadrons. These residual hadrons are attenuated to a negligible number by the absorber. Four alcoves have been excavated in the rock just downstream of the absorber and are used to house three muon monitors allowing measurement of the residual muon flux with three different threshold energies\footnote{The fourth alcove was not instrumented during MINOS running.}. The 240~m of rock following the absorber stops the muons remaining in the beam but allows the neutrinos to pass. After 240~m a cavern has been excavated to house the MINOS Near Detector. The cavern subsequently housed additional experiments such as MINER$\nu$A or ArgoNeuT, taking advantage of the high neutrino flux at that location. The schematic of the NuMI beam is shown in Fig.~\ref{fig:beamschematic}. The individual beam components are described in more detail in the sections below.

\subsection{The Primary Beam Line}
The primary beam line is a transfer line carrying the 120~GeV protons from the Main Injector to the NuMI target. 
There were two central design principles for the NuMI proton beam line \cite{NuMIBeam}: safe and low-loss transmission of a very high-power proton beam and accuracy and stability of targeting. Fractional losses over the 350~m beam line were required to be kept below $10^{-5}$.  The physics of the MINOS experiment required the beam to have an angular stability of $\pm60$~$\mu$rad, and a positional stability of $\pm250$~$\mu$m at the target. Typical operational values achieved were fractional beam loss prior to the target profile monitor of 3$\times10^{-7}$, angular stability of $\pm 15$~$\mu$rad, and positional stability of $\pm 100$~$\mu$m.

The proton beam is extracted from the Main Injector accelerator using ``single-turn'' extraction. 
A single kicker bends the beam a small angle into the primary beam line, also known as an ``extraction channel''.  The magnetic field in the kicker changes from zero to its full value in 700~ns which is less than the length of the extraction gap left in the beam. The entire beam is delivered in 10~$\mu$s, producing a high instantaneous rate in the MINOS Near Detector.  An alternative technique, resonant extraction, would allow a much slower spill (about 1~ms long) but would lead to unacceptable irradiation of the Main Injector tunnel downstream of the extraction point. In resonant extraction the tune of the beam is slightly changed so that the beam expands towards an electrostatic septum on the side, usually an array of a number of wires whose field provides the initial outward deflection of the beam. The mass of the septum is kept as low as possible to minimize proton interactions but it was estimated that the minimum achievable loss would be 1-2~\% of the beam which was much too high to be tolerable. With single-turn extraction, fractional beam loss is maintained at part per million levels, as needed for the 400~kW design NuMI beam. Accordingly, it was necessary to adopt single-turn extraction even though it necessitated more sophisticated electronics in the MINOS Near Detector. Single-turn extraction was accomplished by a kicker consisting of three pulsed magnets with an integrated field of 0.36~T$\cdot$m, followed by three standard Main Injector Lambertson magnets and then a standard C magnet \cite{fnalnumitdh}. 

\begin{figure}
\begin{centering}
    \includegraphics[width=1.0\textwidth]{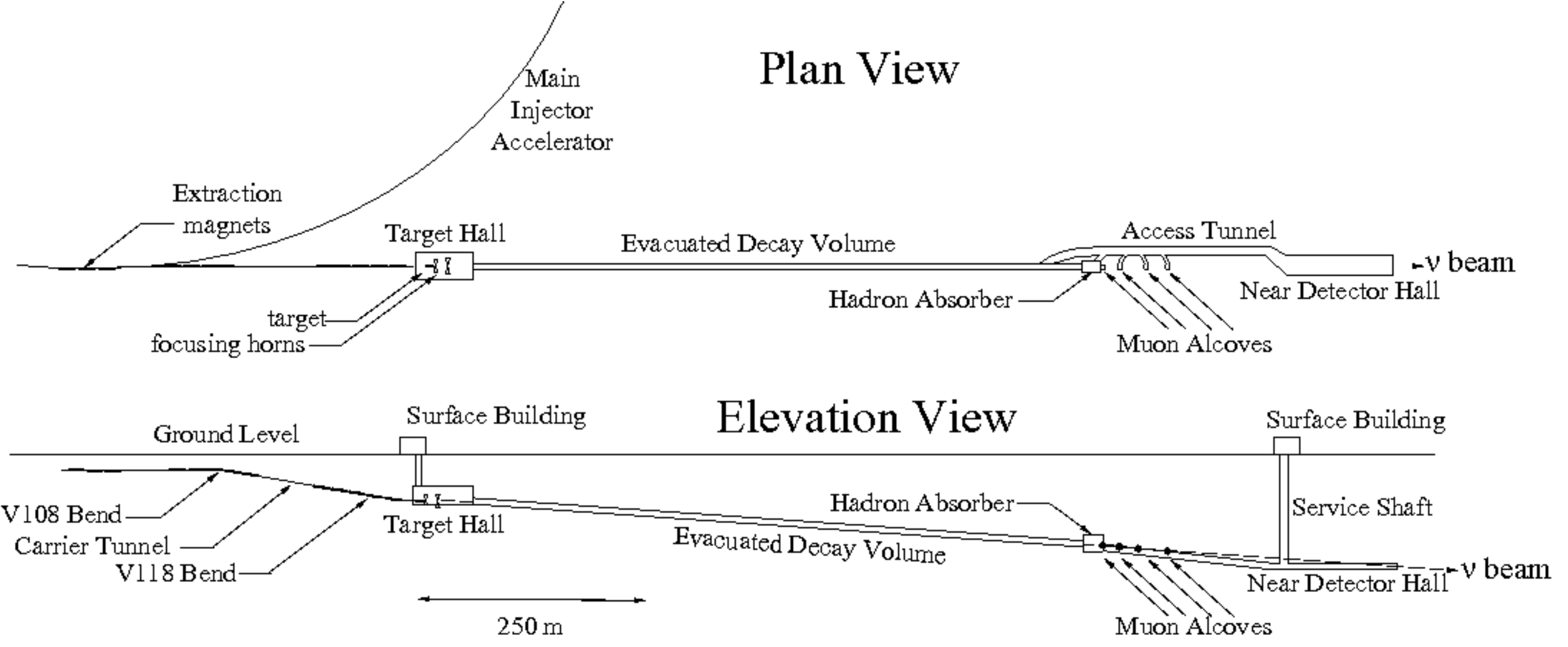}
    \caption{Plan and Elevation Views of the NuMI Beam Facility. The proton beam is directed onto a target, where subsequently the secondary pions and kaons are focused into an evacuated decay volume (later filled with helium) via magnetic horns. Ionization chambers at the end of the beam line measure the secondary hadron beam and tertiary muon beam.  }
\label{fig:beamdescr}
\end{centering}
\end{figure}

To point towards the Soudan Laboratory the proton beam had to be inclined downward by 58~mrad.  One advantage of using the MI-60 location for extraction was that only a 59~mrad horizontal bend was necessary to aim directly at Soudan. The overall trajectory of the beam line for the first 100~m was constrained to a horizontal plane by having to fit between the Main Injector magnets and Recycler magnets. The required trajectory in both planes in this part of the beam line was achieved by six dipoles of varying rotation angles. The subsequent initial vertical trajectory was determined mainly by issues of cost and constructability. It was preferable to do most of the construction in solid bedrock rather than soil overburden, and to minimize the path through the water table. Thus the beam was initially overbent downward by 156~mrad (V108 in Fig.~\ref{fig:beamdescr}), then it was bent back upward by 98~mrad (V118 in Fig.~\ref{fig:beamdescr}) giving a final vertical angle of 58~mrad or 3.343$^{\circ}$ downwards\footnote{The precise number is 3.34349$^{\circ}$, however, for this level of accuracy the location on the Fermilab site has to be specified as the curvature of the earth becomes important. The number 3.34349$^{\circ}$ corresponds to the angle of the beam in the Target Hall in local gravity coordinates.} through the Earth towards the Soudan Mine in Minnesota. The bending of the beam was achieved using six, and then four, refurbished Main Ring dipole magnets for the two stages. The regulation of the power supplies for the dipole magnets was 50~ppm to achieve the beam stability specifications.

The upper 65~m of the beam line that passes through the soil and soil/bedrock interface, referred to as the carrier tunnel, is uninstrumented and contains a vacuum tube. The proton beam line utilizes 21 standard Fermilab quadrupoles whose functions are to match the Main Injector optics, to optimize the transported beam for very low beam loss, and to control the size, divergence and dispersion of the beam at the target. All proton line magnets are ramped for each beam cycle to reduce the overall power consumption. The plan and elevation views of the NuMI beam line are shown in Fig.~\ref{fig:beamdescr}.

\begin{figure}
\begin{centering}
    \includegraphics[width=.96\textwidth]{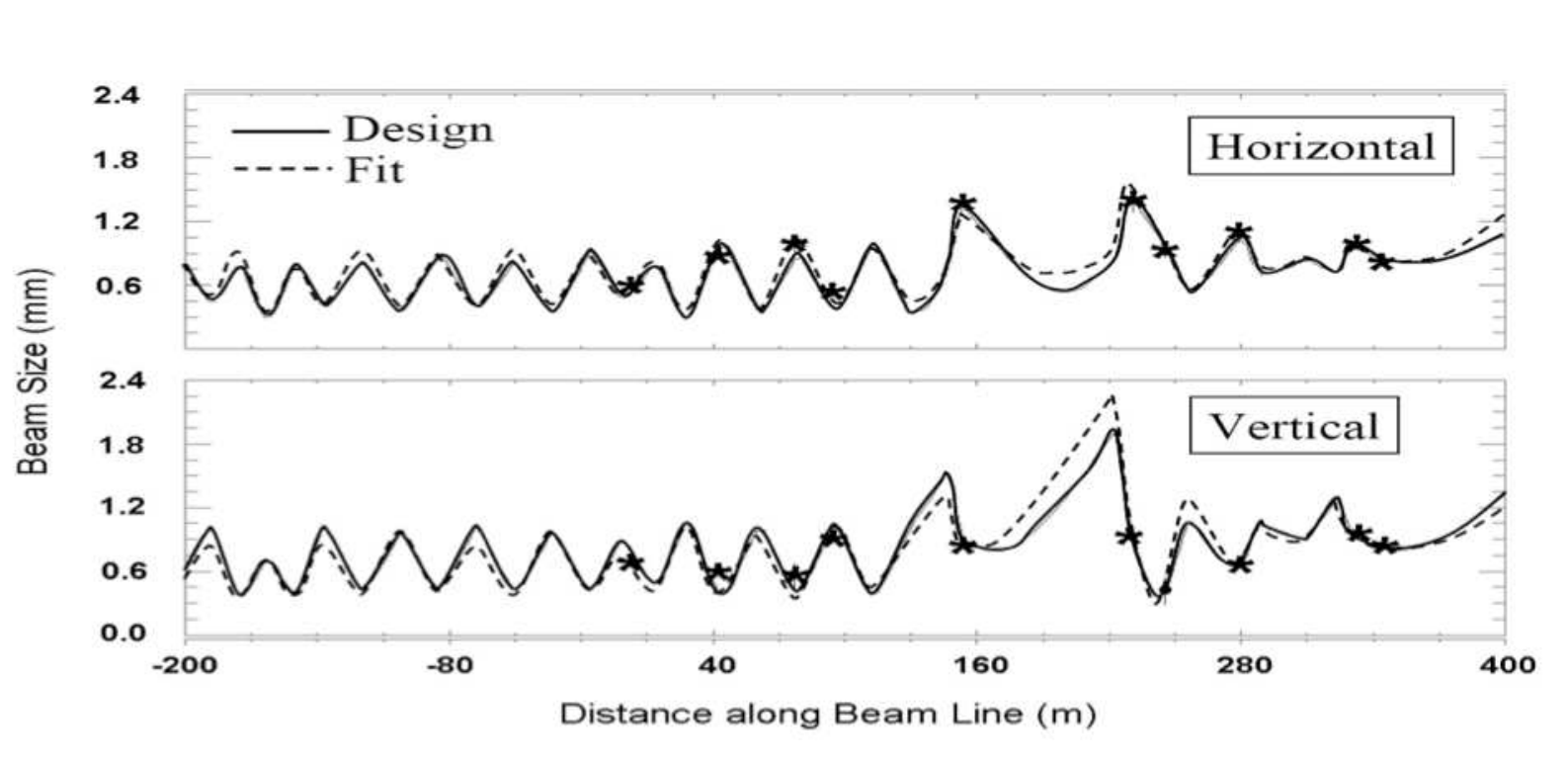}
    \caption{Measured and Calculated Beam Envelopes. The envelopes from the kicker magnets in the Main Injector to the target are shown. The target is located at 356~m in this plot, about 10~m downstream of the last measurement point. The measured envelopes are displayed as data points, and the calculated ones are shown as lines. They are in good agreement with each other. }
\label{fig:figSamC}
\end{centering}
\end{figure}

The very low loss requirement mandates that the NuMI proton beam envelope stays clear of all aperture limitations. Thus, the beam line was designed with larger acceptance than that for the largest beam emittance that could be accelerated in the Main Injector. Furthermore, an automated system measured and corrected the beam trajectory pulse-to-pulse. The beam line was equipped with 19 trim dipole magnets, 10 horizontal and 9 vertical, to provide precise direction corrections along the proton path. An extensive system of beam position monitors, discussed in Section \ref{sec:protmon}, provided the position information for these corrections. The beam spot at the NuMI target is approximately circular with a measured diameter of 1.1-1.2~mm ($\sigma_{x}=1.1$~mm and $\sigma_{y}=1.2$~mm). 
Fig.~\ref{fig:figSamC} shows the calculated proton beam envelope compared with measured beam widths. A more detailed discussion of the proton beam line is provided in \cite{NuMIBeam}.

\subsection{The NuMI Target Hall}
\label{sec:hall}
The NuMI Target Hall consists of a large underground cavern which contains the shielding and all the support modules for the target, horns, and other major beam components.  The design power of 400~kW dictates extensive shielding and that the components withstand high levels of radioactivity, heating, and thermal shock.  In considering the design for an enclosure for these systems, several things needed to be taken into account, including the prompt radiation occurring when the beam strikes the target, the residual radioactivity, and the activation of the surrounding air, soil, and ground water.  In addition, provision had to be made for remote handling of radioactive components for repair and replacement.  Temporary shielded storage is therefore provided in the Target Hall for used beam components before they are moved to more permanent storage locations elsewhere at Fermilab. The NuMI Target Hall has been designed to allow for the safe running and maintenance of all the beam components.

\begin{figure}
\begin{centering}
    \includegraphics[width=.96\textwidth]{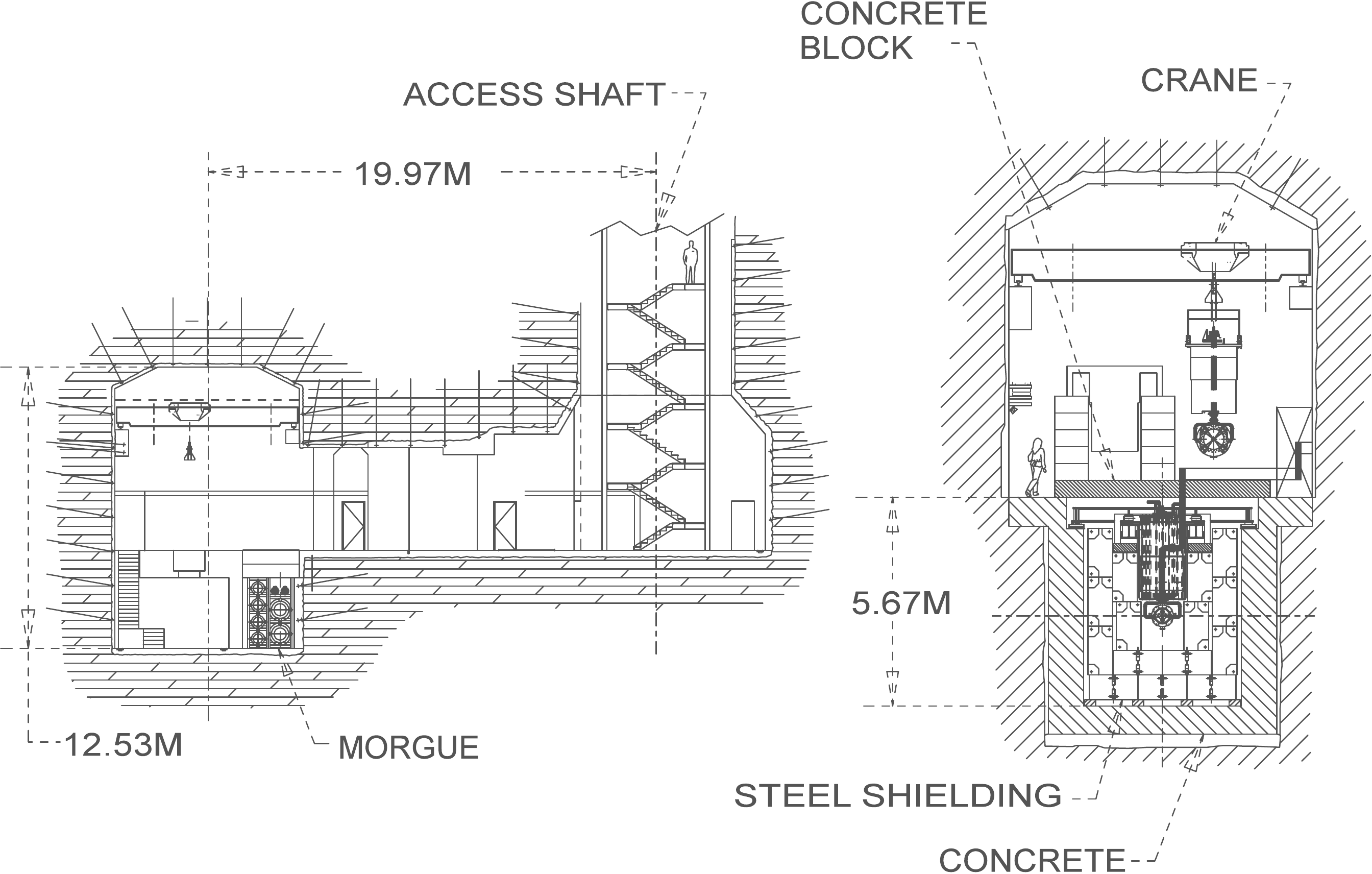}
    \caption{The NuMI Target Hall Cross Section. The left part of the Figure shows the Target Hall cross section including the access shaft, stairway and the auxiliary rooms and the morgue used to temporarily store removed beam components. The right part details the Target Hall itself including shielding, the target chase with the target, and also the crane used for moving beam components.}
\label{fig:targethallxsec}
\end{centering}
\end{figure}

\begin{figure}
\begin{centering}
    \includegraphics[width=.96\textwidth]{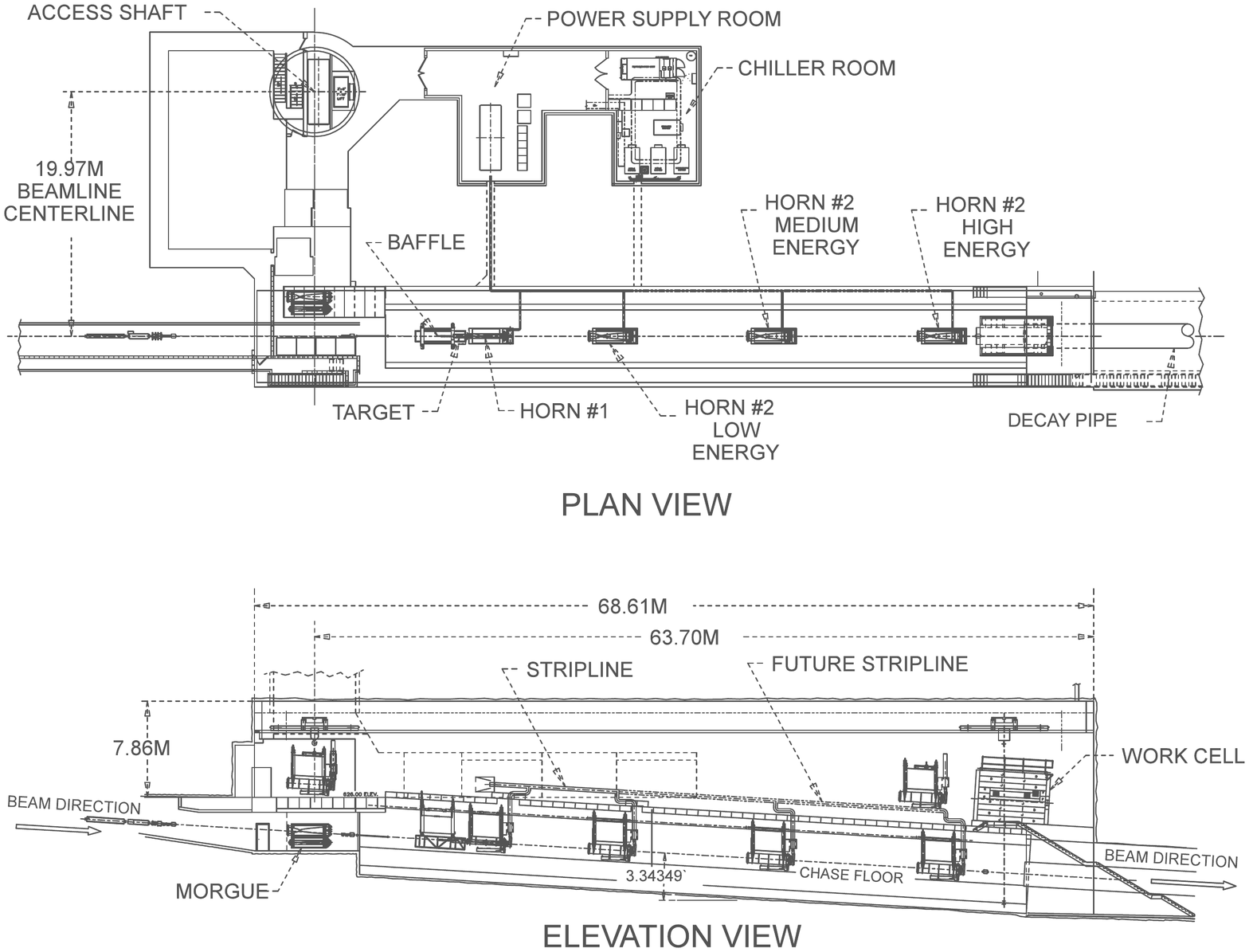}
    \caption{The Plan and Elevation Views of the NuMI Target Hall. The locations of the principal beam elements are shown as well as alternate Horn~2 locations for potential higher energy configurations. Also shown are the locations of the morgue and the Work Cell.}
\label{fig:targethallviews}
\end{centering}
\end{figure}

The Target Hall is located approximately 41~m underground in the dolomite rock formation and is a long domed chamber, approximately 69~m long, 8.1~m wide, and 12.5~m high. This depth was chosen so as to provide at least 6~m of rock foundation for the target and the associated equipment, and enough rock above the Target Hall so the roof would be self-supporting. Fig.~\ref{fig:targethallxsec} shows cross-sectional views of the Hall including its extension at the upstream end, and Fig.~\ref{fig:targethallviews} shows plan and elevation views of the Target Hall complex. The beam line extends through a pit about 6~m wide and 6~m deep running for over 50~m up to the decay pipe, which protrudes 0.35~m into the downstream end of the pit. The pit follows the beam direction at a downward angle of 3.343$^{\circ}$ to the horizontal. 

The pit serves as the housing for the baffle/target and the focusing horns together with their support structures and auxiliary equipment. It is lined at the sides and the bottom with a layer of concrete shielding to provide a tunnel 4.7~m deep and 4.3~m wide. The pit cross-section is shown in Fig.~\ref{fig:targethallxsec}. One of the elements shown is steel shielding, which is installed in this tunnel to form a central ``chase'' with a rectangular cross-section, 1.2~m wide and 1.3~m high (essentially a beam passageway). Beam line components and instrumentation are installed in this chase; the chase also serves as a channel for the recirculation of chilled air from an air cooling system. Multiple configurations of beam components were envisioned, the highest energy configuration having a separation between horns of up to 40~m. Thus a 60~m total length of the Target Hall was developed. The steel shielding is composed of a total of 633 Duratek 10~ton steel blocks\footnote{Duratek BluBlocks cast by Duratek Inc.\ at Oak Ridge National Laboratory from scrap iron and steel.}, with each block having the dimensions 1.33~m $\times$ 1.33~m $\times$ 0.67~m. The steel blocks are stacked in an interlocking manner and layers are staggered with respect to each other to avoid aligned gaps. 46~cm thick concrete covers, in 0.9~m longitudinal segments, are used to provide shielding on top. The Target Hall also contains a stripline for powering the horns.

The calculated heat load from the beam in the normal NuMI beam low energy configuration into the target pile is 158~kW. The removal of this heat and additional heat from horn current and electrical devices is important for the correct functioning of the beam. This function is provided by a recirculating air cooling system designed for 240~kW of cooling. In addition the system provides dehumidification and a filter to remove particulates down to 0.3~$\mu$m in size. Since maintaining alignment of the target and horns with the primary proton beam is critical, the entire chain of mounting those components to the bedrock must control thermal expansion effects. 

The air-cooling equipment, dehumidifiers and filters are installed alongside the downstream end of the chase. Air at rock temperature is supplied at that end to the space between the concrete shield and the steel inner shielding; this air flows through there to the upstream end of the target pile. This maintains the thermal stability of the concrete. The horns and targets are mounted at beam height on water-cooled hangers, which hang on low thermal expansion rods. The rods penetrate through the support modules and are attached at the top of the modules. Those “ears” at the top of the modules rest on I-beam structures that are supported by the concrete. All the shielding steel, by contrast, is supported by the bottom of the target pile, and expands/contracts independently of the target and horn supports. Controlled width cracks are set at the top of the T-block\footnote{A T-block is a specially shaped shielding block that has a steel bar across the top creating a T-shape; this bar is used to rest the T-Block on the shielding walls.} shielding and modules so that some of the circulating air is cooling the stripline, rods, T-blocks etc. The air is directed from the top of the modules into central shielding to minimize radioactive contamination in places used by personnel when they need to work on the modules. The air turns around at the upstream end of the target pile, and then comes downstream through the center of the chase, passing by the target and the horns and removing the heat from the steel shielding.  At the downstream end the air is filtered, chilled and dehumidified, and then pushed by a fan back for another pass. 

The Target Hall walls support a crane of 30~ton load capacity hanging from guiding tracks which allows the lifting and moving of NuMI beam components and shielding. The crane can travel in the direction of the beam between 2.4~m and 64.3~m from the upstream face of the Target Hall. The travel of the crane trolley in the transverse direction is about 5.5~m. A system of ten video cameras allows the safe operation of the crane from a remote location and allows the precision placement and lifting of crane loads. The video signals are transmitted by wireless and wired connections to receivers at the upstream end of the Target Hall.

The remote crane operation is essential for repair and replacement of target and horn assemblies. Used beam elements are moved as a unit with their support modules to the Work Cell using the overhead crane. The Work Cell is a well-shielded facility, located at the downstream end of the Target Hall, that allows installation, inspection, and, if necessary, replacement of beam components. In some cases it has been possible to repair failed targets and horns. The Work Cell is shielded by 0.9~m of concrete on its east and west sides and 0.3~m of steel on the north and south. The steel wall on the south side is remotely movable by sliding it to the east. There are four apertures, three in the east wall and one in the north, which are filled with lead-glass blocks allowing the viewing of the beam elements in the cell during remote operations. 
Removed elements are temporarily stored in a well-shielded location dubbed the ``morgue'', within the target shield pile. The elements can later be recovered after a certain cool-down period, and sent to long-term storage.  Transportation out of the Target Hall or other storage facilities requires specially made shield casks to reduce any incidental radiation exposure.

The primary personnel access to the Target Hall is through the elevator at the upstream end of the Hall. A stairway provides alternative access. The major pieces of equipment are lowered into the Hall with a crane through a 36.6~m deep shaft. The shaft is close to, but not situated directly over the main Target Hall chamber. The carrier tunnel on the upstream end and a 670~m long passageway along the decay pipe provide additional emergency exit paths. The Target Hall is a very high radiation area, especially when the beam is operating, and as such it is under a personnel safety interlock system that is failsafe and redundant. This system prevents personnel access during beam operation to the Pre-target and Target Hall enclosures. The power supplies, the control systems and associated equipment are located in an adjoining enclosure outside of the interlocked system; they are accessible during beam operation. 

\subsection{The Baffle}
The design power of the NuMI beam is so high (up to 400~kW) that even a relatively small mis-steering of the beam could cause significant damage to the beam components. A single pulse of mis-steered beam could destroy a primary beam magnet or a Target Hall component such as a focusing horn. Especially vulnerable are the target cooling and support components and the magnetic horns whose narrowest apertures (referred to as “necks”) are not much larger than the nominal beam size at their locations. The construction of horns, from a few millimeter thin aluminium, makes them fragile enough that they have to be protected from any large sudden energy deposition. To provide such protection a specially designed device, referred to as the baffle, has been constructed and installed just upstream of the target. 

\begin{figure}
\begin{centering}
    \includegraphics[width=.9\textwidth]{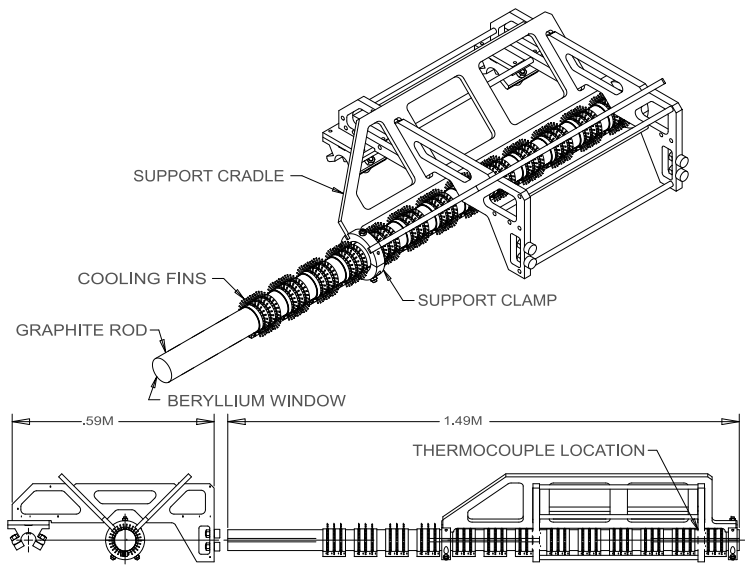}
    \caption{The NuMI Beam Baffle. Isometric and cross-sectional views are shown of the baffle which is 150~cm long, and has a 57~mm diameter graphite core with an 11~mm diameter bore through the middle. It is encased by 60~mm diameter aluminium tube. }
\label{fig:baffle}
\end{centering}
\end{figure}

The baffle consists of a graphite core, 1.5~m long along the beam direction and 57~mm in diameter, encased in a 60~mm O.D. aluminium tube. At the center is an 11~mm diameter hole through which the proton beam passes. The baffle is designed to degrade mis-steered beam enough that the target and horns are not damaged. It is designed to withstand, without damage, the full intensity beam for a short time (a few pulses) until mis-steering can be detected and the beam shut off. Thermocouples installed on the baffle are connected to an interlock which will turn the beam off if significant mis-steering is detected as excessive heat.  This monitoring also provides a measure of how much beam is ``scraping'', or hitting, the inside walls of the baffle.  Both monitoring functions are provided by measuring the temperature at the downstream end of the baffle.  The baffle is designed to be able to operate continuously with up to 3~\% of beam scraping at design luminosity, and measure scraping with an accuracy of at least $\pm1$~\% of the proton beam. The baffle hole is approximately 5~$\sigma$ in terms of the typical beam spot size, so for properly steered beam only non-Gaussian tails of the beam are a source of baffle scraping. During normal operation, the heating of the baffle has corresponded to an estimated 0.6\% of beam scraping, but this is an overestimate as heating from backscatter radiation from the target has not been subtracted.  The baffle is mounted on the same carrier as the target so that they are moved together when the beam configuration is changed. The baffle has beryllium windows that would contain the radioactivated graphite if it powdered.  It also has a ventilation hole which prevents air pressure build-up. A schematic of the baffle is shown in Fig.~\ref{fig:baffle}.

\subsection{The NuMI Target}
\label{sec:target}
The target \cite{Abramov:2001nr} is one of the more delicate elements in the NuMI beam line. The target must be able to withstand the 400~kW design power without disintegrating, while maximizing the produced flux of hadrons and hence the neutrino yield. To maximize the neutrino flux, most of the proton beam should be intercepted in the target in the smallest possible volume so as to minimize the number of subsequent secondary meson interactions. This therefore requires small transverse target dimensions. On the other hand, the target is made more robust by enlarging both the beam size at the target and the target itself, reducing volumetric energy deposition and gradients and maintaining a high fraction of the beam striking the target.  

The target is made of graphite of the type ZXF-5Q (POCO graphite)  with a density of 1.78~g/cm$^{3}$. It consists of 47 fins, each of which is 20~mm long (along the beam direction), 15~mm tall, and 6.4~mm wide; the fins are spaced 0.3~mm apart giving a total target length of 95.38~cm. The fins are brazed in vacuum to two stainless steel pipes which conduct the water coolant; the pipes' external diameter is 6~mm and wall thickness 0.2~mm. In addition to those fins, a $48^{th}$ segment is mounted horizontally in the target canister 15.73~cm upstream of the main target. The vertical alignment for most of the targets was set by centering the beam in the baffle and relying on the surveyed slope\footnote{For the first NuMI target (NT-01) the $48^{th}$ segment was mounted 2.26~mm above its vertical center as the target tube was bent at the connection between the tube and the base shortly before installation. This meant that the cross fin, which is mounted to the main can rather than the tube, was misaligned and 2.26~mm high. For subsequent targets, a protection fixture for target tips was built that would get removed just before installation in the Target Hall; this resulted in no other bent target tubes. Consequently, the misalignment for the second target (NT-02) was only 0.3~mm according to survey, and for the third target (NT-03) it was 0.9~mm.}. This extra fin is used in the beam-based alignment procedure as discussed in Section \ref{sec:beamalign}. For low energy running the target was inserted 50.4~cm into the first horn to obtain the maximum flux possible in the 1-3 GeV range. In this configuration the clearance between the last target fin and the horn conductor is just a few mm.

The target structure was analyzed using MARS \cite{MARSref} for energy deposition and finite-element mechanical modeling. Mechanical stresses, temperatures, and cooling capability were evaluated for the highest proton intensities envisaged. The calculated maximum instantaneous temperature rise was 288$^{\circ}$C giving a maximum temperature after a beam spill of 344$^{\circ}$C. Stresses in the target have been calculated not to exceed 25.6~MPa, compared to 36~MPa, the estimated fatigue limit of the graphite used. The heat load is estimated as 3.04~kW giving a water temperature rise of 12$^{\circ}$C assuming a water flow rate of 3~m/s.

\begin{figure}
\begin{centering}
    \includegraphics[width=.82\textwidth]{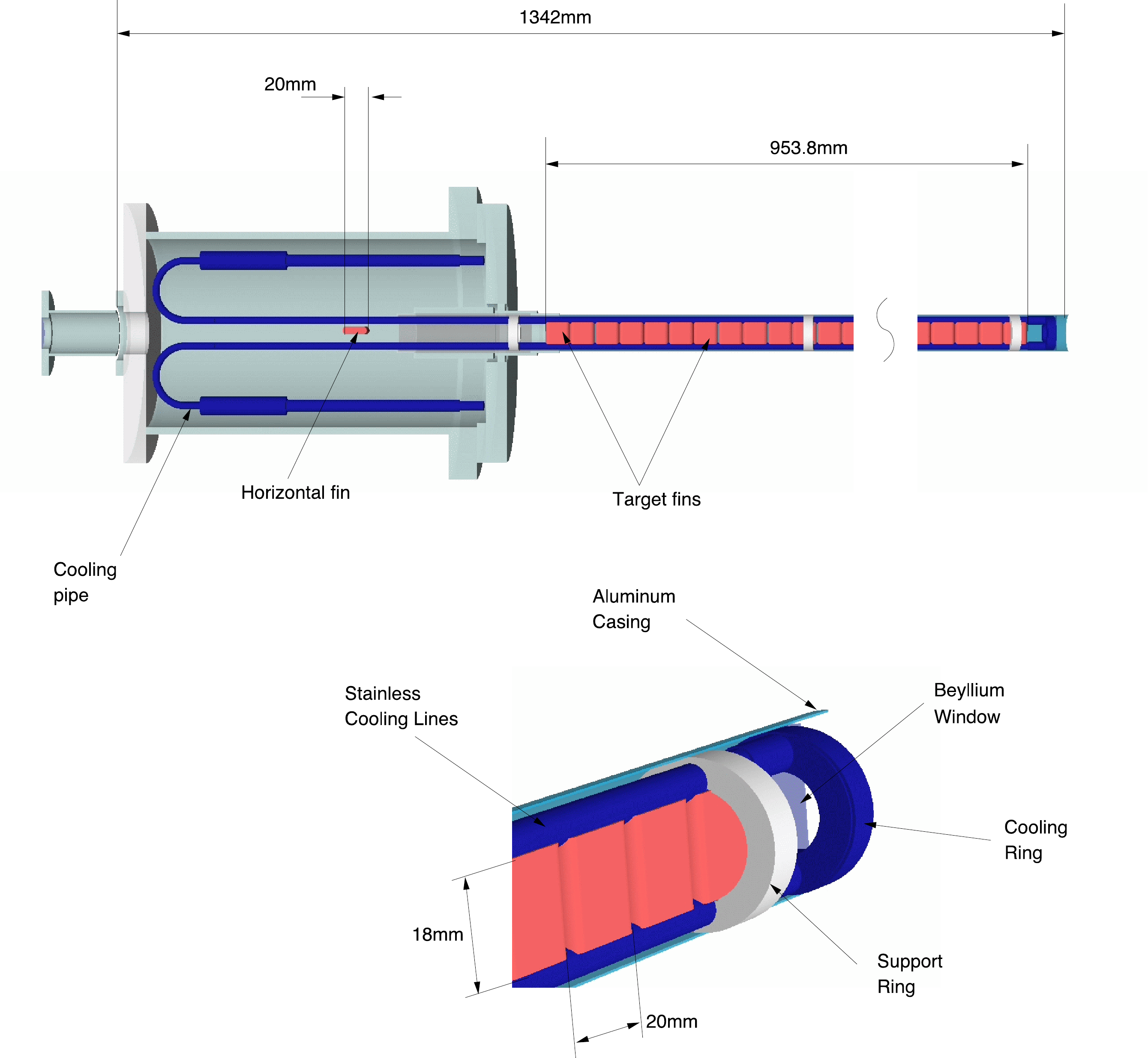}
    \caption{Longitudinal Cross-Section of the NuMI Target and the Target Canister. }
\label{fig:target}
\end{centering}
\end{figure}

Beryllium windows are used at the entrance and exit of the target vessel to protect the graphite structure. The target canister and casing comprise a vessel which provides the primary containment for the target. The target canister is made of a solid piece of aluminium alloy, and surrounds the volume upstream of the target. The target casing is a thin aluminium tube surrounding the target segments. The target can be evacuated, but is normally operated in helium gas somewhat above atmospheric pressure. Anodized aluminium spacers provide electrical isolation of the target fins and the cooling tubes from the target casing and the first horn to prevent any discharge between the target and the horn inner conductor. The target and target canister design are shown in Fig.~\ref{fig:target}.

 A horizontal Budal monitor \cite{Budal} consisting of a wire connected to the cooling tube allows the measurement of delta-ray charge knocked out of the target when it is hit by the beam. %The target fins plus cooling tube are electrically isolated from the casing by ceramic spacers and water line breaks. 
The horizontal Budal monitor allows the beam to be scanned horizontally across the target to check the target position in this dimension. Another Budal signal readout is located on the additional 48$^{th}$ fin and provides a position check for beam scans in the vertical direction. The cooling for this segment is provided by conduction through its clamping plates to the canister. The clamping plates are anodized to provide the required electrical isolation.

\begin{figure}
\begin{centering}
    \includegraphics[width=.8\textwidth]{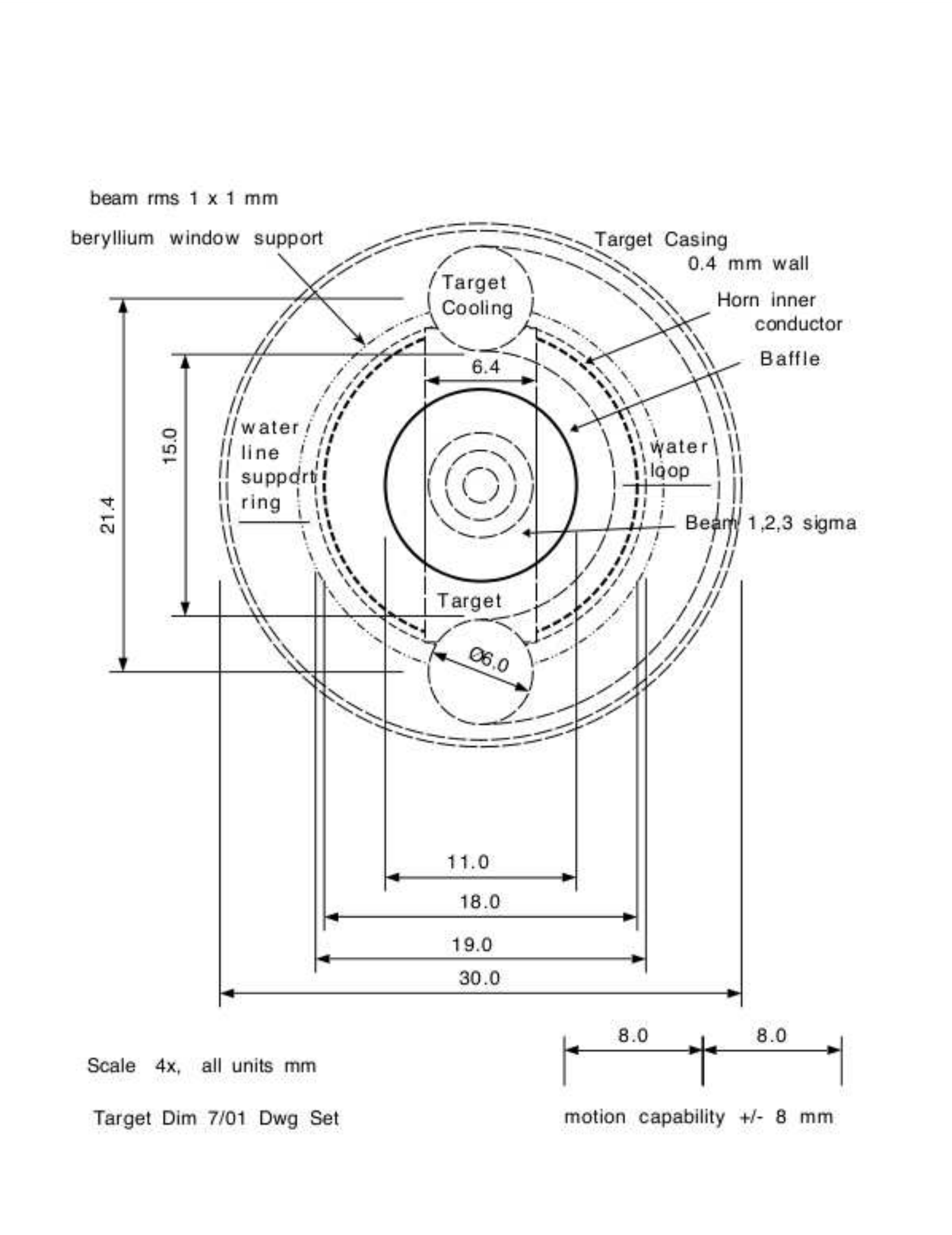}
    \caption{Beam's Eye View of the Baffle Inner Aperture. This figure shows what the proton beam sees as it travels through the NuMI baffle and hits the target. Superimposed on the diagram are the beam spot, the target fin, the horn neck, and the target cooling and support structure. All dimensions are in mm. }
\label{fig:beamspot}
\end{centering}
\end{figure}

Fig.~\ref{fig:beamspot} shows a beam-eye view of the baffle inner aperture superimposed on the beam spot, the target fin, the horn neck and the target cooling and support structure. 

\subsection{The Target/Baffle Carrier}
\label{sec:targetcarrier}
Studies of beam composition \cite{fluxpaper} require the ability to measure neutrino energy spectra in different beam configurations with different energies. Thus it was found very useful to build in a system that would allow energy changes with very little down-time, which in turn implies that the changes would have to be done remotely without access to the radiation areas \cite{VariableEnergy}. The movable carrier system that was adopted accomplishes this with about a 25\% penalty in flux compared to the more extensive beam modifications required for optimized higher energy configurations discussed in the Section \ref{sec:horns} \cite{MarinoMorfin, VariableEnergy}. In this method the two horns remain in their standard low energy positions but the target is moved with respect to the first horn.

\begin{figure}
\begin{centering}
    \includegraphics[width=.9\textwidth]{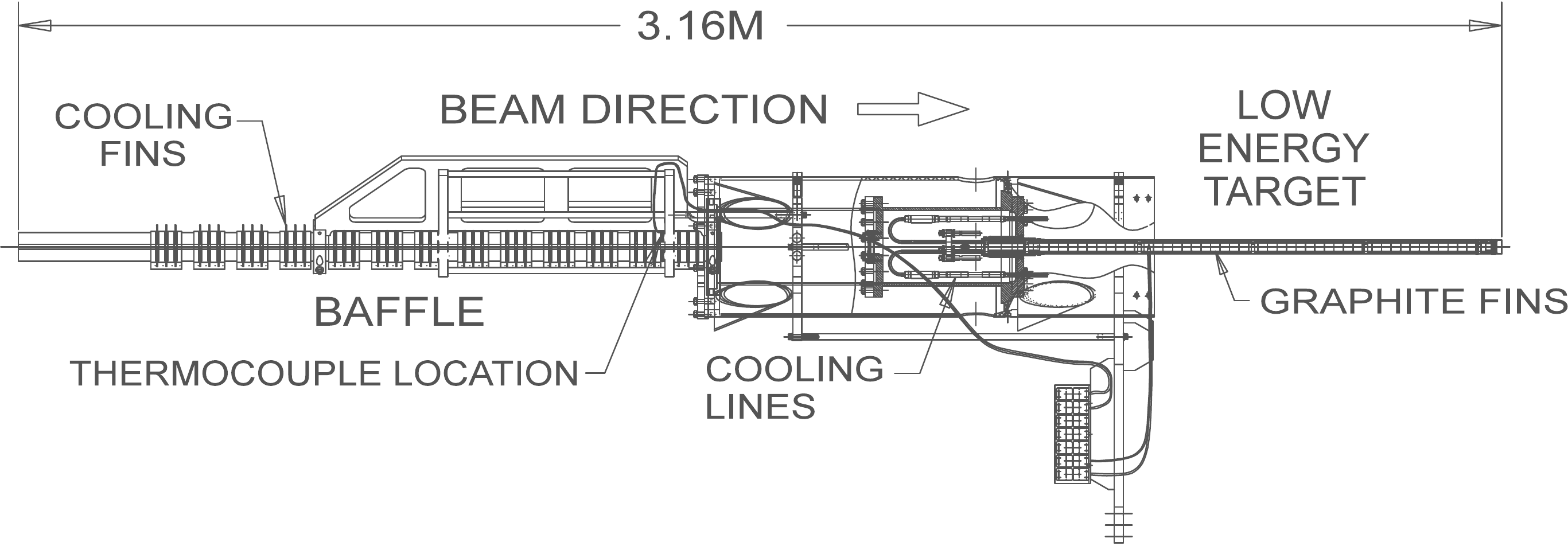}
    \includegraphics[width=.72\textwidth]{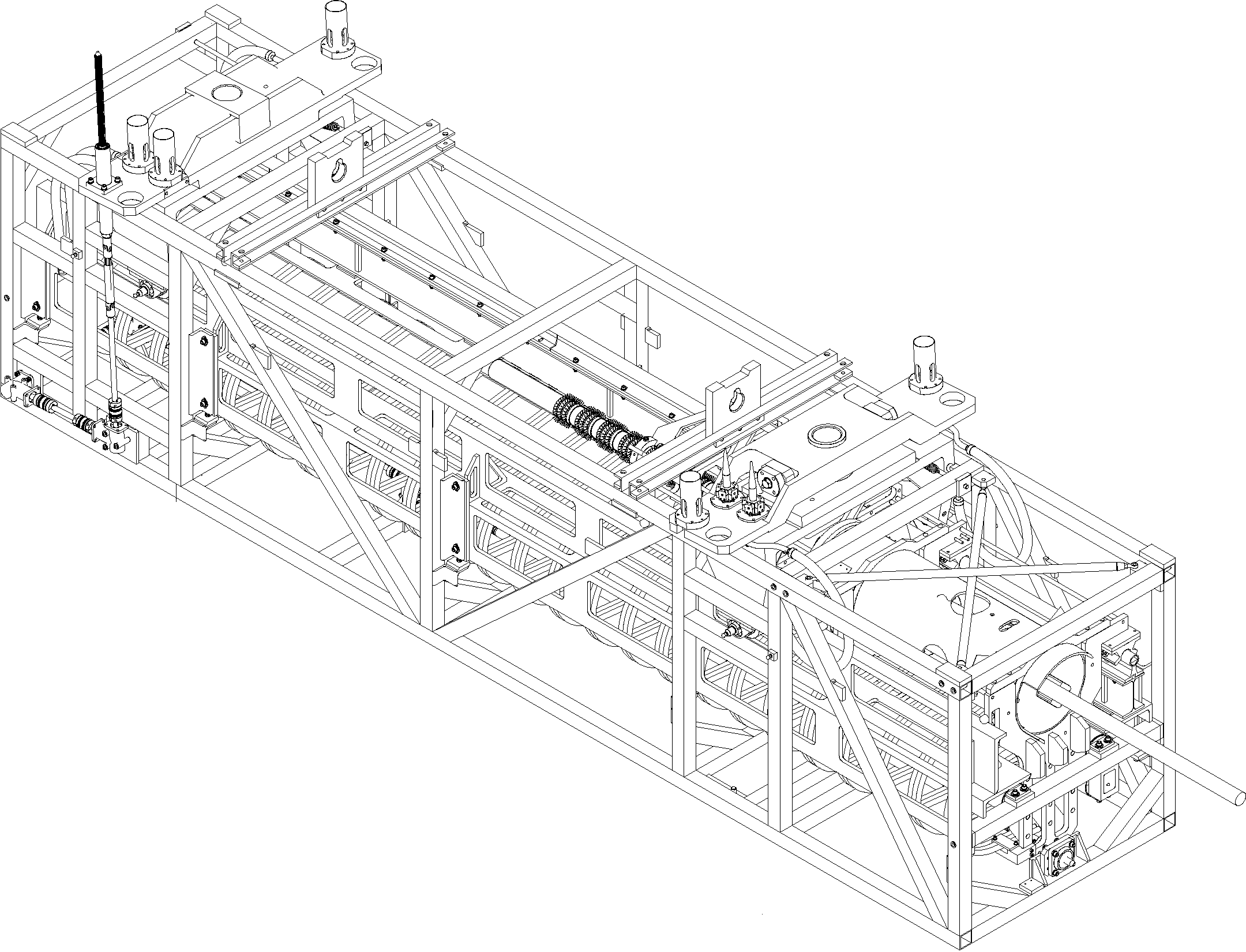}
    \caption{The Target/Baffle Carrier System. The top Figure shows a schematic of how the NuMI target and the baffle fit together in the carrier system. The bottom figure shows an isometric of the carrier system itself which has both the target and the baffle mounted. The target/baffle combination can be extended out into the focusing horn downstream. }
\label{fig:carrier}
\end{centering}  
\end{figure}

The target/baffle carrier system is shown in Fig.~\ref{fig:carrier}. The baffle and the target are mounted 68~cm from each other on cradles that have rollers that ride on aluminium rails coated with tungsten-disulfide for toughness. This allows the baffle and the target to be moved as one unit with respect to the horns when the beam configuration needs to be changed.  Furthermore,  the carrier supports target and baffle utilities such as the target cooling water supply, the target water return line, the target vacuum line, the eight thermocouple electrical lines and the two Budal target monitor electrical lines. Those different lines are looped below the carrier and are attached to the carrier at the upstream end; the downstream ends of the lines move with the cradles. The carrier hangs from two shafts that penetrate through a heavy shielding module. Positioning motors mounted on top of the module allow motion control of the target $\pm8$~mm horizontally, and $+8$/$-200$~mm vertically. For maintenance, the module and carrier are moved to the Work Cell.

When a new target is required it is first installed on the carrier in the medium energy position so that the target tip and baffle are not sticking out of the ends of the carrier during the step of module insertion into the target pile. The longitudinal positioning and readback of location of the target is all done optically after insertion. Subsequently a beam scan is performed by slowly steering the beam through a sequence of closely spaced horizontal locations and the carrier is aligned transversely with the beam line using the module motor drives. The target and baffle system is then moved on the carrier to the desired longitudinal position. Another beam scan is then performed to see if additional transverse adjustment is needed. The same procedure is also followed when it is desired to change the energy by moving the target with respect to the horns. The typical beam down-time for this whole operation is about one day. 

The target can be positioned with an accuracy of 0.5~mm transversely. Longitudinally, it can be positioned to an accuracy of 1~cm, and then the position can be surveyed with 0.3~cm accuracy. The baffle/target carrier system was designed to maintain position to within 0.5~mm transversely and 1~mm longitudinally under beam heating conditions created by up to $4\times10^{13}$~POT striking the target every 1.87~s. The NuMI carrier systems used over the course of the MINOS experiment were designed to survive radiation doses of up to $10^{11}$~rad/year for up to 10 years. 
%Interestingly, 2.5~m was not the limit that the target could be moved out of the horn. Removing the heavy shielding by crane from the top of the target module, and then moving the whole module further back and restacking the shielding, would have made it possible to achieve a total target range of at least 3.65~m, thus peaking the beam spectrum at an even higher energy, but this configuration was never needed during the MINOS data taking run.

\subsection{The Magnetic Horns}
\label{sec:horns}
\begin{figure}
\begin{centering}
    \includegraphics[width=.9\textwidth]{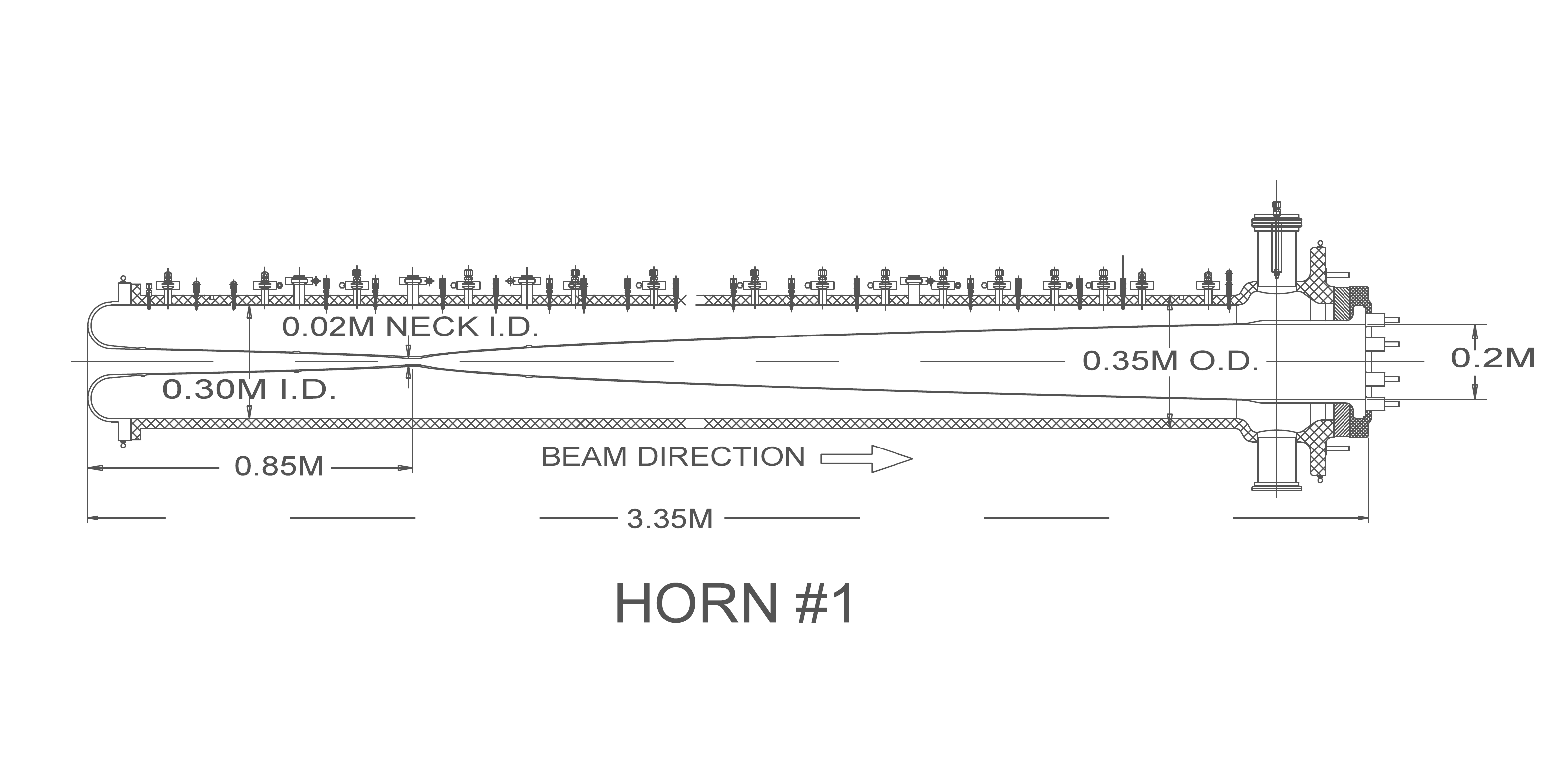}
    \includegraphics[width=.9\textwidth]{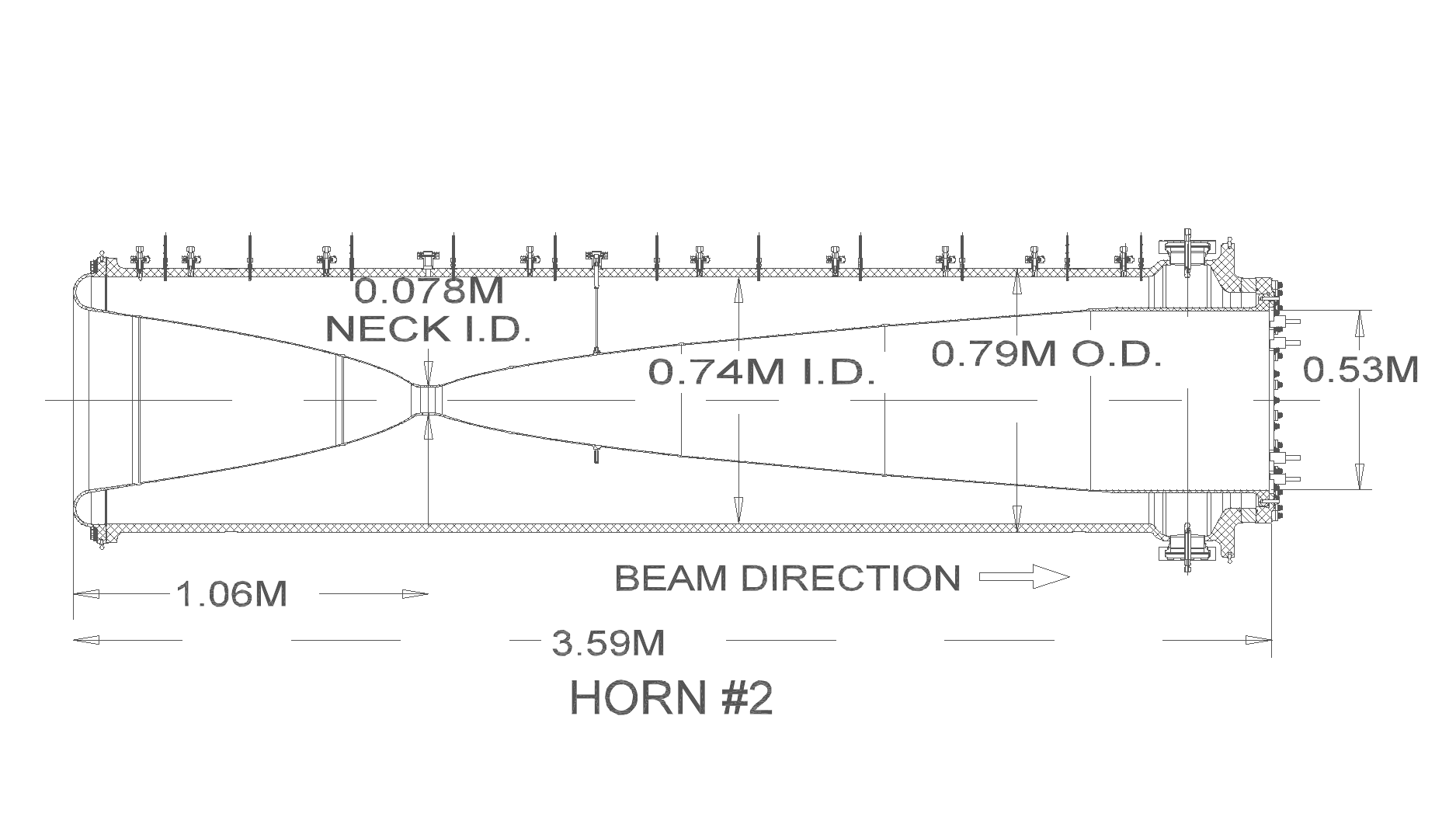}
    \caption{Schematic of the NuMI Horns' Cross-Section Views. The top illustration shows the shape and dimensions of Horn~1, and the bottom illustration shows the same for Horn~2.}
\label{fig:horns}
\end{centering}
\end{figure}

\begin{figure}
\begin{centering}
    \includegraphics[width=.82\textwidth]{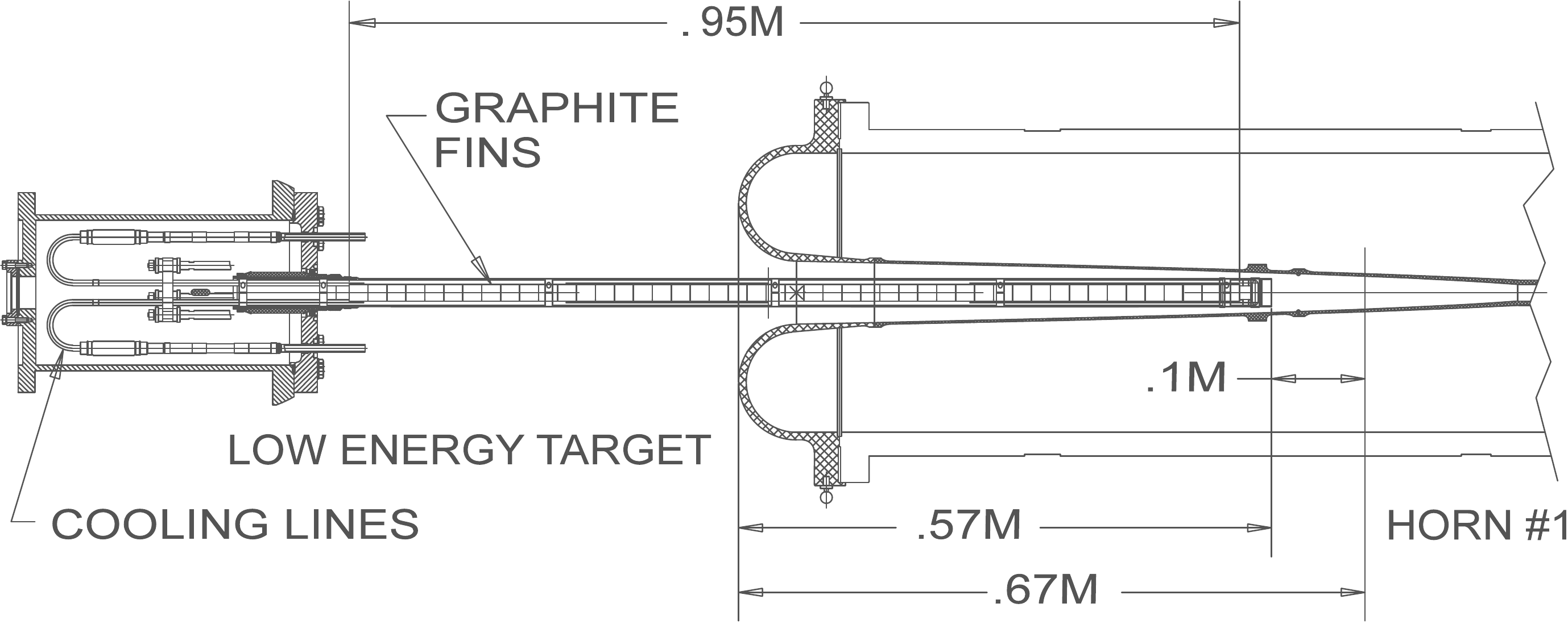}
    \caption{Low Energy Target with Front Half of Horn~1. The target is shown retracted by 10~cm upstream from the maximal design insertion. This corresponds to the nominal low energy beam configuration in which most of the MINOS data was taken.}
\label{fig:hornmodule}
\end{centering}
\end{figure}

The secondary mesons produced from the target are focused by two magnetic horns \cite{hornsref}, Horn~1 and Horn~2, which essentially act as hadron lenses. The horns are illustrated in Fig.~\ref{fig:horns}. The horns significantly increase hadron flux in the desired energy range and provide flexibility in choosing that energy. The target to horn distance is flexible and the separation between the two horns can also be changed. The design accommodates three potential Horn~2 positions of 10~m, 23~m, and 37~m downstream from the zero position (taken to be the upstream end of Horn~1), corresponding to low, medium and high energy respectively, coupled with appropriate target movement upstream. In the MINOS experiment the option to move Horn~2 was never exercised given the prevailing wisdom on $\Delta m^{2}$ by the time NuMI turned on\footnote{Horn~2 remained in the low-energy position during the MINOS experiment. The substantial shielding modifications needed to move the horn would have resulted in an unacceptably long down period.  Furthermore, physics considerations generally preferred running with Horn~2 in the low-energy position so as to maximize the amount of data at the neutrino oscillation peak.} and the MINOS medium and high energy configurations were achieved by moving the target with respect to Horn~1 and adjusting the horn current. The resulting MINOS ``pseudo'' medium and high energy runs were short special runs used for beam studies. Fig.~\ref{fig:hornmodule} shows a schematic of the target inserted into Horn~1 for the low energy configuration used in the MINOS experiment.

The NuMI horn inner conductors have a parabolic radial profile, such that they act as linear lenses and can be treated in the thin-lens approximation when the target is not too close to Horn~1. By Ampere's law the magnetic field between the inner and outer conductors should fall as 1/$R$ and should be zero at radii smaller than the inner conductor. The field measurements of the first horn verified the 1/$R$ dependence to a high degree of accuracy. 
Both the transverse and axial field components were essentially zero everywhere along the beam axis except at the neck where the transverse component was 30~gauss, 0.1\% of the maximum transverse field of 3~T \cite{hornmap, horn710}. The parabolic shape causes the path length of particle trajectories in the magnetic field region to approximately scale with the square of the radius at which the particle enters the conductor. The transverse momentum kick from the horn increases linearly with $R$. Thus the horn appears to the incoming positive hadrons as a focusing lens with a focal length proportional to their momentum. The position of the target determines the energy range of the hadrons focused by the horns. 

\begin{figure}
\begin{centering}
    \includegraphics[width=.76\textwidth]{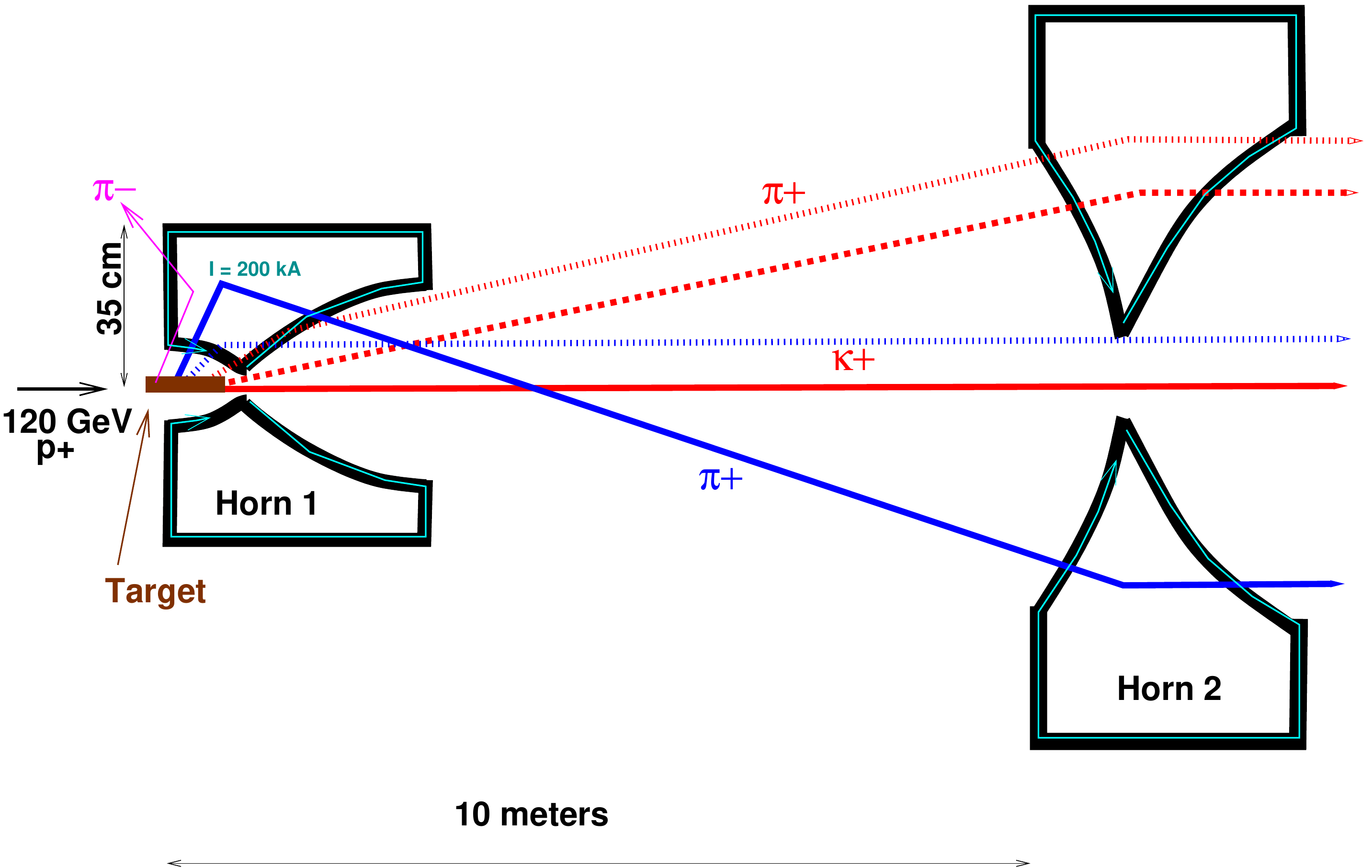}
    \includegraphics[width=.76\textwidth]{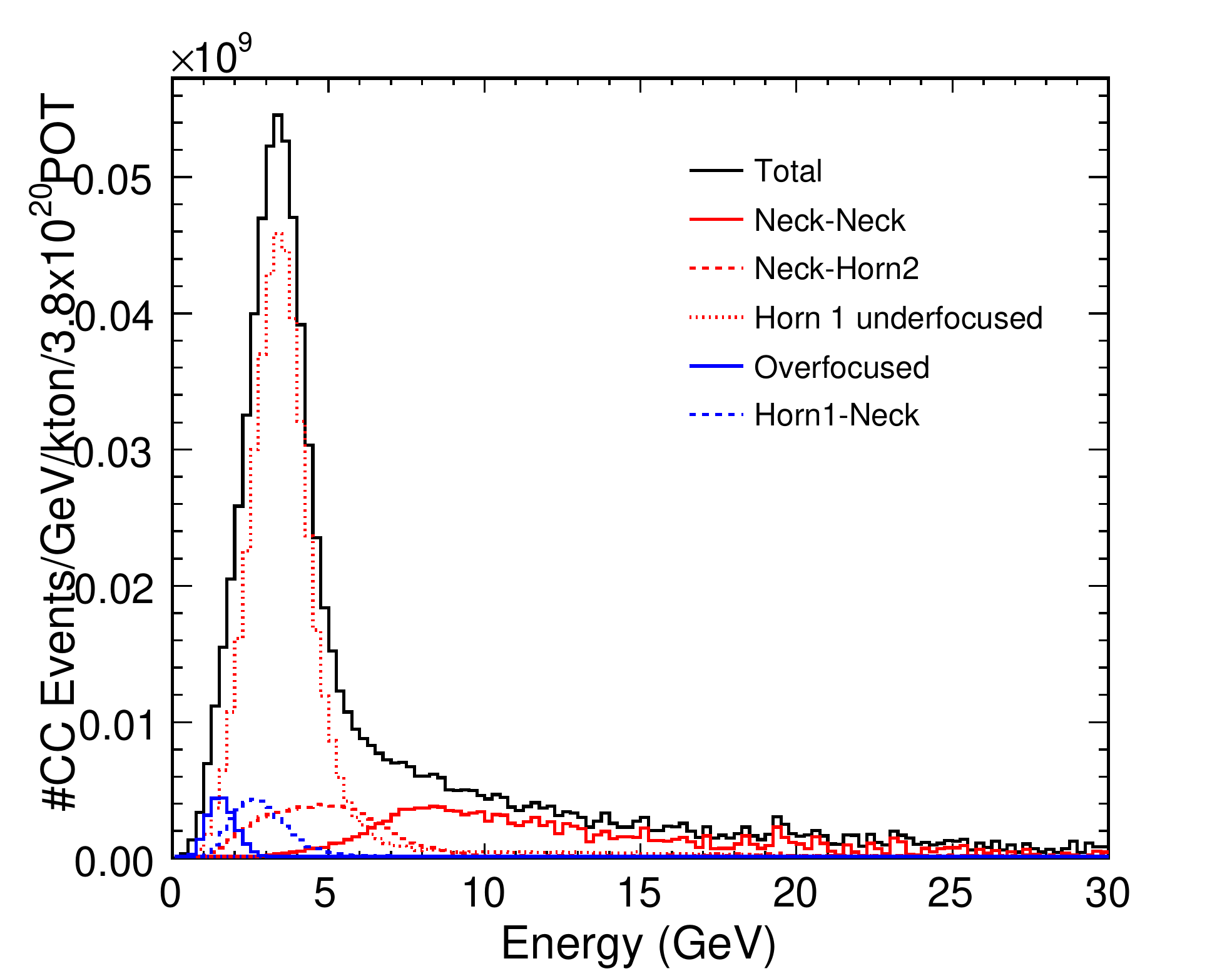}
    \caption{Hadron Trajectories Through the Two Horns. The top diagram illustrates possible trajectories through the two NuMI horns. Hadrons that are underfocused or overfocused by the first horn are further focused by the second horn. The bottom graph illustrates the composition of the low energy NuMI spectrum from the different hadron trajectory classes through the horns. }
\label{fig:hornfocus}
\end{centering}
\end{figure}

\begin{figure}
\begin{centering}
    \includegraphics[width=.66\textwidth]{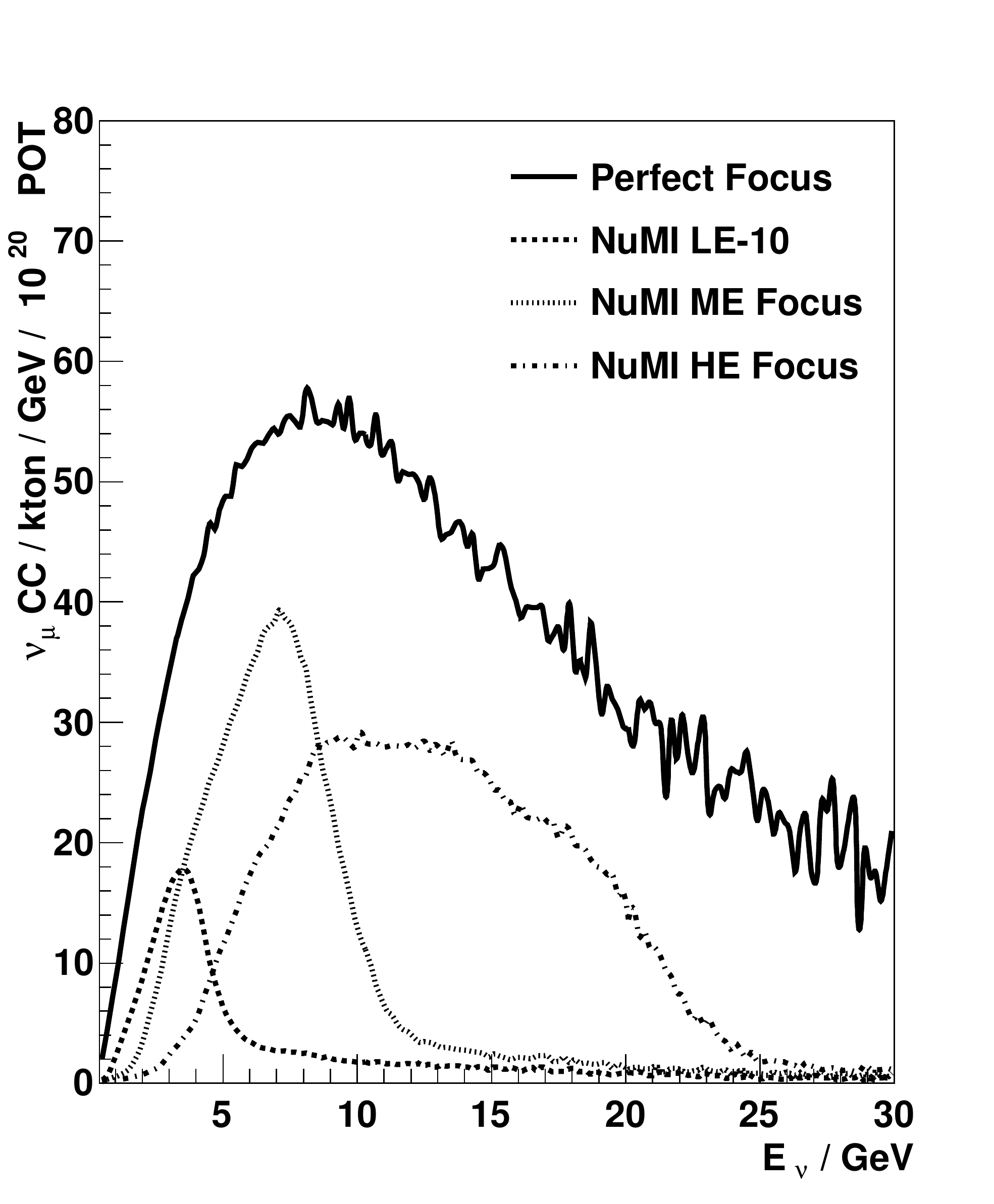}
    \caption{Variation of Neutrino Spectra for Different Focusing Schemes. This plot illustrates the differences between ``perfect'' focusing, that is if every pion produced in the target was focused precisely along the beam axis, and what can be achieved using the actual focusing horns. The curves are the NuMI event rate (flux $\times$ cross-section) as a function of neutrino energy at the Soudan mine in Minnesota for each instance; the solid black line shows the ``perfect'' focusing, and the dotted lines show the actual focusing for different NuMI beam configurations. The three different beam configurations correspond to the true low energy (LE), medium energy (ME), and high energy (HE) beam configurations where both the target and the horn positions are changed (rather than the ``pseudo'' ME and HE where only the target is moved).}
\label{fig:horn2effect}
\end{centering}
\end{figure}

The inelastic collisions which produce mesons impart a transverse momentum peaking at approximately 0.35~GeV/c, with only slight dependence on the meson longitudinal momentum. Therefore, the typical production angle of mesons is inversely proportional to meson momentum. Hadrons produced in the target along the beam axis pass through the horns unaffected. Hadrons that were well focused by the first horn are generally not affected by the second horn. A large fraction of hadrons that were either over- or under-focused by the first horn are focused in the second horn thus increasing the efficiency of the focusing system by about 50\%. Different initial production angles with trajectories through the horns are illustrated in Fig.~\ref{fig:hornfocus}. The different trajectory classes give rise to distinct portions of the neutrino spectrum. For comparison, a hypothetical ``perfect'' neutrino flux is defined as one where every pion is directed precisely along the beam axis. Fig.~\ref{fig:horn2effect} compares this ``perfect'' neutrino flux with the actual ones.

The two horns are connected in series with the first horn conductor closer to ground so as to minimize the potential difference between the target and the horn.  The horns are pulsed with a half-sine wave having a duration of 2.3~ms to produce toroidal magnetic focusing fields of up to 3~T. The maximum design current is 205~kA with a repetition rate of 1.87~s, but during the MINOS run the typical value used was 185~kA with a repetition rate of 2.2~s. To overcome the inductance of the horns and striplines the typical voltage produced by the power supply is 680~V. The horn current can be reversed by reversing the power supply voltage, thus allowing sign selection of the hadrons focused to produce an antineutrino-enhanced ($\bar{\nu}$-mode) beam instead of a predominantly neutrino beam (the normal neutrino-dominated forward horn current running is sometimes refered to as $\nu$-mode running). 

\begin{figure}
\begin{centering}
    \includegraphics[width=.66\textwidth]{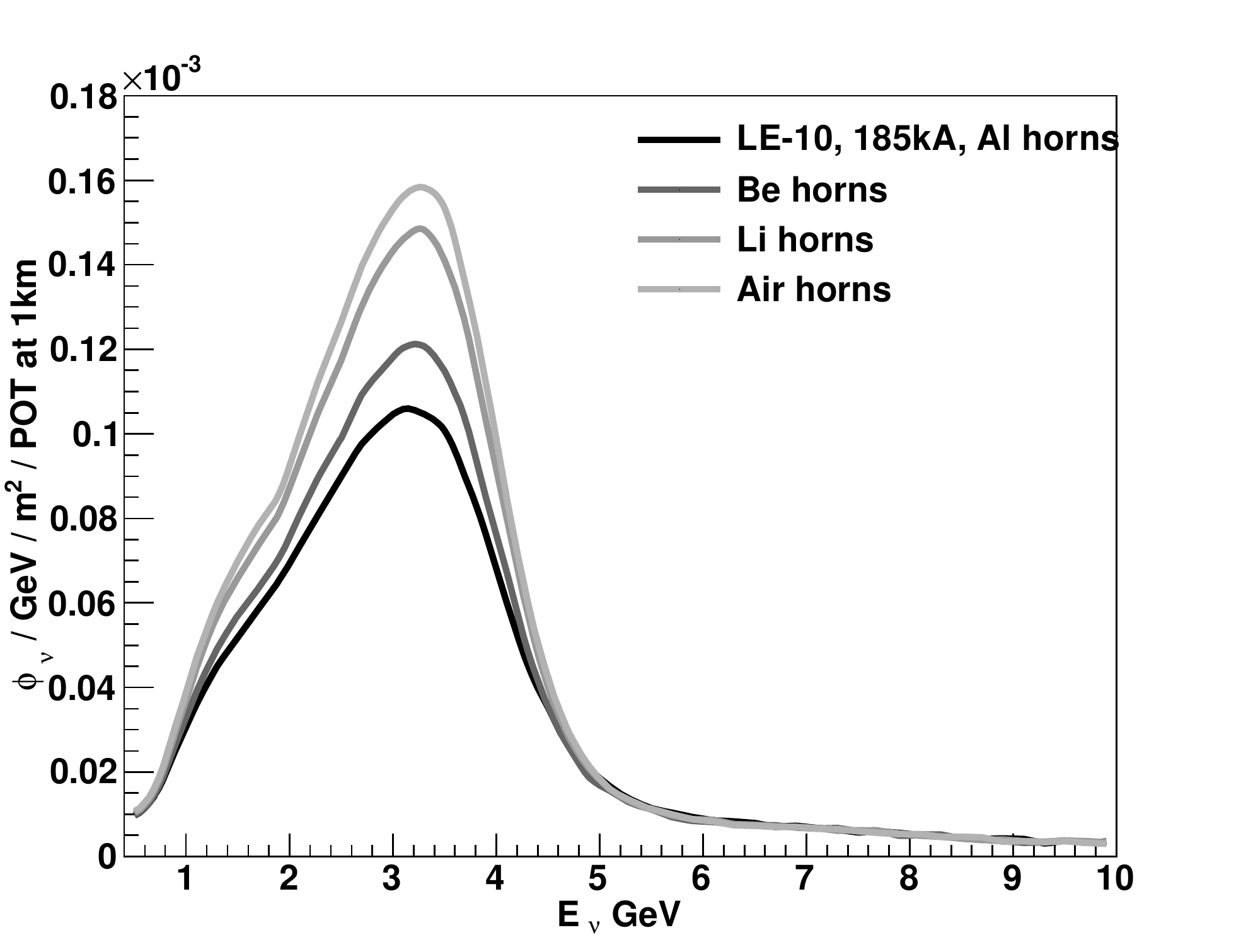}
    \caption{Impact of Horn Material on the Neutrino Flux. Simulated comparison of neutrino flux as a function of energy in the low energy beam configuration resulting from normal aluminium horns (black line) with the flux coming from idealized horns where the horn material has been changed to beryllium, lithium or air (different gray shades).}
\label{fig:airhorns}
\end{centering}
\end{figure}

Both NuMI horns are 3~m long. The inner conductors of both horns are constructed out of 6061-T6 aluminium, 2~mm thick for Horn~1 and 3 mm for Horn~2, except for the neck regions where the thickness is 4.5~mm. The inner conductor of the first horn is TIG (Tungsten Inert Gas) welded from seven longitudinal sections. To minimize meson absorption and scattering in the conductors, the aluminium thickness was reduced to what could reliably withstand the mechanical stresses and fatigue due to years of 205~kA pulses. For the same reason, no flanges were used inside the inner conducter and the latter was specially welded. The absorption of pions in the horn conductors reduces the neutrino flux. This reduction is illustrated in Fig.~\ref{fig:airhorns} which shows simulation of the flux with horn material changed to lower density materials.
The aluminium inner conductors of the horns are continuously sprayed with water to cool them and remove heat deposited by the beam and electrical resistance to the current pulse. The design dimensions and shapes of the two horns are tabulated in Tables \ref{table:horn1} and \ref{table:horn2}. Conductors of both horns are extended at their downstream ends with straight sections of lengths equal to the inner conductor diameter for current equalisation. This allows the current supplied from the four stripline taps to redistribute to a uniform azimuthal sheet. The horn inner conductor is electroless Nickle coated, and the outer conductor is anodized.  The horns interiors are flushed with Argon gas during operation to reduce corrosion and to remove oxygen and hydrogen from dissociation of water by ionizing particles.

The magnetic fields in the horns are monitored using a set of three pickup (Bdot) coils in each horn, which are mounted through ports in the outer conductors at the neck longitudinal positions (narrowest points) of each horn at 120$^{\circ}$ azimuth separations around the horn axis; one coil is located at top of the horn and the other two are located toward the bottom, 120$^{\circ}$ from the top one. Readout of the voltage induced in the pickup coils by the pulsing of the horn provides a measurement of the magnetic field. 

\begin{table}
\centering
\begin{tabular} {c c c c c c}
\hline
\vspace{0.5em}
&\multicolumn{2}{c}{Upstream}&Neck&\multicolumn{2}{c}{Downstream} \\
\vspace{0.5em}
%\hline
Z(cm) & 0-44.047 & 44.047-80. & 80.-83.982 & 83.982-95.128 & 95.128-300. \\
\hline
\vspace{1em}
$R_{in}^{IC}$(cm) & $\sqrt{\frac{92.8484 - z} {7.0483}}$ - 0.2  &  $\sqrt{\frac{85.7091 - z} {7.0483}}$  & 0.90 & $\sqrt{\frac{z-82.2123} {2.1850}}$ & $\sqrt{\frac{z-80.} {2.1850}}$-0.2 \\
$R_{out}^{IC}$(cm) &\multicolumn{2}{c}{$\sqrt{\frac{92.8484 - z}{7.0483}}$} &1.35 & \multicolumn{2}{c}{$\sqrt{\frac{z-80} {2.1850} }$ } \\
%$R_{out}^{IC}$(cm) & $\sqrt{\frac{92.8484 - z}{7.0483}}$ &  & 1.35 & $\sqrt{\frac{z-80} {2.1850} }$   \\
$R_{in}^{OC}$(cm)  &  & &14.92 & &\\
$R_{out}^{OC}$(cm)  &  &  &17.46 & &\\
\hline
\end{tabular}
\caption {Idealized Dimensions of Horn~1 in cm. Subscripts IC and OC refer to inner and outer conductors, respectively,  and in and out to the inside and outside surfaces of each conductor.}
\label{table:horn1}
\end{table} 

\begin{table}
\centering
\begin{tabular} {c c c c}
\hline
\vspace{0.5em}
&Upstream& Neck& Downstream \\ 
\vspace{0.5em}
%\hline
Z(cm) & 0-97.617 & 97.617-104.803 & 104.803-300. \\ [0.5ex] 
\hline
$R_{in}^{IC}$(cm) & $\sqrt{\frac{100 - z} {0.1351}}$ - 0.3  &  3.90 & $\sqrt{\frac{z-100} {0.2723}}$-0.3 \\
$R_{out}^{IC}$(cm) & $\sqrt{\frac{100 - z}{0.1351}}$ & 4.40 & $\sqrt{\frac{z-100} {0.273} }$ \\
$R_{in}^{OC}$(cm)  &  & 37.0 & \\
$R_{out}^{OC}$(cm)  &  & 39.54 &\\
\hline
\end{tabular}
\caption {Idealized Dimensions of Horn~2 in cm. Subscripts IC and OC refer to inner and outer conductors, respectively,  and in and out to the inside and outside surfaces of each conductor. }
\label{table:horn2}
\end{table} 

\begin{figure}
\begin{centering}
    \includegraphics[width=.48\textwidth]{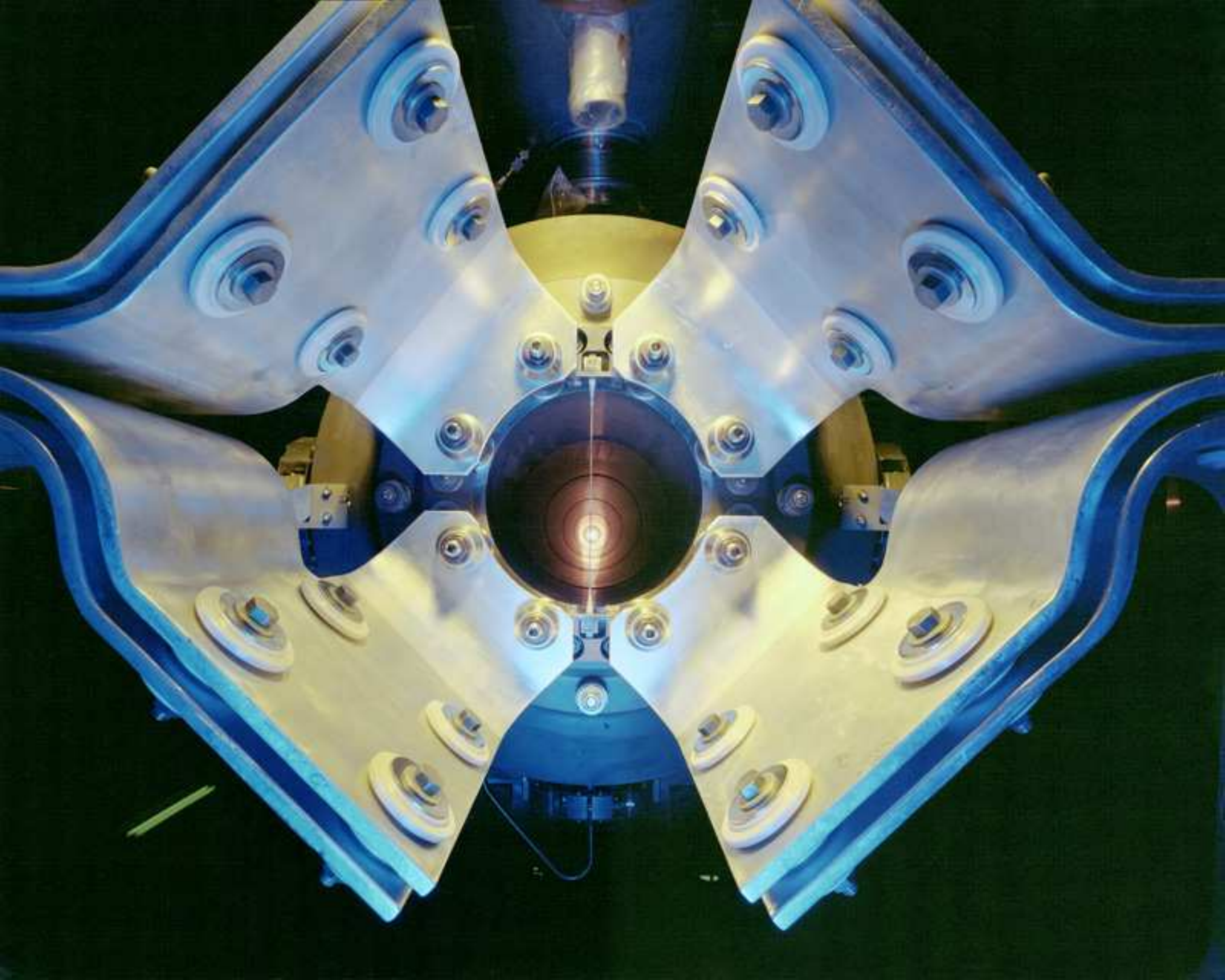}
    \includegraphics[width=.48\textwidth]{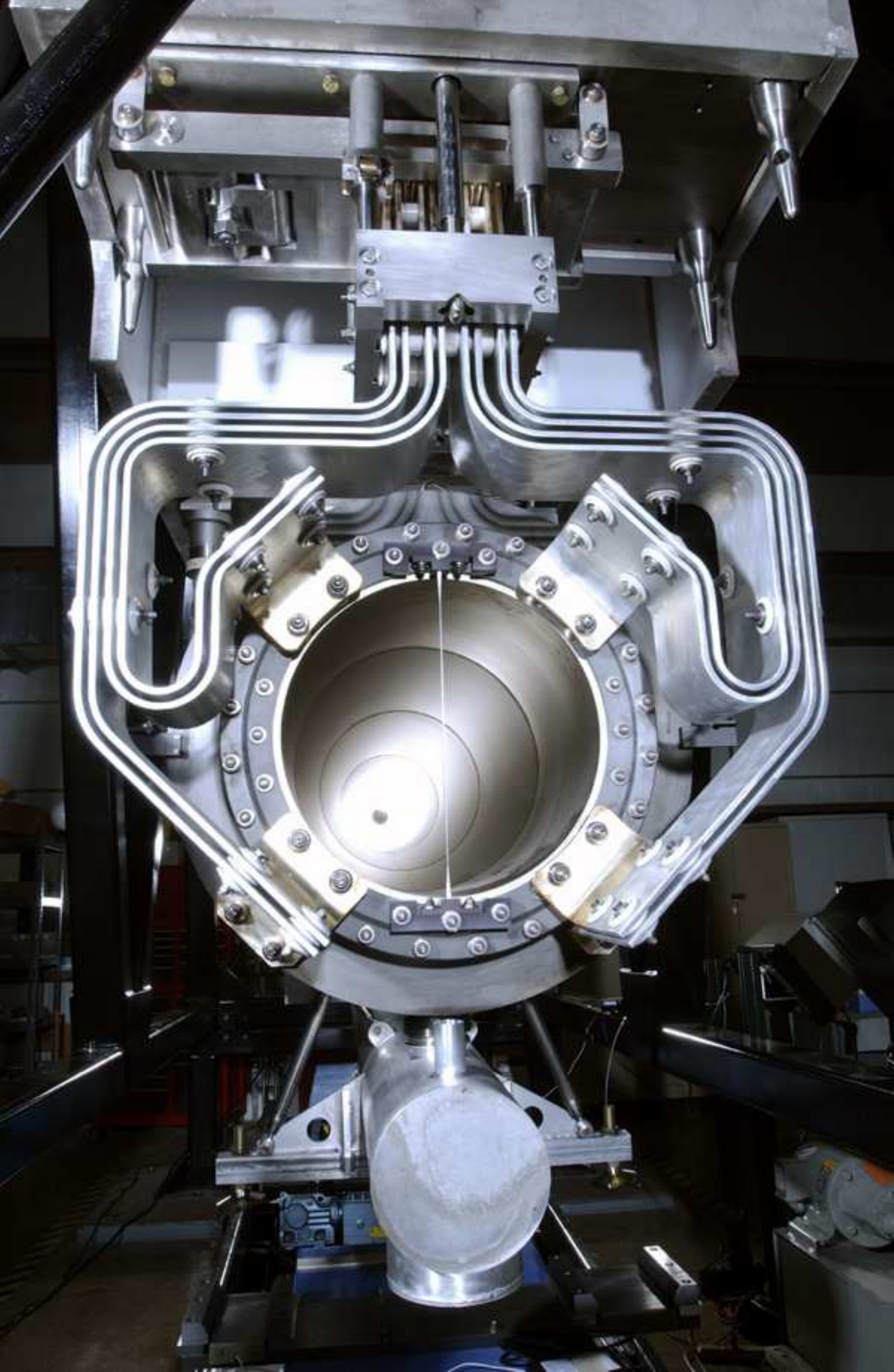}
    \caption{Photographs of Downstream Ends of Magnetic Horns. Horn~1 is shown on the left, and Horn~2 is shown on the right. The aluminium cross hairs used to align the beam are visible in both pictures.  }
\label{fig:hornpics}
\end{centering}
\end{figure}

The NuMI horns are initially aligned with optical survey methods using line-of-sight penetrations through the shielding. The horns have an additional alignment system consisting of aluminium strip cross hairs attached to the downstream end of Horn~1 and both the upstream and downstream ends of Horn~2, and also beam loss monitors located downstream of each horn. The loss monitors are mounted away from the beam axis and detect secondary particles from beam proton interactions in the cross-hair material during the alignment process discussed in Section \ref{sec:beamalign}. This system allows the locations of the NuMI magnetic horns to be measured relative to the incident proton beam center line.  Fig.~\ref{fig:hornpics} shows pictures of both horns with their alignment cross hairs. The cross hairs remain in place during running. 

Similar to the target, the horns are hung from horn support modules which provide mechanical support, cooling water, instrumentation connections and precision alignment capabilities. Horn 1 can be remotely controlled and transversely adjusted by $\pm$3~mm in both the horizontal and vertical directions. The Horn 2 position cannot be adjusted within the module after it has been placed in its final location, and so becomes the fixed point that beam and other components are adjusted to match. There is provision for adjustment of Horn~2 by re-shimming the carriage that the module sits on, but that feature has never been exercised.

\subsection{The Decay Pipe}
\label{sec:pipe}
The function of the decay pipe is to provide a vacuum or low density environment for the mesons to propagate and decay after being focused by the horns.  The mesons decay into tertiary mesons, charged leptons and neutrinos, thus producing the neutrino beam. Mass along the trajectory causes some fraction of the mesons to interact, reducing the neutrino flux. It also introduces Coulomb scattering that somewhat alters the mesons' paths.

The decay length for a 10~GeV pion, which would produce a 4.2~GeV neutrino is about 700~m. Thus for beams in this energy range it is not cost effective to make the decay pipe much longer. 
The divergence of low-energy mesons is greater, and thus a wider decay pipe radius would accommodate more meson trajectories and produce more neutrinos.
%The divergence of mesons after the horns is greater for low energy particles and thus for low energies there is a greater premium in making the transverse dimension as large as possible to keep the off-angle mesons in the decay volume before they strike the walls of the pipe. Thus the main issue in the design is the relative investment in length vs area and that compromise depends very much on the energy desired. 

The NuMI decay pipe is 2~m in diameter and 675~m in length within a larger excavated tunnel. At the time when the design had to be frozen the optimum energy was still uncertain since a large range of $\Delta m^{2}_{32}$ could account for the atmospheric neutrino observations. Thus no detailed optimization was possible. %Furthermore the computing power available at the time (1995) made very detailed simulations impractical. Today, the optimization for MINOS would most likely be different with larger radius and shorter length.

The NuMI decay pipe runs through the whole length of the decay tunnel. The pipe is made from steel pipe sections which are 12.2~m long and have a 2~m inner diameter. The steel thickness is 0.95~cm. Each pipe section is reinforced with five encircling stiffener rings, also 0.95~cm thick, and 12.7~cm tall in the radial dimension, to prevent buckling when the decay pipe is evacuated and to provide structural integrity against the surrounding concrete shielding. The pipe runs through the approximate center of the decay tunnel, which was excavated with a tunnel boring machine 6.7~m in diameter. Various sections of the tunnel were enlarged using ordinary drilling excavation, to allow for thicker surrounding shielding in the upstream sections and also a passageway outside of the shielding along the entire length of the pipe. The steel pipe sections were held above the floor of the tunnel by steel support structures. The pipe sections and supports were installed one after another, downstream to upstream, surveyed to confirm correct placement around the beam center line, then welded together. The supports also served to hold the pipe down against buoyant forces applied when concrete for shielding was poured in.

The decay pipe starts 46~m downstream of the NuMI target. A thin two-component steel-aluminium window closes the upstream end of the decay pipe. 
The upstream window is thin to minimize meson interactions, while thick enough to provide sufficient strength to avoid any possibility of rupture. The adopted solution utilized the fact that the meson beam leading to observed neutrinos does not cover the full 2~m-diameter beam pipe aperture but only the central part corresponding to the beam chase dimensions. The decay pipe window consists of a central 1.6~mm thick aluminium disk that is surrounded by a 0.95~cm thick steel annular window. The downstream window is made from 6.35~mm steel. 

\begin{figure}
\begin{centering}
 \includegraphics[width=.48\textwidth]{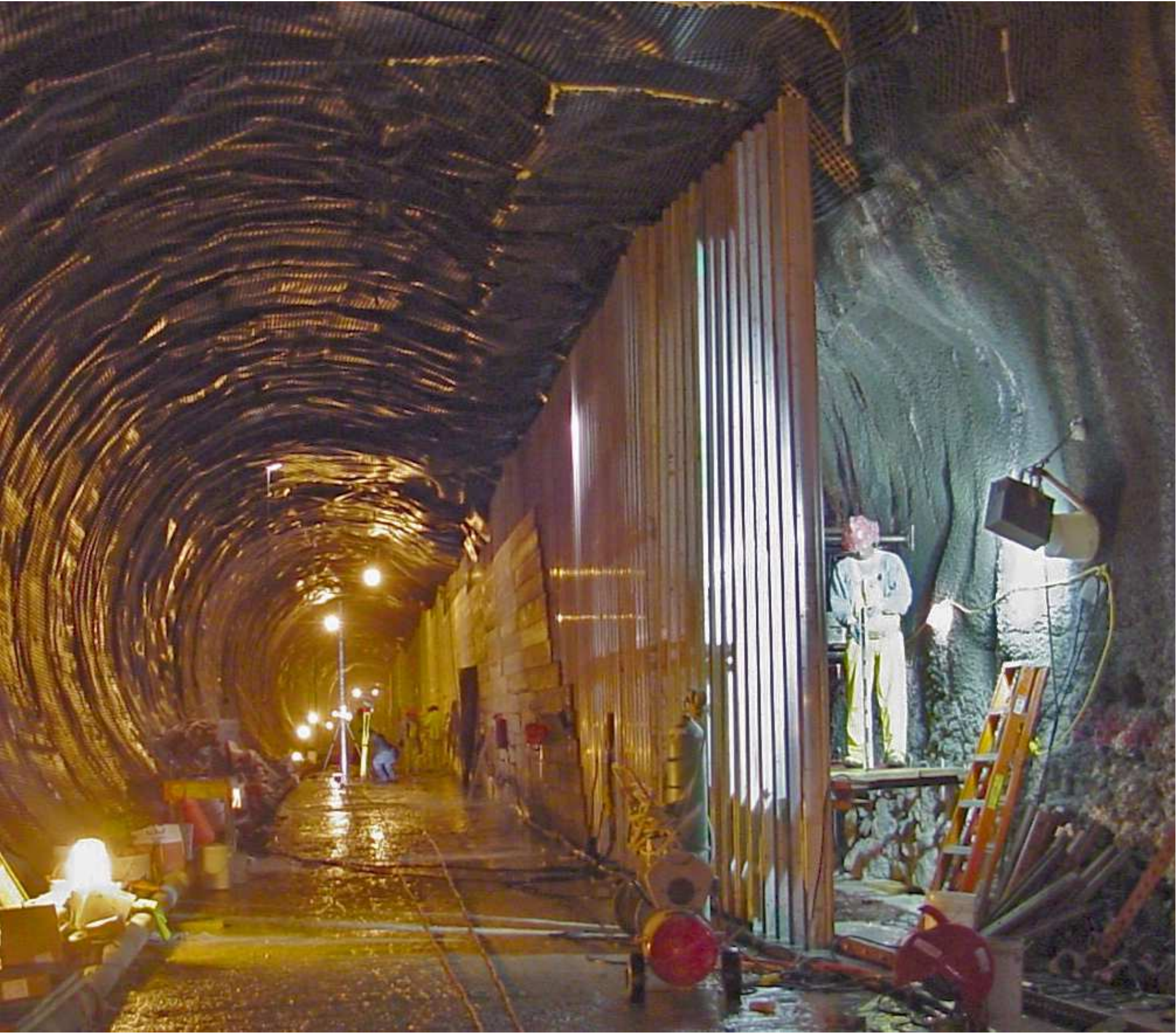}
 \includegraphics[width=.48\textwidth]{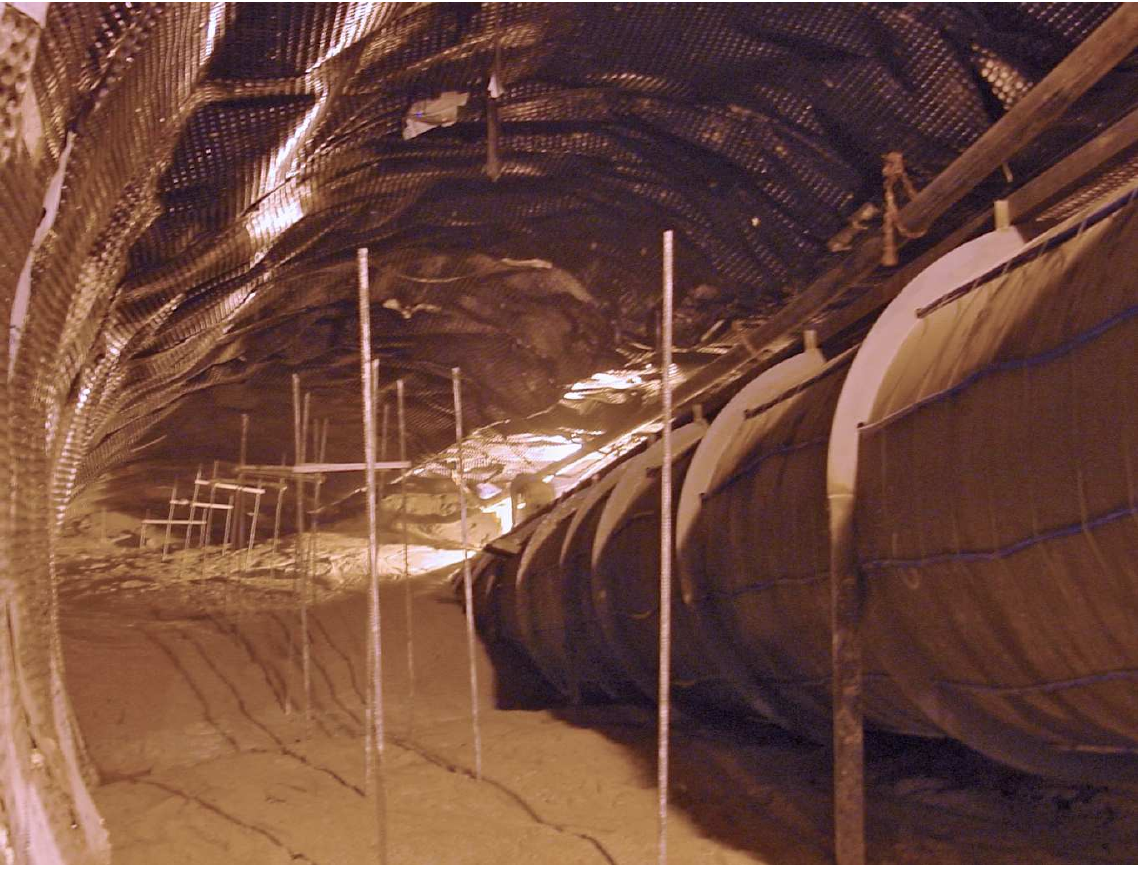}
    \caption{The NuMI Decay Pipe Tunnel. The left photograph shows the decay tunnel before the decay pipe was installed. The wall was the form for the concrete shielding and the worker shown was standing in the egress passage. The right photograph shows the installed decay pipe with concrete shielding being poured. The stiffening ribs around the pipe are also visible as well as the copper cooling lines running along the length of the pipe. }
\label{fig:decaypipepic}
\end{centering}
\end{figure}

The decay tunnel was backfilled with concrete around the steel decay pipe to provide shielding for the groundwater from particles impacting the pipe walls. A 1.27~m wide egress passage was maintained along the east side of the tunnel; this passage provides emergency access between the upstream Target Hall and the downstream MINOS areas. 
A picture of the decay tunnel during construction is shown in Fig.~\ref{fig:decaypipepic}. As mentioned above, the concrete shielding itself is thicker upstream than it is downstream due to the higher number of upstream interactions and ranges from 2.1~m at 150~m, to 1.4~m at 425~m, along the decay pipe. At the beginning of the MINOS experiment running in 2005 the decay pipe was evacuated to a pressure of 0.5~Torr to avoid the hadrons reinteracting in air and being lost. In 2007 the decay pipe was filled with helium to a pressure of 13.2~PSIA due to possible corrosion of the decay pipe entrance window which could compromise its mechanical strength. The effect of this change on the neutrino spectrum was of the order of a few percent loss as will be discussed in Section \ref{longtermhistory} (mostly in the focusing peak).

The decay pipe and the surrounding concrete are heated by the energy deposition of off-angle particles. Simulations have shown that the total energy deposition in the pipe in the low energy beam configuration is 63~kW, and an additional 52~kW deposition in the surrounding concrete. Such a large energy deposition would give rise to an increase in temperature of the steel such that the resulting thermal expansion could crack the surrounding concrete. To alleviate this situation, before the concrete was poured, 12 copper cooling lines  were installed symmetrically around the circumference of the pipe along its full length. Those water cooling lines were connected to heat exchange systems with 150 kW capacity at each end of the decay pipe. The calculated temperature of the pipe with this arrangement was 50$^{\circ}$C.

An additional focusing system named the ``hadron hose''\cite{hadronichose} had been contemplated early in the NuMI project. The hadron hose would be a conductor carrying 1~kA along the entire axis of the decay pipe. The current would induce a toroidal field to focus the secondary mesons and cause them to follow a spiral path along the axis of the conductor, and thus increase their probability of decaying before hitting the walls of the pipe. The 1~kA current would be pulsed for 0.5 msec in synchronism with the beam and result in roughly a 30\% increase in neutrino flux. In addition, such a system would reduce the Near/Far flux differences thus decreasing the need for detailed reliance on Monte Carlo simulations.  Extensive studies were made of mechanical, electrical and thermal issues and the system appeared workable for those parameters. A 14~m long prototype was constructed and tested successfully without beam. This device was eventually not pursued due to financial constraints.

\subsection{The Absorber}
\begin{sidewaysfigure}
\begin{centering}
    \includegraphics[width=1.0\textwidth]{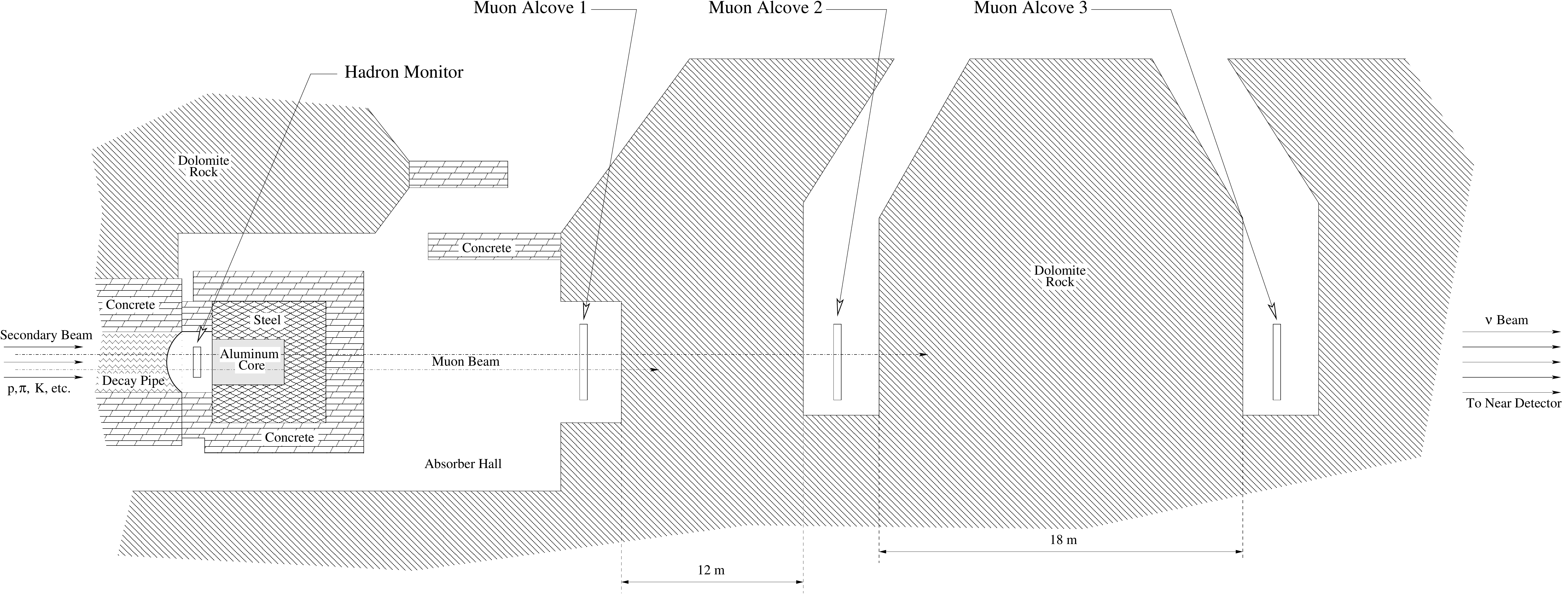}
    \caption{Schematic Representation of the Beam Area Downstream of the Decay Pipe. The Figure indicates the location of the hadron absorber, the hadron monitor, and the three muon monitors. The fourth alcove was not instrumented during the MINOS run and is not shown in the Figure.}
\label{fig:absorberenclosure}
\end{centering}
\end{sidewaysfigure}

The NuMI absorber is a massive aluminium, steel and concrete structure just downstream of the decay pipe whose function is to absorb the residual beam. A schematic of the beam area downstream of the decay pipe which includes the hadron absorber, as well as the location of the muon monitors, is shown in Fig.~\ref{fig:absorberenclosure}. According to NuMI beam simulations \cite{absorberconcept}, approximately 1/6 of the primary beam power reaches the end of the NuMI decay pipe in the form of various particles. Of the particles interacting in the absorber, the principal component (approximately 80\%) is the proton beam that has not interacted. The remainder are mainly mesons which have not decayed in the pipe or secondary protons. A small fraction (around 4\%) is due to electrons, neutrons, and gammas. Muons and neutrinos deposit little or no energy in the absorber.

The absorber has several functions. First, it stops most of the particles still remaining in the beam. Their energy is deposited through ionization and the resulting heat is transferred to circulating water through thermal conduction. Second, it serves to protect groundwater from irradiation. Third, it also limits the levels of radiation in tunnel regions accessible to personnel under all running conditions. Fourth, it limits residual radiation to people accessing the absorber hall under beam-off conditions. The absorber design requires that it meet its principal functions not only during the regular operation of the beam but also in accidental situations (for example accidental mis-steering of the beam) when the primary proton beam does not interact in the target for a short time and its full power is deposited in the absorber.

\begin{figure}
\begin{centering}
    \includegraphics[width=.66\textwidth]{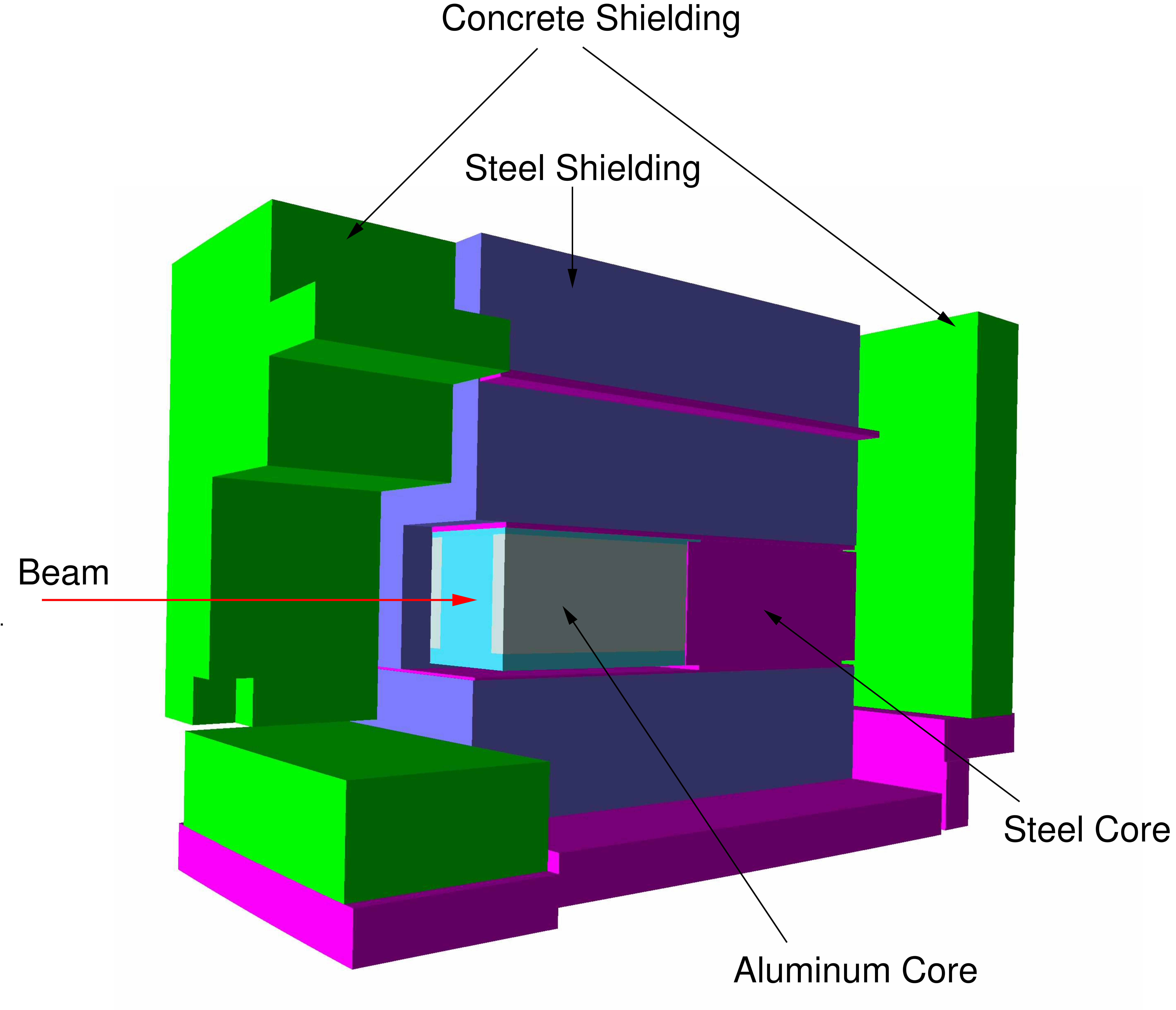}
    \includegraphics[width=.50\textwidth]{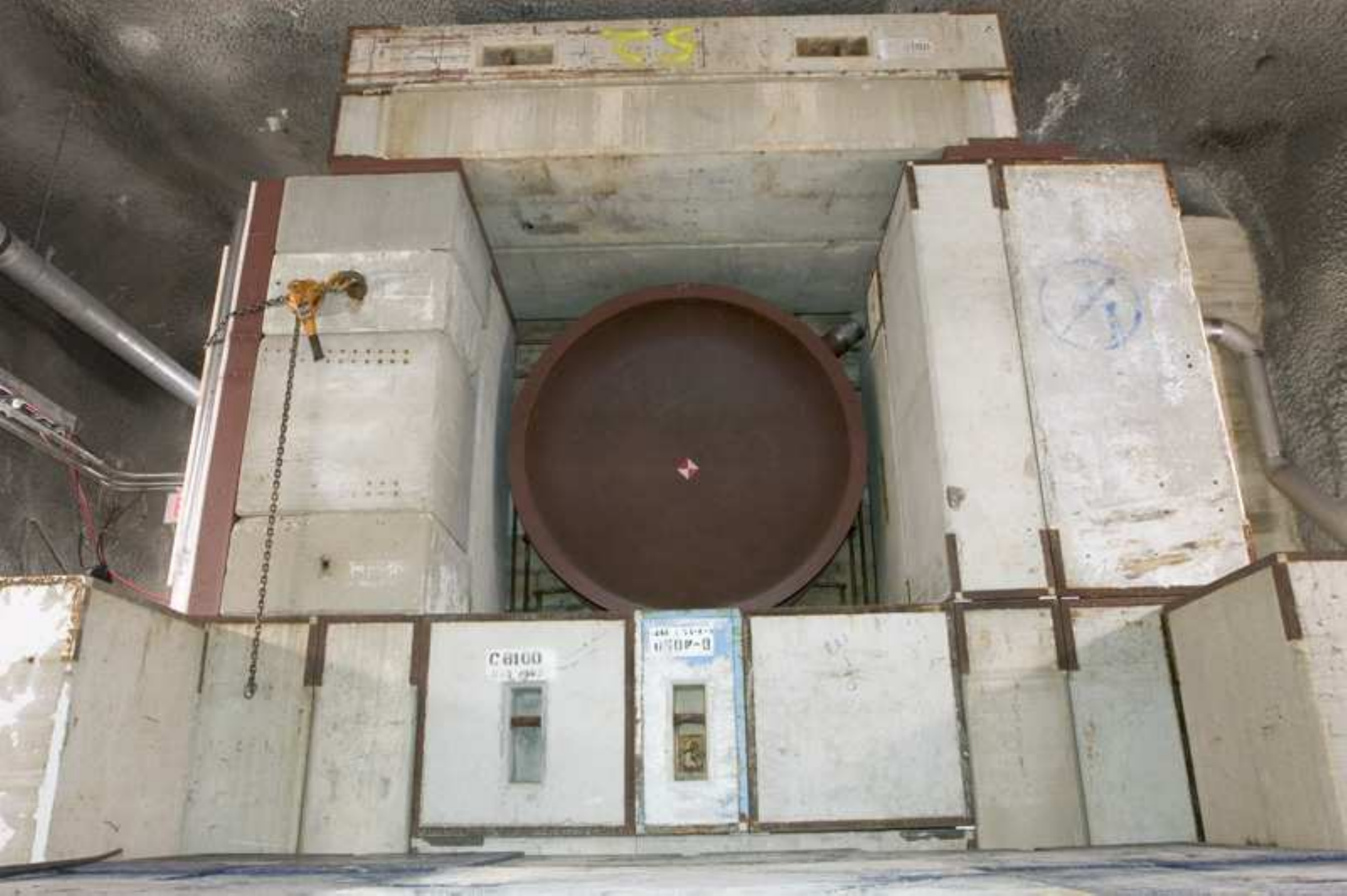}
    \includegraphics[width=.35\textwidth]{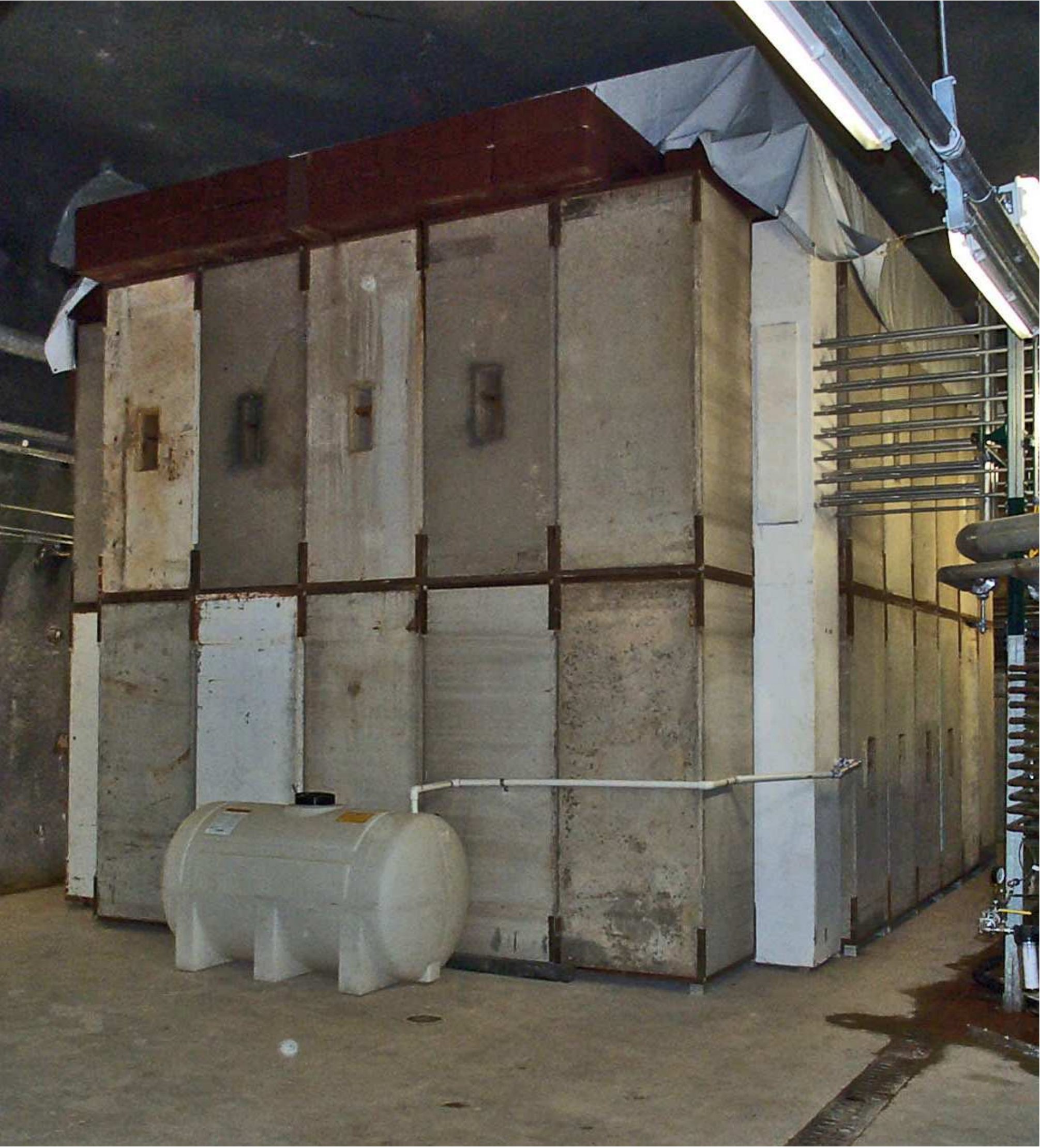}
    \caption{The NuMI Beam Hadron Absorber. The top Figure shows a schematic of the hadron absorber.  The left bottom picture shows concrete block shielding around the downstream end of the decay pipe before the installation of the hadron absorber. The bottom right photograph shows the completed hadron absorber with the back wall of concrete blocks, and the steel roof blocks. }
\label{fig:absorber}
\end{centering}
\end{figure}

A schematic of the hadron absorber itself, as well as a picture of the absorber during assembly, are shown in Fig.~\ref{fig:absorber}. The absorber \cite{docdb2769} is essentially a box approximately 5.5~m wide $\times$ 5.6~m tall $\times$ 8.5~m long.  It is housed in the absorber enclosure, an underground excavated hall with dimensions 8.2~m wide $\times$ 6.1~m high $\times$ 15.2~m long. The central section of the absorber, the core, has dimensions 1.3~m wide $\times$ 1.3~m high $\times$ 4.75~m long. It is composed at its upstream end of eight 1.3~m wide $\times$ 1.3~m high $\times$ 0.3~m long machined aluminium blocks placed longitudinally to the beam, followed by ten 1.3~m high $\times$ 1.3~m wide $\times$ 0.23~m long flame-cut steel blocks. The aluminium is 6061-T6 aluminium, an alloy that is 98\% aluminium with density 2.70~g/cm$^{3}$. The steel in the core\footnote{Continuous Cast Salvage (CCS) from US Steel - Gary Works.} is a grade of steel from the areas of the batches, or heats, where the steel does not conform to ASTM standards due to its chemical composition. It has essentially properties of steel ASTM-836 grade and is about 98\% iron. The density of these blocks is very uniform with an average of 7.8416~g/cm$^{3}\pm 0.0004$~g/cm$^{3}$. 
Each of the eight aluminium blocks is cooled by a pair of drilled water cooling loops, each loop with its own water supply and return pipe for redundancy. The water pipes are aluminium, and welded to the blocks rather than using a fitting. A total of 32 cooling water pipes run longitudinally through the core blocks in 3.8~cm diameter holes. All the cooling pipes are made out of single continuous lengths to reduce any possibility of leaks. 

The absorber core is surrounded on the sides, above and below, by 88 Duratek steel blocks. Their dimensions are nominally 1.32~m $\times$ 1.32~m $\times$ 0.66~m but their inherent non-uniformities in dimensions (of the order of 5~mm) create small gaps between neighbouring blocks. The function of the steel blocks is to catch the tails of the hadronic showers created in the core. There are additional steel plates 46.3~cm thick placed on top of the BluBlocks. The final, outer layer of the absorber is composed of concrete whose main function is radiation shielding, specifically absorption of thermal and low energy neutrons for which iron is relatively transparent. Concrete blocks are stacked around the downstream end of the decay pipe and on the sides of the steel blocks. 

% additional bit of the caption if we want it: On the floor is the tank into which the secondary containment pan drains. At right the core pipes emerge and are connected to manifolds located along the wall. The white structure where the pipes emerge is the plywood box, filled with poly-beads. This box can be moved if access to the rear of the absorber pile is necessary.

\subsection{The Muon Shield}
The muons remaining in the NuMI beam after the decay pipe and absorber are ranged out in the so-called muon shield, which simply consists of 240~m of solid dolomite rock between the absorber and the MINOS ND hall. This distance was originally specified for COSMOS \cite{cosmosref} so there would only be muons in their detector from neutrinos interacting upstream. However, without a muon shield, MINOS would also suffer from muons from hadron decays in the beam pipe as the Near Detector electronics would be overloaded and pattern recognition likely made impossible.

\section{NuMI Beam Line Instrumentation}
\subsection{Proton beam line monitors}
\label{sec:protmon}
The instrumentation in the NuMI proton beam line serves several purposes. It monitors the extraction of the NuMI proton beam from the Main Injector, provides precision position information for automated control of the beam transport and targeting, measures the number of protons delivered to the NuMI target on a pulse-by-pulse basis, and provides input to the beam permit system which inhibits the beam in case of malfunctions. The high power of the proton beam imposed the requirement to keep beam losses to a very low level and made extensive instrumentation necessary for safe and efficient operation. The design criterion chosen, to satisfy all constraints, was for the sustained proton loss fraction never to exceed $10^{-5}$ in the extraction region and over the length of the proton beam line, and also to preclude further extractions if single pulse losses at this level or larger do occur. Typical fractional beam loss is maintained at a factor of 30 below this limit. 

The number of protons delivered on target are monitored on a pulse to pulse basis using two toroids which encircle the proton beam and are positioned at two different places along it. TOR101 is located in the upstream NuMI beam line, and TORTGT about 10~m before the NuMI target. Each toroid's integrated signals are digitized with 14 bit precision. The toroids are calibrated absolutely by a high precision current source whose calibration is checked regularly. In addition the two toroids have been compared with the readings of a DC current transformer which measures the intensity of the Main Injector beam and should give the same value as the two toroids for pulses entirely devoted to NuMI. Fractional beam extraction losses are maintained below part-per-million levels. The two toroids agree to $0.2$\% and the absolute accuracy is $0.5$\%.

NuMI also has two resistive wall monitors (RWMs) with 1~GHz bandwidth. One in the Main Injector is used to examine the bunch structure of the beam at injection and before extraction. There is also a dedicated RWM in the beam line, which is able to show the effects of the kicker rise and fall times.

The beam line is instrumented with two types of beam loss monitor. One set of those are four total loss monitors (TLM’s) located in the downstream part of the beam line. A TLM is a coaxial cable, with Ar-CO$_{2}$ gas between the coaxial conductors. Each TLM is about 75~m in length, positioned in a cable tray adjacent to the beam line, with 800 volts potential on the center conductor. The total ionization current from beam loss seen over the TLM length is then recorded on a pulse-by-pulse basis.

The other type of beam loss monitor is a set of 48 sealed gas ionization chambers generally mounted on magnets and installed over the length of the beam line. Their threshold readings of $<$0.002~rad/s correspond to a fractional loss of 6$\times10^{-8}$ each. Logarithmic amplifiers used to read them out provide six decades of dynamic range. Readouts from all beam loss monitors are then input into the beam permit system, which inhibits the next beam pulse for any beam loss above preset thresholds. Besides these two sets, there is one safety system ionization chamber located in the upstream part of the NuMI beam line set to remove the beam permit and bring down the safety interlock (i.e.\ shut down the beam) if more than 1.7 rads/s are seen. 

The beam is also instrumented with two sets of monitors to measure the position and profile of the beam at various locations. One set consists of 13 horizontal and 11 vertical beam position monitors (BPMs). Each BPM consists of two cylindrical electrodes, with the difference in charge induced on each electrode by the beam giving a precise measurement of the beam position in the horizontal or vertical. These are relatively standard Fermilab devices in use now for some time except for the electronics, which have been modified so as to read out charge during each one of the Booster batches. Because they involve no material in the beam, they are able to remain in their positions during all operations.

\begin{figure}
\begin{centering}
    \includegraphics[width=.48\textwidth]{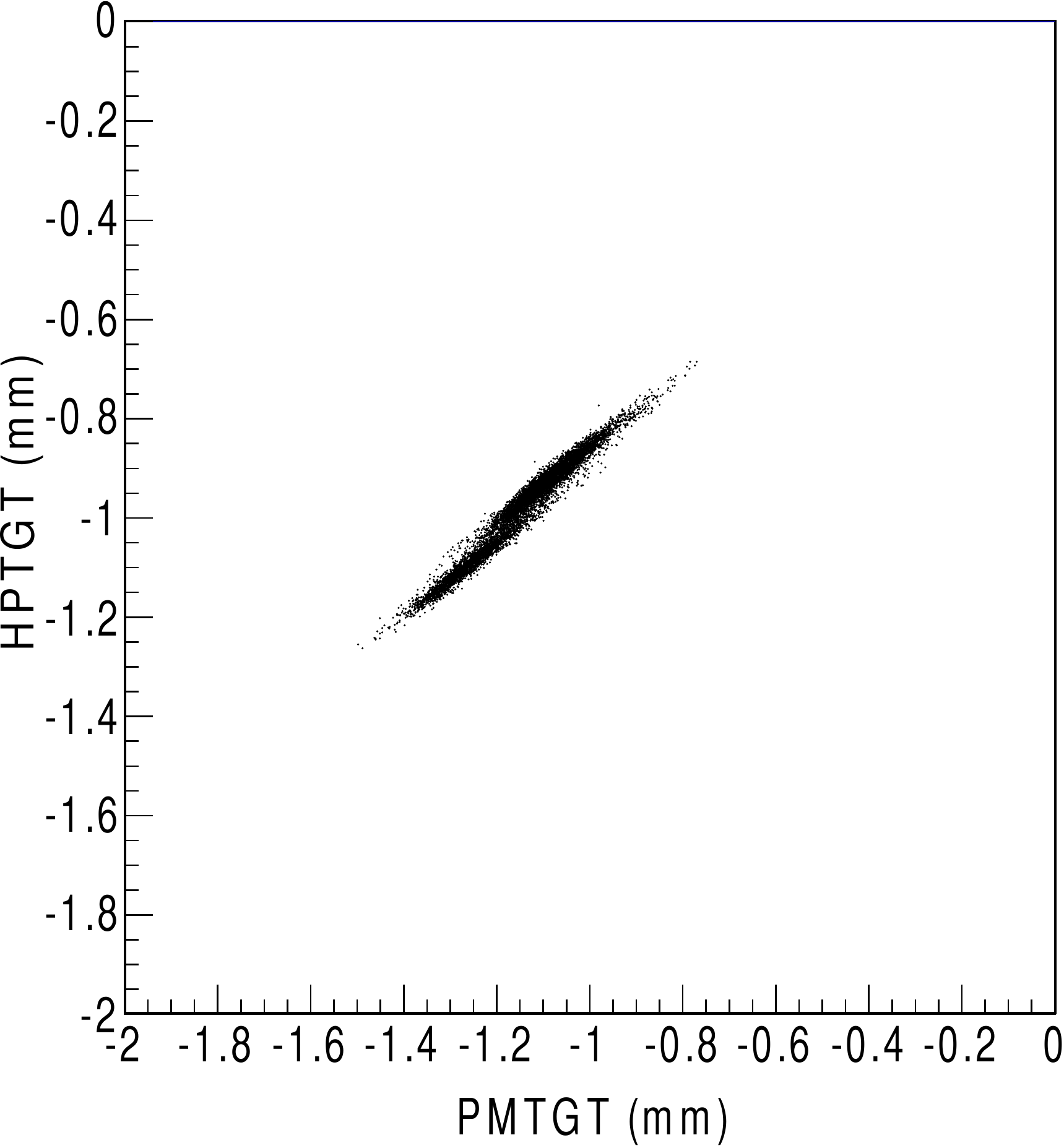}
    \includegraphics[width=.48\textwidth]{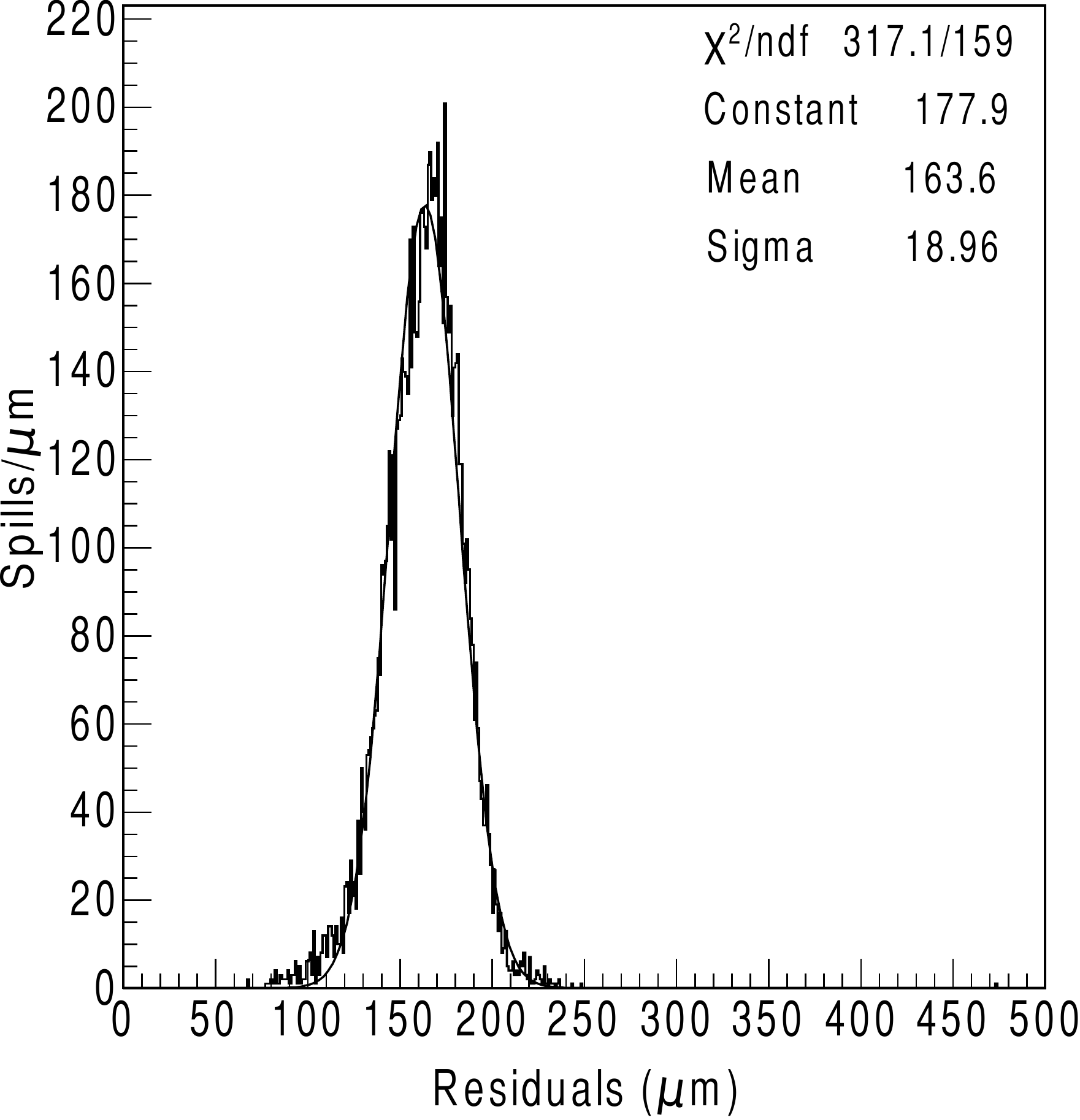}
    \caption{Comparison of the Position Measurement in the BPM and Profile Monitor. The left graph shows the scatter of the position measured in the BPM labeled HPTGT compared to that in the adjacent profile monitor, labeled PMTGT. The right graph shows the histogram of distances of points from the diagonal of the left plot. }
\label{fig:fig12-1669}
\end{centering}
\end{figure}

Besides the BPM’s, the beam is instrumented along its length with ten secondary emission profile monitors (SEM’s) \cite{BeamMonitors2}. These utilize either 5~$\mu$m thick titanium foils or 25~$\mu$m diameter titanium wires, with 1~mm pitch for the upstream monitors and 0.5~mm pitch for the two near the target. The SEM closest to the target remained positioned in the beam while those along the beam transport line were designed so as to be easily moved out of the beam during data taking so as to minimize material in the beam. A comparison of beam positions recorded by the BPM and SEM close to the target shown in Fig.~\ref{fig:fig12-1669} gives 19~$\mu$m sigma for the sum in quadrature of the BPM and profile monitor position resolutions.

%DISCARDED TEXT
%The program uses a matrix to relate the measurements of position detectors (in each plane) to currents in all the trim magnets. The inverse of the matrix provides trim current values required to change the beam positions obtained in the measurements to their desired values. 

% Some of these describe potential subsequent deviations of the secondary proton beam, like for example a displaced Main Injector beam orbit or a wrong flat-top ramp current for any magnet. Those can be detected before beam extraction and the extraction can be prevented. Others, such as a significant loss in the beam line, or a too large fraction of the beam striking the baffle, prevent the extraction on the following pulse.

\begin{figure}
\begin{centering}
    \includegraphics[width=.48\textwidth]{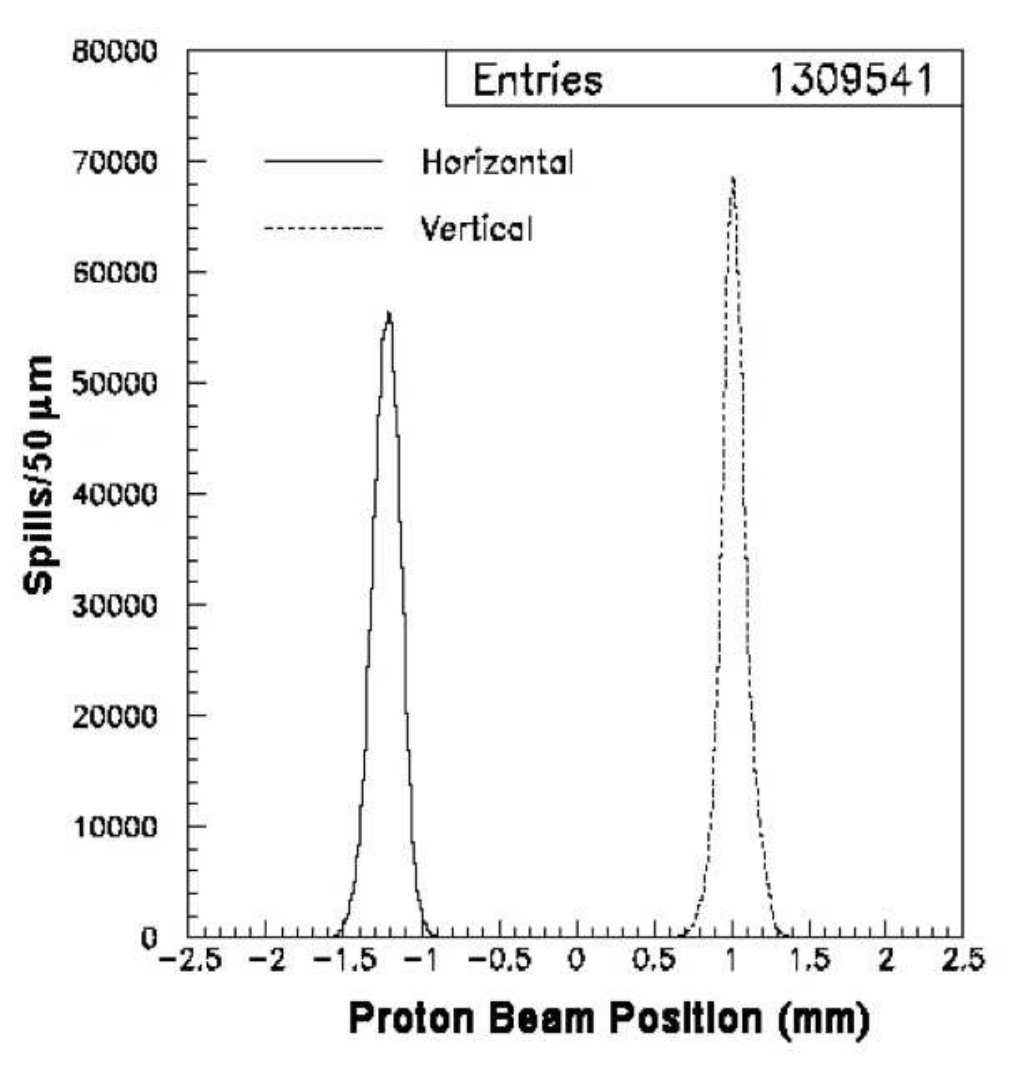}
    \caption{Distribution of Proton Beam Positions in Vertical and Horizontal Planes. The proton beam positions are determined by the target BPMs. The beam is centered on the target, which is located at ($x$,$y$)=(-1.2,+1.0)~mm. The beam is stable to an RMS variation of approximately 90~$\mu$m over the 1.3 million spills shown. }
\label{fig:figSamC2}
\end{centering}
\end{figure}

The currents in the trim magnets in the NuMI beam line are adjusted on a pulse-by-pulse basis as necessary by the program Autotune, which was developed and has been used extensively at Fermilab \cite{Ref14in1669}. The ability to maintain precision control of beam positions on target has consistently been very robust. Fig.~\ref{fig:figSamC2} shows plots of vertical and horizontal proton beam positions at the target for all beam spills over a one month period. 

\begin{figure}
\begin{centering}
    \includegraphics[width=.9\textwidth]{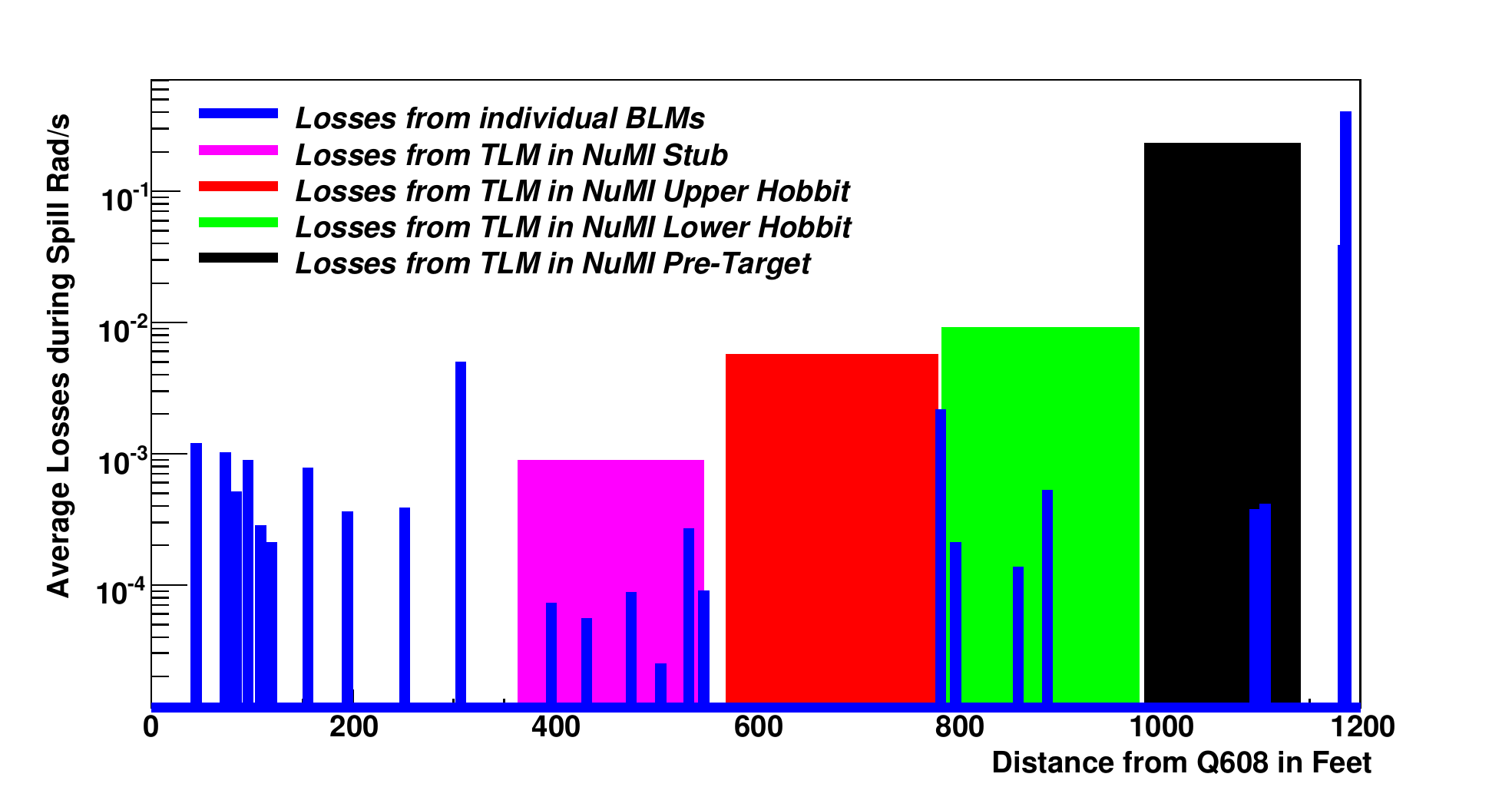}
    \caption{Average Losses along NuMI Beam Line during the 10~$\mu$s Beam Spill. The losses are shown for the beam in NuMI Mixed Mode (see Section \ref{sec:oper}) in January 2006 and are well below the design limit all along the full length of the beam line. There are several sections of the NuMI primary beam line before the proton beam reaches the target. Those are in sequential order the NuMI Stub, the NuMI Upper Hobbit, the Lower Hobbit, and the Pre-Target section. They are shown in different colors in this graph. }
\label{fig:beamloss}
\end{centering}
\end{figure}

To guard against any potential malfunction, a comprehensive Beam Permit System (BPS) has been implemented which inhibits the beam when any malfunction is detected. 250 different inputs are provided to this system. The information from the BPM's about the position of the beam constitutes a very important input to this system. Single-point beam losses during operation, including through the NuMI beam extraction channel, are normally maintained at fractional levels of less than one part-per-million. The beam losses along the proton beam line are displayed in Fig.~\ref{fig:beamloss}. This very low beam loss environment is needed for robust environmental protection, and ensures negligible residual activation for proton beam-line components. Both Autotune and the BPS are discussed in more detail in Section \ref{sec:day2day}.

\subsection{Monitors in the Target Hall}
Measuring devices have also been installed on the baffle, on the target, and on and next to the horn system. The baffle system gives notice if an unusual fraction of the beam is mis-steered (or the beam is exceptionally large). Under normal operation only a sub-percent fraction of the beam strikes the baffle\footnote{Because of radiation backscatter one can only set a limit of less than 0.6\% of the beam scraping the baffle.}. When some beam protons strike the baffle, its temperature rises and is measured by a thermocouple mounted on its downstream end. The relationship between the temperature rise and the number of protons striking the baffle has been calibrated by steering a low-intensity beam into the baffle. This temperature is one of the inputs to the BPS. 

The target is equipped with two Budal monitors which measure the net charge kicked out of the target by the proton beam and have been discussed in Section \ref{sec:target}. The magnetic field of Horns 1 and 2 is monitored by a set of three pickup coils in each horn as has been previously described. The other relevant instrumentation associated with the horns is the cross hair system which is used for alignment of horns with the beam. Cross hairs are vertical strips of aluminium mounted on the downstream end of Horn~1 and on both ends of Horn~2, 2.5~mm away from the beam center line (one on each side for Horn~2).  The Horn~1 cross hair is 12~mm deep and the Horn~2 cross hair strips are 36~mm deep on the upstream end and 12 mm deep on the downstream end. During horizontal scans of the beam when the protons sweep across the hairs the extra proton interactions in the cross hairs increase the size of the signal in the associated BLM’s which are mounted downstream of the horn and away from the beam. The calculated precision of the alignment with this technique is $\pm0.5$~mm. Each cross hair also has an additional horizontal “nub” that is 1~mm in vertical height and extends horizontally 3.5~mm towards and across the center of the beam line. It is used in the scans for vertical alignment. The loss in neutrino interaction rate due to this extra cross hair material in the beam is 0.2$\%$.

\subsection{Hadron and Muon Monitors}
\label{sec:muonmon}
The location of the hadron and muon monitors \cite{BeamMonitors} is shown in Fig.~\ref{fig:absorberenclosure}. Both those monitors are based on the ionization chamber technology but they are somewhat different mechanically and have different functions. The hadron monitor is installed at the end of the decay pipe 80~cm in front of the absorber. Thus it can be used to measure mainly the residual hadron flux consisting of both uninteracted protons and secondary hadrons that did not decay or previously interact. It is used to track the proton spot and the integrity of the NuMI target. It is in a very high radiation environment, up to $10^{9}$ charged particles/cm$^2$/spill during beam operation, consisting predominantly of 120~GeV protons which have not interacted in the target or in the decay pipe. In addition, the monitor sees $2\times10^{9}$ neutrons/cm$^{2}$/spill, largely as a result of splashback from the hadron absorber. The particle fluences result in 1.3~Grad/year at the monitor center and 1.0~Grad/year at its edge. Initially, because of budgetary reasons, the plan was that the hadron monitor would be used only for low-intensity beam tuning and beam alignment and probably not survive lengthy exposure in full-power beam. Accordingly the initial construction anticipated that fact and provision was made for access to the hadron monitor so that later designs for a removable and replaceable system could be incorporated. The hadron monitor was found to be very useful as discussed in Section \ref{sec:beamalign} and after the first unit failed two more were built with special mounting and a removal system designed so that the chamber could be moved out of the beam in case of failure. 

\begin{figure}
\begin{centering}
    \includegraphics[width=.76\textwidth]{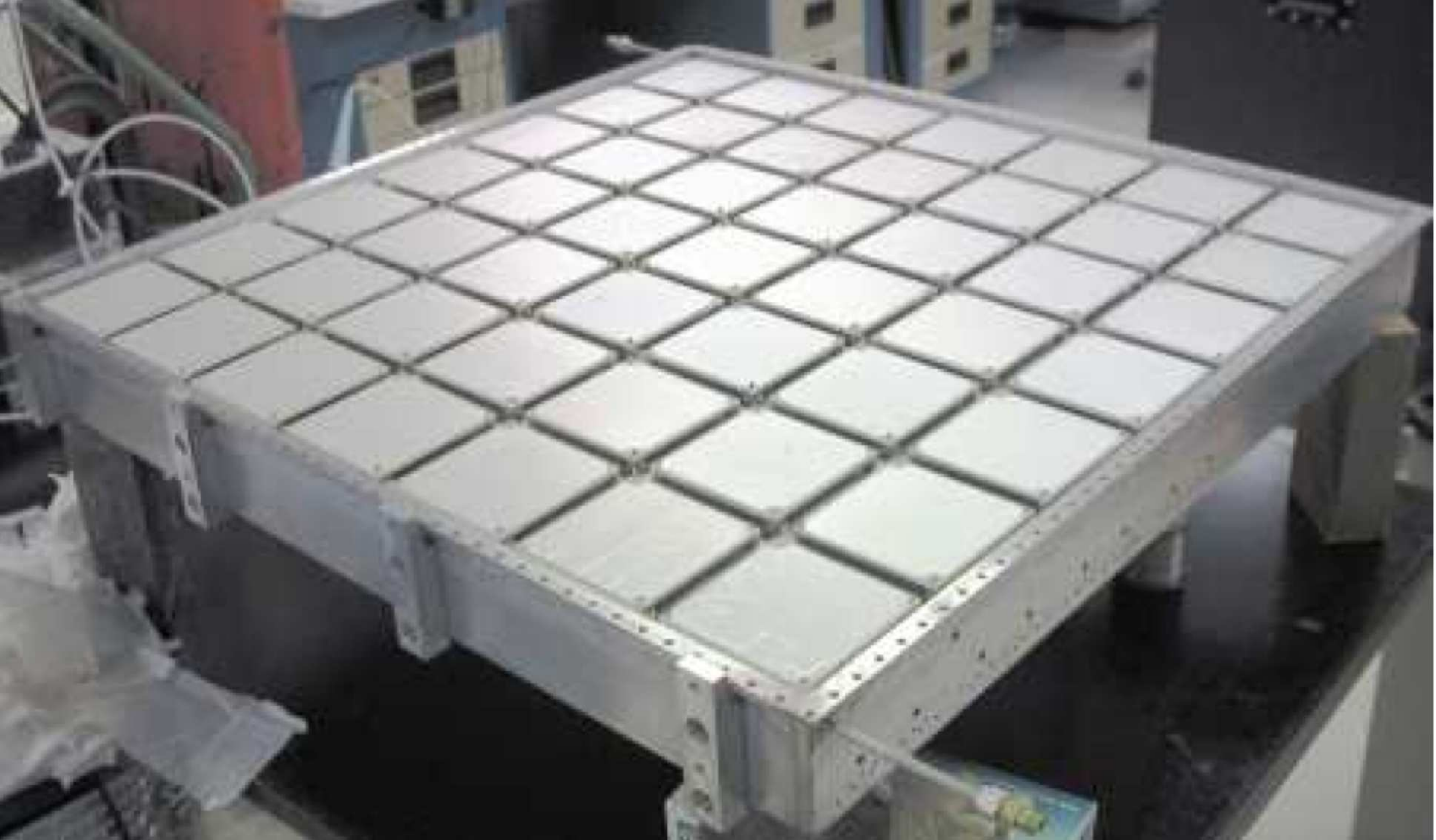}
    \caption{Photograph of the Hadron Monitor Ionization Chamber Array during Assembly. The array has 49 chambers and has a total area of approximately 1~m $\times$ 1~m. The hadron monitor chamber array uses helium as the ionization medium. }
\label{fig:hadmon}
\end{centering}
\end{figure}

The hadron monitor is a square array of 49 chambers each one 10.2~cm on the side. The individual chambers are parallel plates made from ceramic wafers with Ag-Pt electrodes separated by a 1~mm gap. They are mounted on the rear wall of a single aluminium chamber sealed with Pb-Sn wire. The signal and HV cables are transmitted to the monitor edge through ceramic feedthroughs to custom-made cables with aluminium core, ceramic tube insulators and aluminium sheath shields. There the cables are spliced to a kapton-insulated coaxial cable. A photograph of the ionization chambers during assembly is shown in Fig.~\ref{fig:hadmon}. This chamber array, as well as the muon ones, uses helium gas as the ionization medium.

The three muon monitors are in specially-excavated muon alcoves in the dolomite downstream of the NuMI hadron absorber. In their locations the only charged particles which survive are muons and thus they are not exposed to large amounts of radiation. The three locations correspond to detection thresholds due to muon range-out of 4~GeV, 10~GeV, and 20~GeV. By providing a 2-dimensional profile of the muon beam at these locations, the quality and relative intensity of the beam can be monitored on a pulse-to-pulse basis. 

\begin{figure}
\begin{centering}
    \includegraphics[width=.8\textwidth]{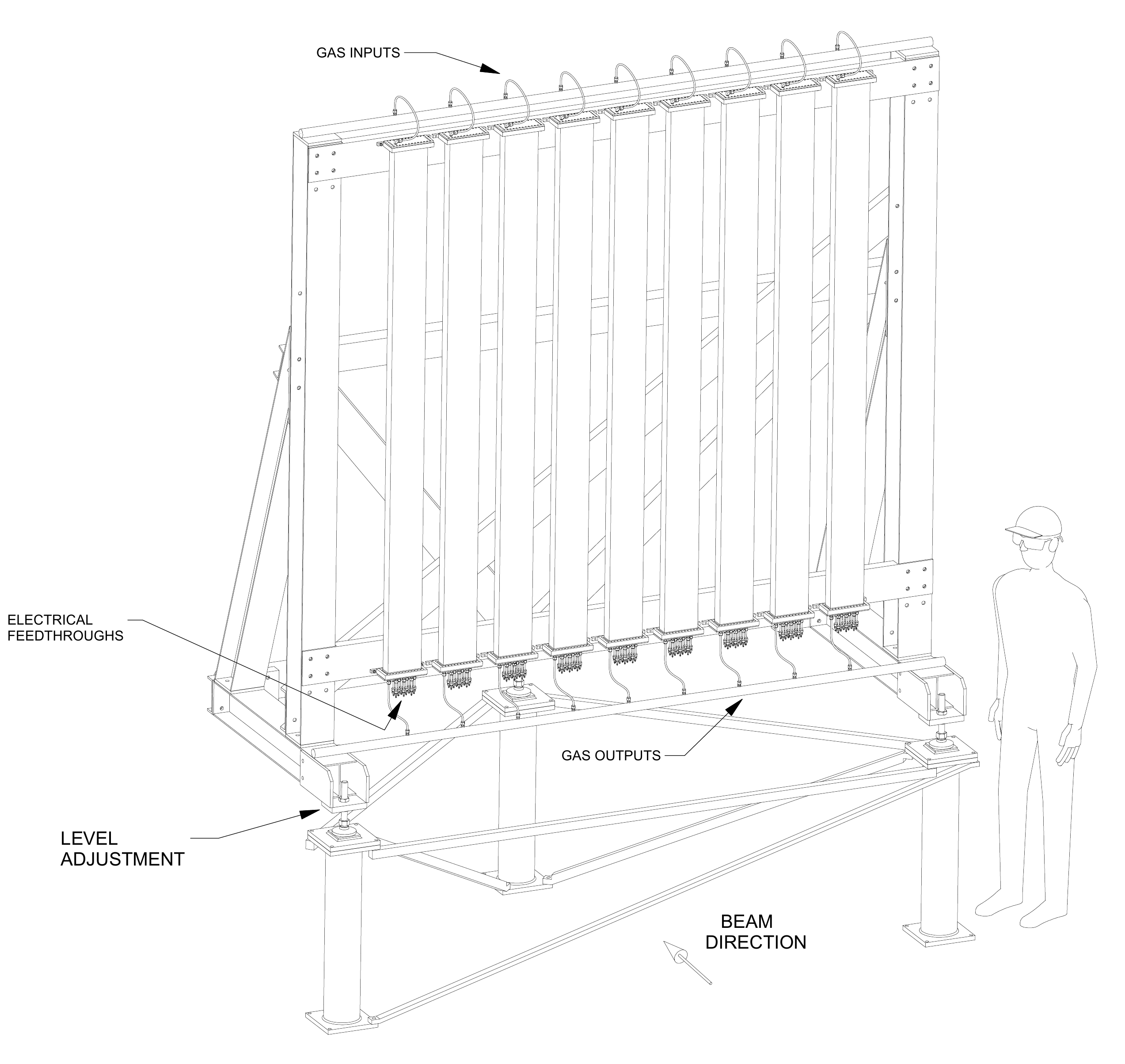}
    \caption{ Schematic Drawing of a  NuMI Beam Muon Monitor. The mounting of the nine tubes which contain the nine ionization chambers is shown to illustrate the construction and size of the muon monitors.  }
\label{fig:muonmon}
\end{centering}
\end{figure}

\begin{figure}
\begin{centering}
    \includegraphics[width=.8\textwidth]{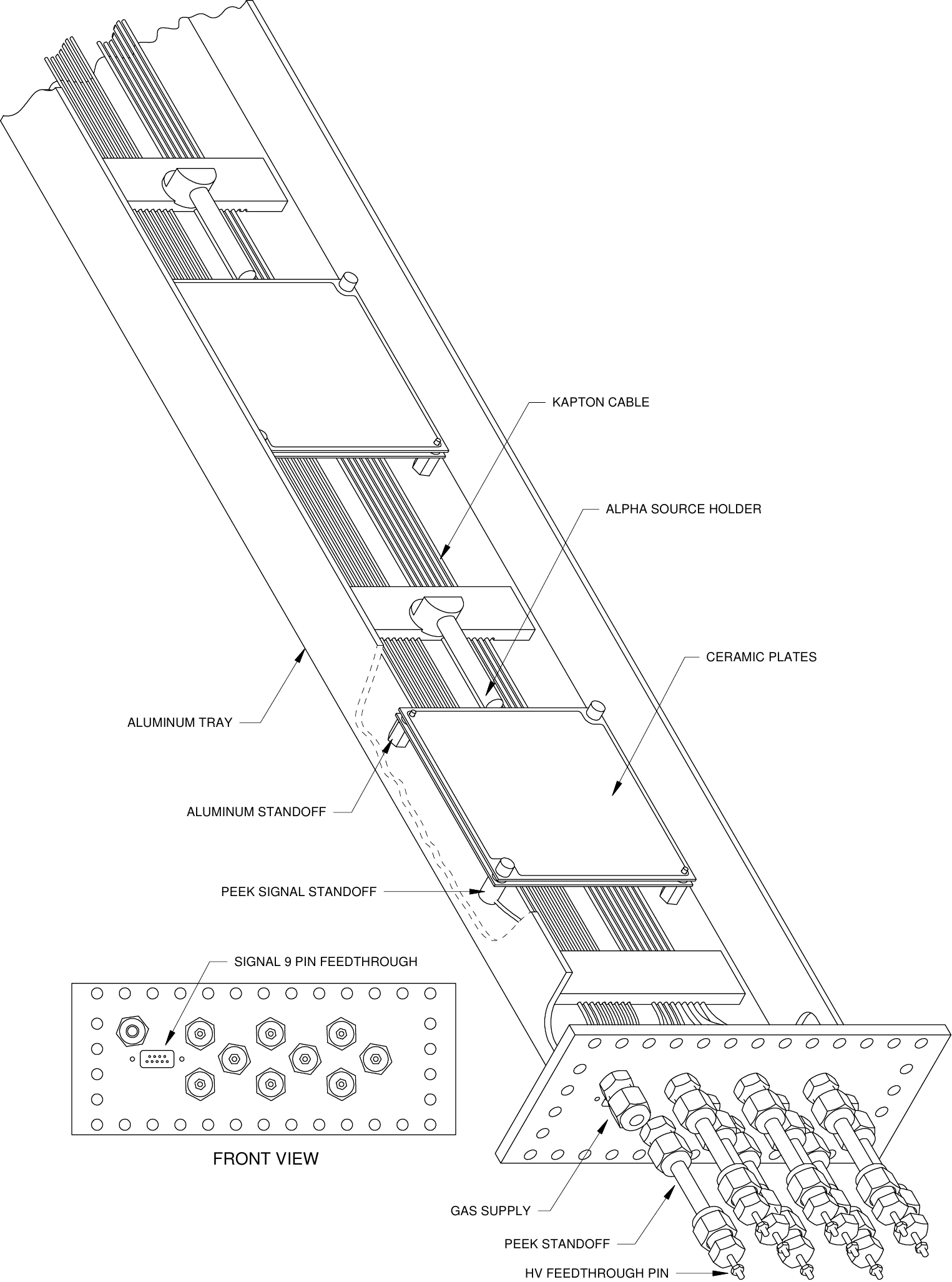}
    \caption{ View of a Portion of a Muon Monitor “Tray”. The individual chambers are mounted to an aluminium channel which sits in the tray. The cables run down the length of the channel from the chambers to the feedthroughs in the flange at the end of the tray. The front view of the flange is shown on the left. }
\label{fig:muonmon2}
\end{centering}
\end{figure}

The muon chambers are all of the same construction and are 2.3~m $\times$ 2.3~m in size. They consist of $9\times9$ orthogonal arrays of individual ionization chamber modules which are 10.2~cm on the side and have a 3~mm gap between their electrodes. Their support structure consists of 9 vertical tubes 15.24~cm wide and 228.6~cm high and made of 0.37~cm thick aluminium. Each vertical column of 9 chambers is located in a tray, which slides into one of these aluminium tubes, with signal and HV routed to the end of the tube via shielded Kapton insulated cables. Fig.~\ref{fig:muonmon} shows a schematic drawing of a muon monitor used in the NuMI beam line. Fig.~\ref{fig:muonmon2} shows a schematic view of a muon monitor ``tray'' with more detail.

\begin{figure}
\begin{centering}
    \includegraphics[width=.66\textwidth]{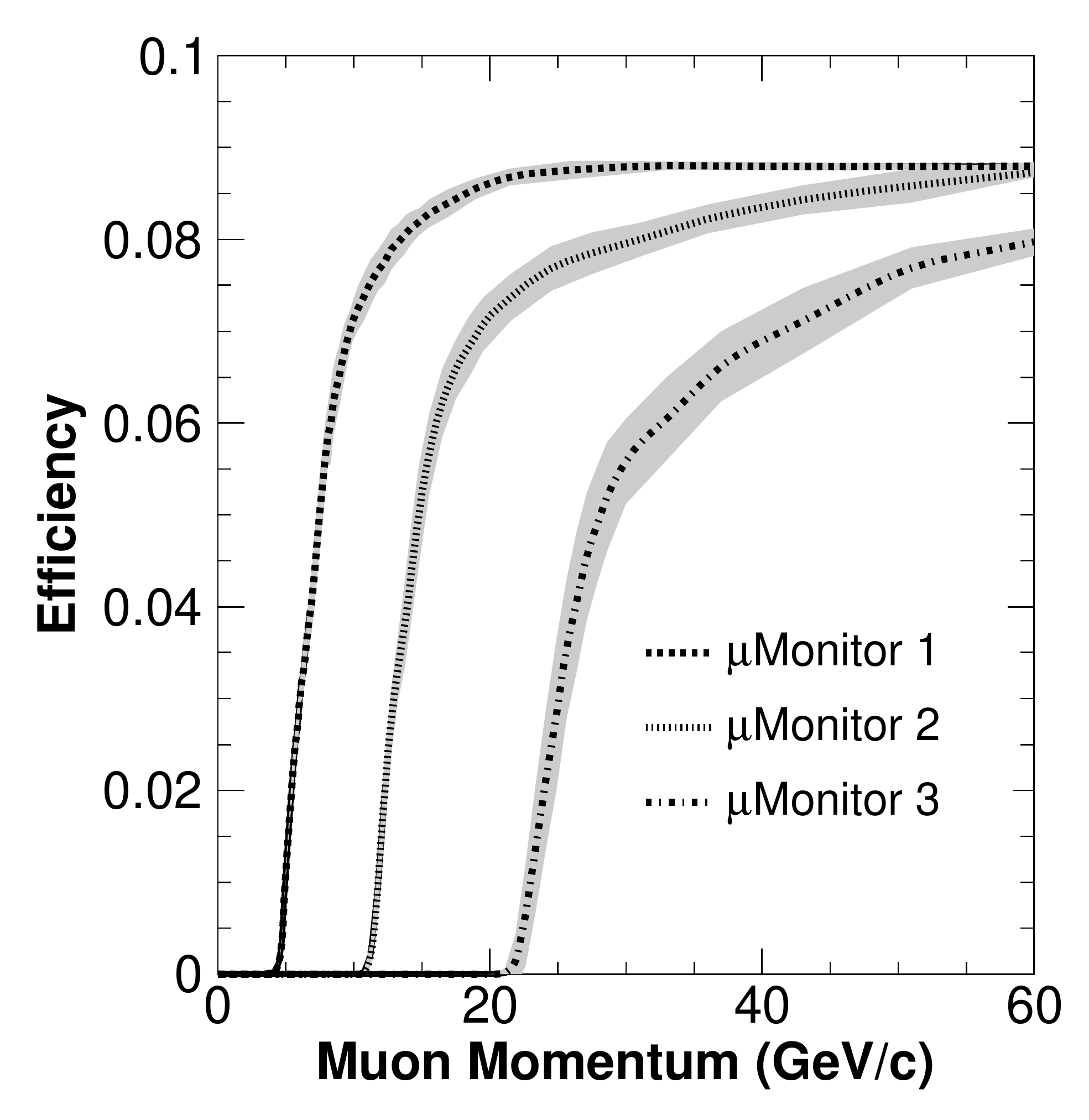}
    \caption{The Detection Efficiency for Muons in the Three Muon Monitors. The curves show the fraction of muons exiting the downstream end of the decay pipe which reach the three muon monitors as a function of their momentum at the end of the decay pipe. The curves were calculated using a Monte Carlo simulation with detailed material composition and absorber simulation. It can clearly be seen that the muon monitors that are separated from the decay pipe by more rock see muons with higher momentum as the lower momentum muons are stopped before ever reaching the monitors. }
\label{fig:muonrates}
\end{centering}
\end{figure}

Beam tests at Brookhaven National Laboratory \cite{HadronMuonMonitors} and at Fermilab \cite{HadronMuonMonitors2} indicate that the response quantifying the number of muons going through is linear up to a NuMI muon flux corresponding to at least $2.5\times10^{13}$ protons per spill. The chambers in the downstream alcoves show a large drop in rates. This is illustrated in Fig.~\ref{fig:muonrates}, which indicates the muon momentum threshold for each alcove.

Muon chambers have been used exclusively to study the profile of the downstream muon beam. Since muons and neutrinos are produced in the same pion or kaon decays, muon distributions reflect reasonably well the neutrino distributions except that they are somewhat modified by Coulomb scattering of the muons in the intervening rock. In principle the muon rates can also be used to obtain absolute neutrino intensities and an effort has been made to make progress on this front \cite{LauraThesis, ZarkoThesis}. The issue is quite complex, however, largely because of difficulties in evaluating the delta-ray contribution to the ionization deposited in the chambers.

\subsection {MINOS Near Detector}
Strictly speaking the MINOS Near Detector is not part of the beam instrumentation but rather a part of the MINOS experimental neutrino detection apparatus. It does provide important beam monitoring functions so it seems appropriate to include a few comments about it in this section. It is a magnetized segmented iron-scintillator calorimeter located 1.04~km downstream of the NuMI target. The location is behind a sufficient amount of rock that the only particles entering the ND are neutrinos and muons and possibly parts of hadronic showers from neutrino interactions just upstream of the detector. It has been described in detail in \cite{NIMdetectorpaper}.

Both the strength and the weakness of the Near Detector as a beam monitor lie in the fact that it deals with final neutrino interactions and hence higher-level information. One of the implications is that neutrino event statistics need time to accumulate and thus longer time scales may be necessary for identifying potential problems. The other implication is that much higher quality information can be obtained, for example from the full fit of the interactions, so that a more precise diagnosis of a problem can be made. A good example here is the damage of Target number 2 (discussed in the next section) where information on the neutrino spectrum was crucial for understanding the the nature of the problem. However such information comes in days or weeks rather than on a pulse-by-pulse basis.

\section{Beam Operations}
\label{sec:operations}
This section is concerned with various issues connected with the operations of the NuMI beam. The initial subsection deals with the beam-based alignment procedures that were followed whenever there was a major change in the beam elements, for example the insertion of a new target or a new horn. Normal operations then resumed only when satisfactory alignment of the key elements was achieved. The next subsection discusses the normal mode of operation of the beam and how ``slip-stacking'' techniques are utilized to increase the number of protons delivered to the NuMI target. Daily NuMI beam operations are discussed along with an overview of the performance of the beam during the seven year long run of the MINOS experiment.

\subsection{Beam-Based Tuning and Alignment}
\label{sec:beamalign}
Beam-based alignment is a procedure to align the beam components to the proton beam position as measured by the primary beam instrumentation, in order to limit systematic errors that would arise from any deviation.  The neutrino spectrum of the NuMI beam is quite sensitive to the relative alignment of the target and of the horn system. Sub-millimeter precision is required in these components, particularly the target, protective baffle, and two focusing horns. Small misalignments of the components will not only change the overall rate of the beam, but also distort the spectrum; they do so differently in the Near and Far Detectors, potentially biasing a neutrino oscillation measurement.  In this subsection the method used to achieve this relative alignment is described. This discussion is very much based on material presented in \cite{BeamAlignment}, where a more detailed discussion of these procedures is given. 

An optical survey was performed of all the beam components prior to initiation of the beam. The accuracy of this survey for the beam components in the Target Hall was $\pm0.5$~mm. The absorber, 725~m away from the target, was surveyed with relative accuracy of $\pm2.5$~cm. An initial beam-based verification of the optical survey was performed prior to installation of the target. The proton beam passed unimpeded through the decay volume to the hadron monitor 732~m away from the two most downstream SEM’s,  which measured the beam position and direction at the target location. The agreement between the actual measurement and the predicted location was better than 2~cm in both directions, verifying the primary beam direction obtained from the optical survey at the level of 30~$\mu$rad.

As discussed previously in Section \ref{sec:targetcarrier}, the target and baffle are rigidly mounted on the same assembly and are thus fixed in position and angle with respect to each other. Their positions in the beam line were initially measured with the optical survey.  Beam-based alignment is performed after all the shielding is assembled by rastering a low intensity proton beam across physical features of the beam components. The primary beam optics allow for transverse movement of the proton beam while remaining parallel with the central axis. For these measurements, the proton intensity was significantly reduced to 4-8$\times 10^{11}$ protons per pulse; higher intensities would risk damage to beam components such as the target cooling lines and horn conductors. The initial beam alignment took place on March 3 and April 25, 2005. The beam dimensions during the first alignment run were $\sigma_{x}\times \sigma_{y} = 0.7\times 1.4$~mm$^2$. During the second alignment run they were $0.9\times0.9$~mm$^2$. The latter beam aspect ratio corresponded closely to the nominal beam spot used for data taking runs at full intensity.

These alignment procedures have been repeated whenever the target has been moved or replaced. The method takes advantage of the facts that the proton beam is attenuated and generates additional radiation as it passes through material, and that the RMS width of the  beam increases due to multiple scattering. Variation of the intensity and spatial distributions measured in the hadron monitor and loss monitors associated with the horn crosshairs are used to determine the positions of the elements being studied. The measured data are correlated to the proton beam position and are fit to a set of empirical functions to obtain quantitative determination of device positions and angles.

\begin{figure}
\begin{centering}
    \includegraphics[width=.66\textwidth]{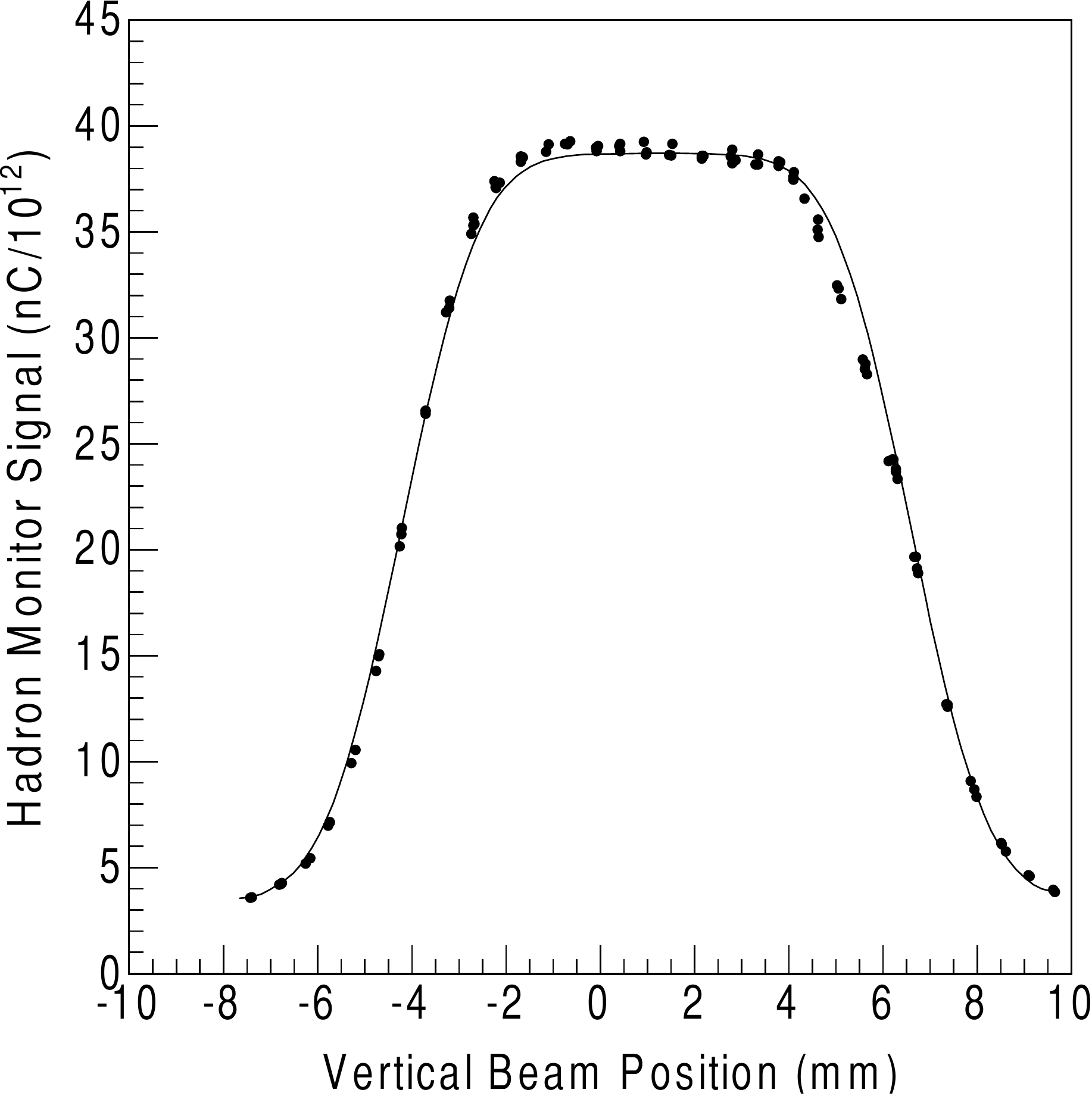}
    \caption{Vertical Scan of the Proton Beam Across the Target and Baffle. Plotted is the total amount of charge collected in the Hadron Monitor on March 5, 2005, normalized by proton beam intensity, as a function of proton beam position at the target.  The edges on each side correspond to the edges of the baffle passage.}
\label{fig:fig10-2099}
\end{centering}
\end{figure}

Fig.~\ref{fig:fig10-2099} shows the charge collected in the hadron monitor as the proton beam is scanned vertically across the target and baffle. In the central part of the graph the proton beam passes through the target and is attenuated by 2.0 interaction lengths of the target graphite. The signal drops at either side corresponding to the beam entering the baffle and experiencing 3.1 additional interaction lengths of graphite in the baffle. Horn focusing was disabled for this measurement. The data indicated that the position of the baffle was at $+1.2\pm0.1$~mm with respect to the proton beam axis. 

\begin{figure}
\begin{centering}
    \includegraphics[width=.66\textwidth]{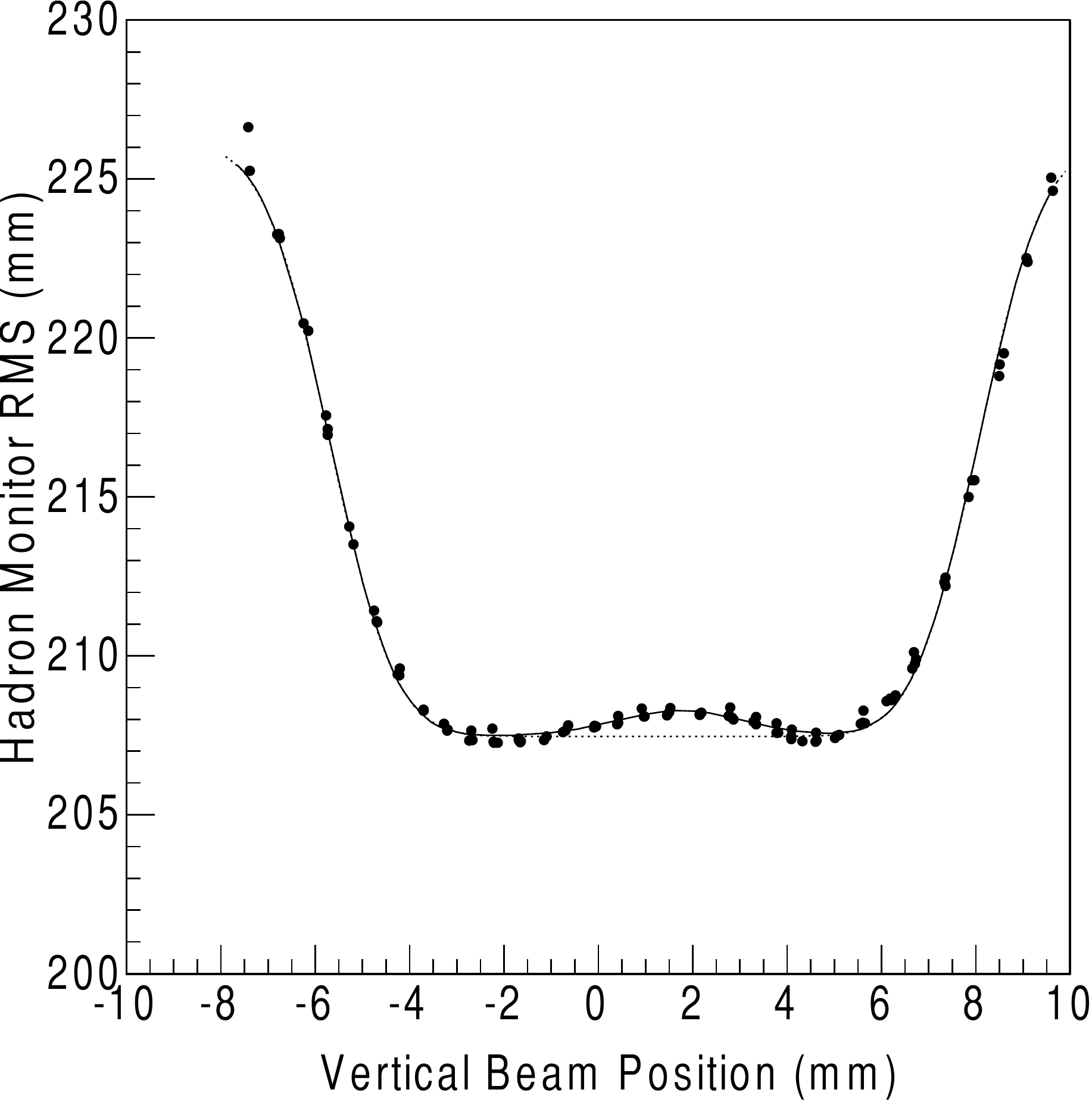}
    \caption{Vertical Scan of the Proton Beam Across the Target and Baffle. Plotted is the RMS of the measured distribution in the Hadron Monitor shown as a function of proton beam position at the target. The position of the horizontal fin is indicated by an increase in RMS due to scattering. The solid line is a fit to a simple model of absorption, with the dashed line showing the fit without the effect of the horizontal fin.}
\label{fig:fig11-2099}
\end{centering}
\end{figure}

The vertical position of the target can be obtained from the increased RMS width of the beam due to the presence of the horizontal fin which provides additional mass in the path of the beam.  The RMS distribution is shown in Fig.~\ref{fig:fig11-2099}. The RMS is higher in the wings than in the center due to scattering while passing through the baffle. The horizontal fin causes the small increase near the center. A Gaussian fit to this structure gives the value for the center of $+1.8\pm0.1$~mm. 

\begin{figure}
\begin{centering}
    \includegraphics[width=.66\textwidth]{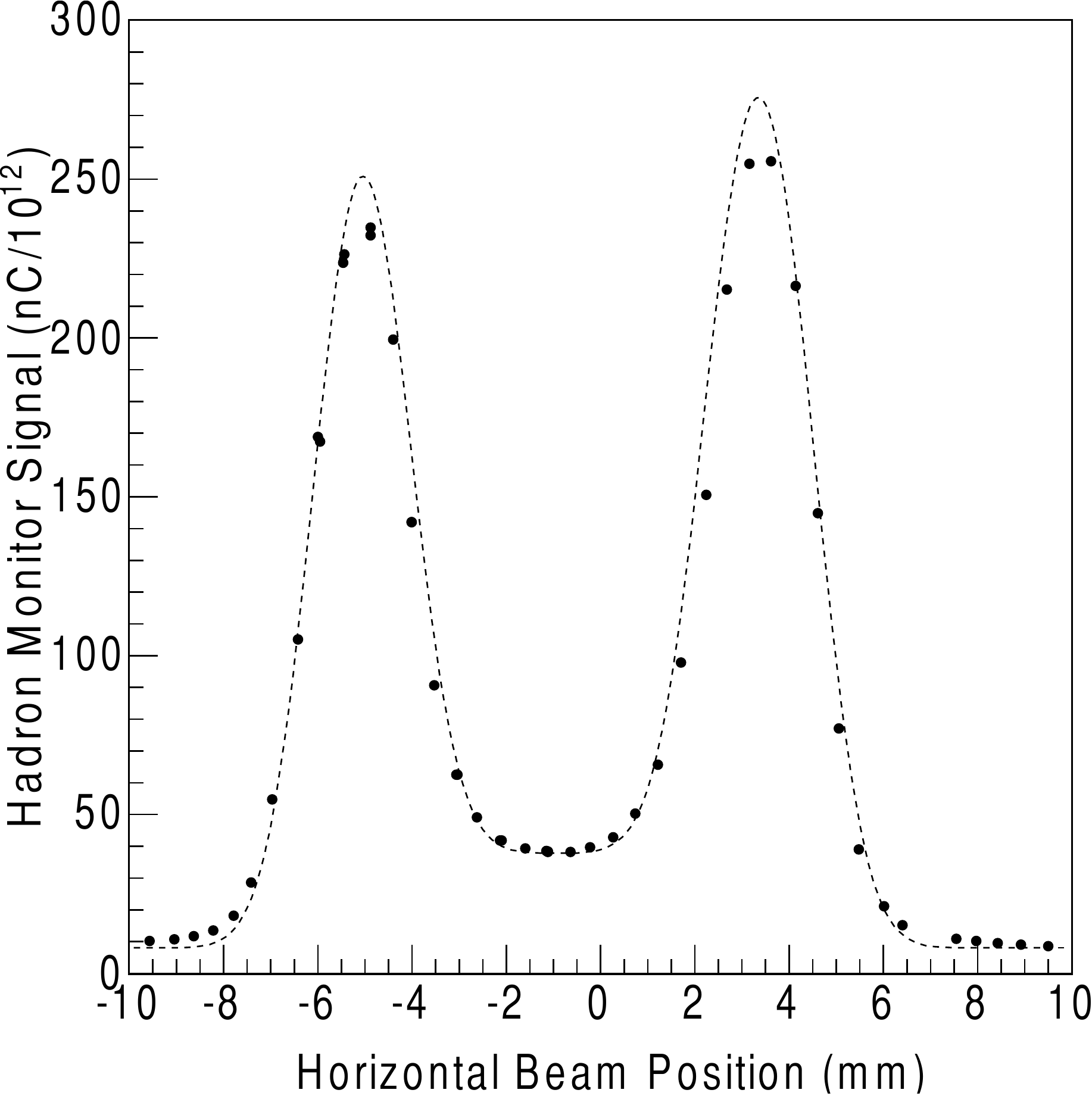}
    \caption{Horizontal Scan of the Proton Beam across the Target and Baffle. Data is shown from a single horizontal scan that establishes the horizontal positions of the target and baffle. Plotted is the total Hadron Monitor charge normalized by proton intensity and displayed as a function of the horizontal beam position at the target. The edges of the target are found by fitting to the central dip; the inner edges of the baffle are found by fitting to the outer dips.}
\label{fig:fig12-2099}
\end{centering}
\end{figure}

The horizontal beam scan measurement is displayed in Fig.~\ref{fig:fig12-2099}. The beam is scanned horizontally and it passes first through the baffle, then through the gap between the baffle and the target,  then through the target and then again through the gap, and finally the baffle. The beam is unattenuated in the gap and attenuated heavily in the baffle that is longer than the target. The fit to the baffle width gives 10.7~mm, suggesting that it was off-center relative to the beam by as much as 1.3~mm vertically, or at an angle up to 200~$\mu$rad.

\begin{figure}
\begin{centering}
    \includegraphics[width=.66\textwidth]{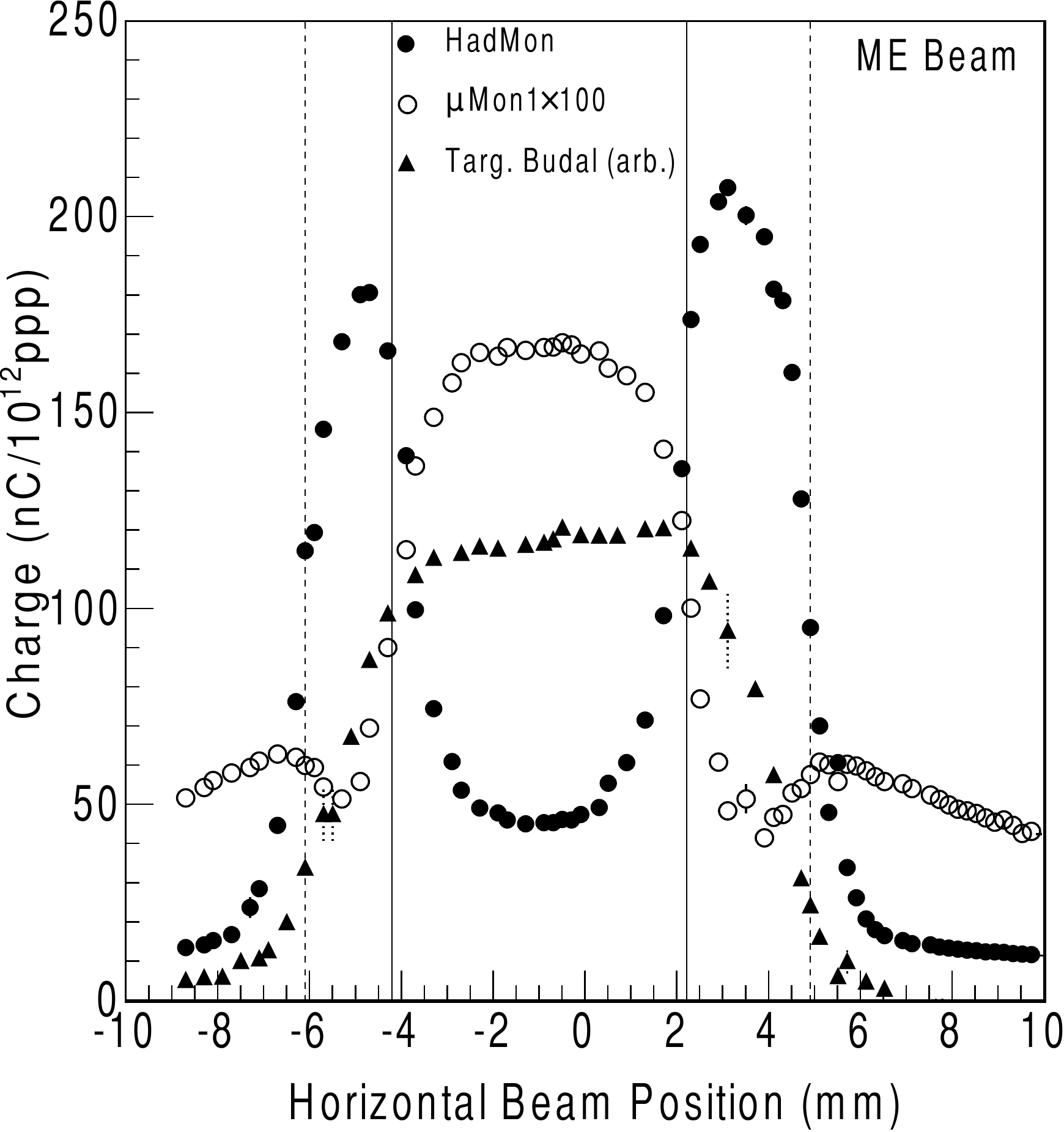}
    \caption{Horizontal Scan of the Proton Beam Across the Target and Baffle. This graph shows data from a single horizontal scan of the target showing the pulse height seen in the Hadron Monitor, first Muon Alcove, and the target Budal monitor. The solid (dashed) lines indicate the edges of the target (baffle) as indicated by the previous beam alignment scans using the Hadron Monitor. }
\label{fig:fig13-2099}
\end{centering}
\end{figure}

In principle, both the muon monitors and the Budal monitors (see Sections \ref{sec:target} and \ref{sec:muonmon}) can also be used to obtain positions of the target and the baffle. They were not used in this analysis because of the inherent systematic difficulties that are more severe than with the hadron monitor. Their qualitative results, however, are  consistent  with the hadron monitor measurement as can be seen in Fig.~\ref{fig:fig13-2099} showing the profiles for the hadron monitor, first muon monitor, and the Budal monitor from a horizontal scan.

\begin{figure}
\begin{centering}
    \includegraphics[width=.66\textwidth]{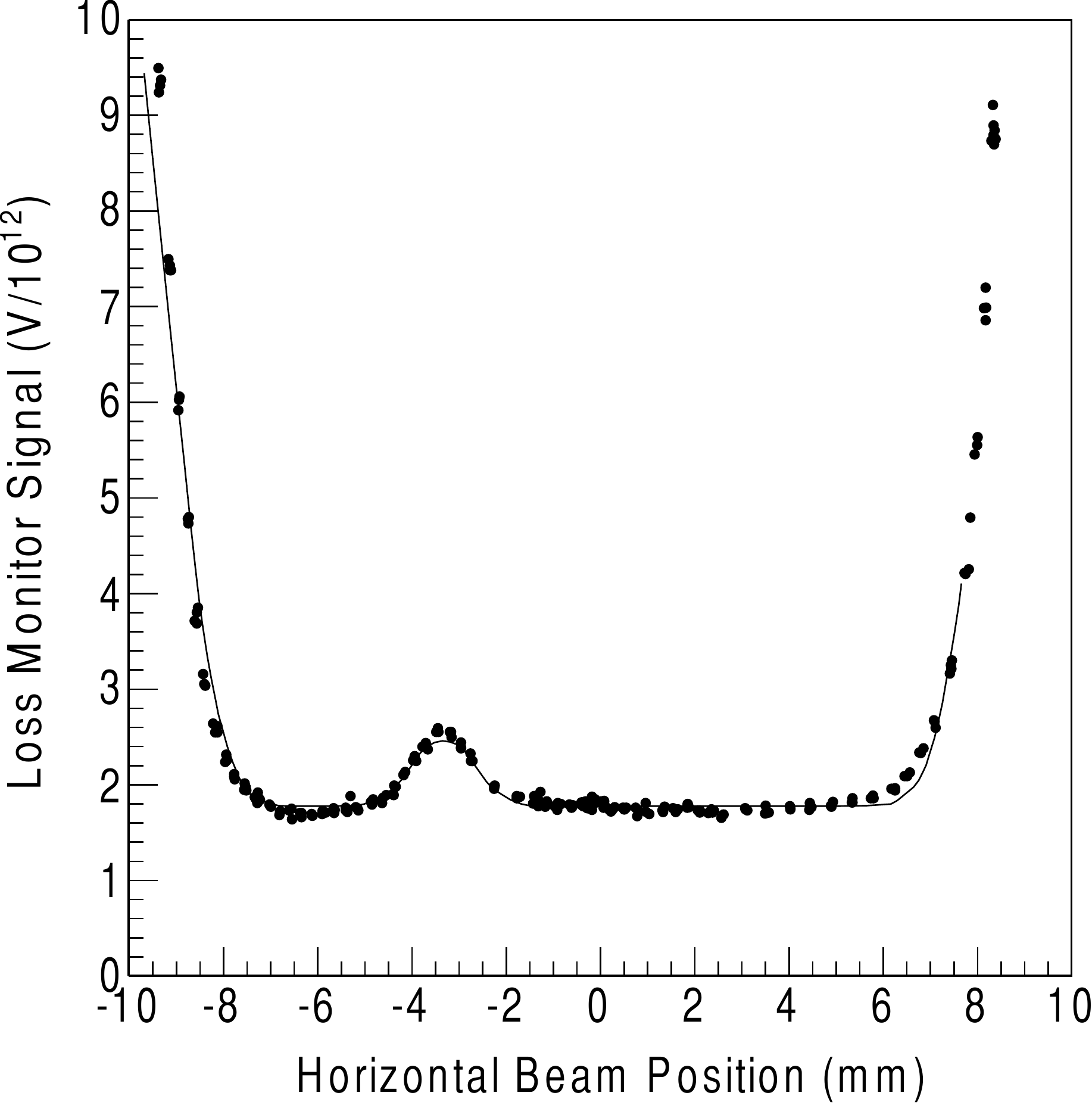}
    \caption{Horizontal Scan of the Proton Beam Across Horn~1. The signal is shown for the Horn 1~loss monitor as a function of horizontal beam position in mm. Superimposed on the data is the fit used to determine the center of the horn neck from the outer peaks, and the crosshair position from the central peak. }
\label{fig:fig15-2099}
\end{centering}
\end{figure}

The alignment of horns was performed with similar techniques. The target-baffle carrier was removed for these studies to minimize the proton beam size at the horn locations. When the proton beam passes through the cross hairs mounted on the horns, secondary particles and radiation produce signals in ionization loss monitors adjacent to the horns. The location of the horns is established by variation of this signal as the beam position is changed. Fig.~\ref{fig:fig15-2099}, showing the signal on Horn~1 loss monitor from a horizontal scan, illustrates the technique. The steep edges on each side correspond the the horn neck (minimum inner radius of 18~mm). They allow determination of the position of the center of the horn neck as -0.46~mm, compared to 0~mm for perfect alignment. The bump near -4~mm is due to the cross-hair and gives its position as -3.36~mm which should be -2.5~mm for perfect alignment.  The position and angles of the horn are determined by the two independent measurements of horizontal and vertical positions. 

Variation between different measurements of the same parameters and the uncertainties in the fit results allow estimation of the position uncertainty as $\pm0.3$~mm for the baffle, target and Horn~1 and $\pm0.5$~mm for Horn~2. The horn angle uncertainty is estimated as $\pm0.2$~mrad. The results of the fits for the different measurements are summarised in Table \ref{tab:tab1-2099}.

\begin{table}
\begin{centering}
  \begin{tabular}{|c|c|c|c|c|}
    \hline
    {\bf Device} & {\bf Direction}
    & {\bf Offset (mm)} & {\bf Angle (mrad)} \\
    \hline
    Baffle & Horz & -0.75 & $<$ 0.14 \\
    \cline{2-4}
    Baffle & Vert & +0.14 &  \\
    \cline{1-4}
    Target & Horz & -0.95 & $<$ 0.14 \\
    \cline{2-4}
    Target & Vert & -0.90 & \\
    \hline
    Horn 1 & Horz & -0.65 & -0.18 \\
    \cline{2-4}
    Horn 1 & Vert & -0.33 & +0.20  \\
    \hline
    Horn 2 & Horz & -1.01 & -0.11 \\
    \cline{2-4}
    Horn 2 & Vert & -1.61 & -0.42 \\
    \hline
  \end{tabular}
    \caption{Positions and angles of Target Hall components as measured with the beam-based alignment. There is only one position measurement in the vertical direction for the target and baffle, so there is no angle information.}
\label{tab:tab1-2099}
\end{centering}
\end{table}

Several of the deviations from the ideal alignment would have led to unacceptably large biases or systematic errors for the neutrino oscillation measurements.  Particularly, the target horizontal offset and baffle offsets would have caused a deviation greater than 2\% in the calculated ratio  between measurements in the Near and Far Detectors, potentially spoofing or masking an oscillation signal, or giving erroneous values of the oscillation parameters.  Some of the offsets were corrected by redirecting the beam. During the data running the beam was steered to x = -1.2~mm and y = +1.0~mm at the target. The neutrino spectrum is not very sensitive to the vertical position on the target because of its vertical orientation.  Accordingly, it was chosen to steer the beam to the vertical center of the baffle to minimize scraping.  The reduced offsets and angles are listed in Table \ref{tab:tab2-2099}.  Additionally tabulated are the maximum systematic distortions of the Near-to-Far ratio of the energy spectra in the MINOS detectors corresponding to the residual errors.

The beam based alignment is done with all the shielding in place, while the optical survey is done before placement of the top concrete covers.  The survey monuments in the target hall were seen to move when the concrete shielding covers were added by an amount that indicates that  much of the difference between the optical survey and the beam based alignment may be due to deformation of the target pile walls under the load of the covers. 

\begin{table}
\begin{centering}
   \begin{tabular}{|c|c|c|c|c|c|}
    \hline
    {\bf Device} & {\bf Dir.} & {\bf Offset} & 
    {\bf Effect \%} & {\bf Angle} & {\bf Effect \%} \\
    \hline
    Baffle & Horz & 0.0 mm & $<$ 0.1 & -0.1 mrad & $<$ 0.1\\
    \cline{2-6}
    Baffle & Vert & +0.1 & $<$ 0.1 & -0.7 & $<$ 0.1 \\
    \hline
    Target & Horz & -0.2 & 0.4 & -0.1 & $<$ 0.1 \\
    \cline{2-6}
    Target & Vert & -0.9 & $<$ 0.1 & -0.7 & 0.3 \\
    \hline
    Horn 1 & Horz & -0.0 & $<$ 0.1 & -0.2 & 0.3\\
    \cline{2-6}
    Horn 1 & Vert & -0.2 & $<$ 0.1 & +0.3 & 0.4\\
    \hline
    Horn 2 & Horz & -0.6 & 0.2 & -0.2 & $<$ 0.1 \\
    \cline{2-6}
    Horn 2 & Vert & -0.9 & 0.4 & -0.4 & $<$ 0.1 \\
    \hline
  \end{tabular}
    \caption{Positions and angles of Target Hall components, after repositioning of the target and proton beam. Also listed are the calculated systematic effects on the Near-to-Far ratio of the neutrino energy spectrum in MINOS detectors for that offset or angle.  Several of the individual effects would exceed the 2\% error budget without the alignment procedure. }
\label{tab:tab2-2099}
\end{centering}
\end{table}

\subsection{NuMI Mode of Operation}
\label{sec:oper}
The Fermilab Main Injector, which is used to create the NuMI beam, receives 8~GeV protons from the Fermilab Booster and accelerates them to 120~GeV. The full acceleration cycle starts with a Cockroft-Walton (replaced by an RF quadrupole after the MINOS experiment was completed) which accelerates H$^{-}$ ions to an energy of 750~keV. Subsequently a 201.2~MHz linac accelerates them to 400~MeV and sends them through a thin (typically 400-600~$\mu$g/cm$^{2}$) carbon foil to the Booster. The foil strips two electrons from each ion and converts the ions from H$^{-}$ to H$^{+}$. The Booster then accelerates those protons to 8~GeV. The Booster is a rapid cycling synchrotron with a sinusoidal magnet ramp with a frequency of 15~Hz, i.e.\ a nominal cycle time of 67~ms. The Booster has a harmonic number\footnote{The harmonic number is defined as the ratio of accelerator RF frequency to its revolution frequency, in other words it takes protons 84 RF cycles to do a lap of the 472~m circumference of the Booster. The harmonic number is also the maximum number of bunches that can be stored and accelerated in a given synchrotron.} of 84 and an accelerating RF that ramps up the frequency from 37.8~MHz at injection to 52.8~MHz at extraction to match the RF frequency of the Main Injector. At low energy, a three-bucket wide ``notch'' is formed by removing three bunches with a dedicated kicker, leaving a gap to allow for the rise time of the extraction kickers, and a ``batch'' of typically 81 proton bunches to be delivered to the Main Injector.

\begin{figure}
\begin{centering}
      \includegraphics[width=.9\textwidth]{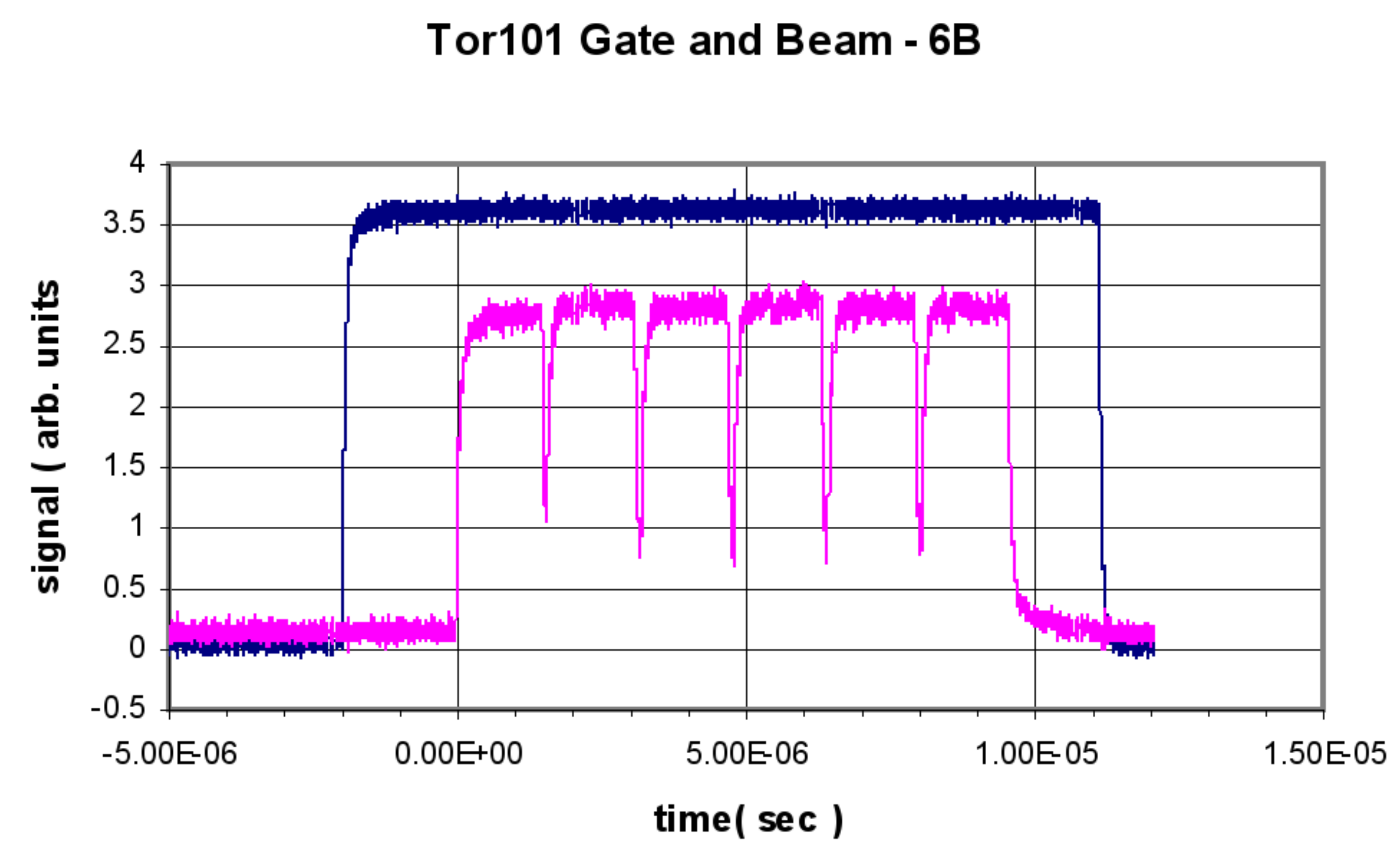}
      \caption{An Oscilloscope Trace of the Proton Current as Seen by the NuMI Toroid in NuMI-Only Mode. This is shown as a pink line, and the dark blue line corresponds to the beam trigger window. The Booster batch structure is clearly visible. The first batch is usually sent to the anti-proton source and is then lost to the NuMI beam. The remaining 5 batches are extracted and sent to NuMI. }
\label{fig:tor101}
 \end{centering}
\end{figure}

The circumference of the Main Injector is seven times that of the Booster which allows the injection of seven Booster batches\footnote{The Main Injector has a harmonic number of 588 (7$\times$84).}. However, one slot must stay empty so as to allow the extraction kicker to ramp up, so a maximum of six proton batches can be accelerated in the Main Injector. At the beginning of the MINOS run there were two alternative modes of operation. The one used most of the time, ``mixed mode'', involved first injecting two batches into the Main Injector and then coalescing them via the slip-stacking technique described below. Subsequently five more batches were injected and all six resulting batches were accelerated to 120~GeV. The double bunch was then extracted and sent to the Accumulator for antiproton production and the remaining five were sent to the NuMI beam line. An alternative running mode was the ``NuMI-only mode''. It was used when the Accumulator could not accept protons; in this mode all six batches from the Main Injector were sent to the NuMI beam line. This was the mode of operation after the Tevatron shutdown. This multi-batch injection required development of synchronisation between the Booster and the Main Injector which had never been required before at Fermilab \cite{bobthesis}. Fig.~\ref{fig:tor101} shows the Booster batch distribution. %Fig.~\ref{fig:numispills} shows a comparison of the mixed mode batch spill structure and the NuMI-only mode spill structure as seen by the MINOS Near Detector. 

\begin{figure}
\begin{centering}
    \includegraphics[width=.7\textwidth]{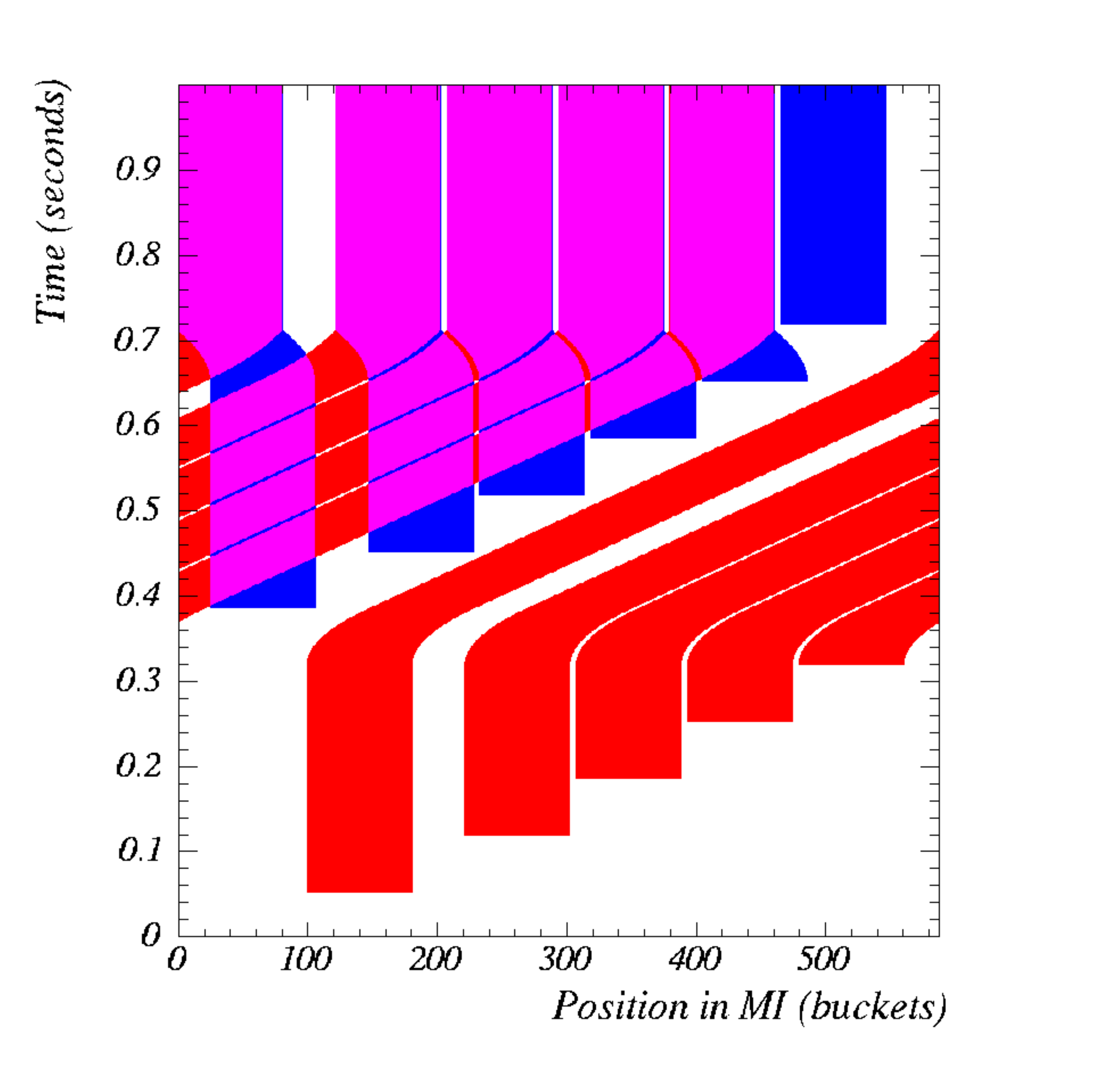}
    \caption{Schematic of the Slip-Stacking Technique for 11 Batch Operation. The red bands correspond to the initially injected batches, the blue ones to the second set. The purple bands are double (slip-stacked) batches. The first slip-stacked batch is destined for antiproton production and thus a short gap is created after it to allow for the extraction kicker rise times. }
\label{fig:slipstack}
\end{centering}
\end{figure}

During the course of the MINOS experiment, the ``slip-stacking'' technique \cite{Brown:2013idd, SlipStacking}, which had been previously used to increase the intensity of the batch of protons devoted to antiproton production, was extended to increase the beam intensity for NuMI through the injection of additional batches into the Main Injector. It became operational in January 2008. In this mode, five batches of beam were injected into the Main Injector and deccelerated slightly, causing them to begin to slip around the circumference with respect to the nominal 8~GeV revolution marker. A second 53~MHz RF system was turned on at the nominal injection frequency, and a further five batches were injected into the gap left in the first set of batches. The frequency separation of these two RF systems was chosen to cause the two sets of batches to ``slip'' by a little more than one batch per 67~ms Booster cycle. Once the two sets of batches overlap in space, the full 1~MV of RF is switched on, recapturing two slipping bunches in one large 53~MHz bucket, and so producing five double-intensity batches. One additional, single intensity batch is injected from the Booster into the remaining gap, giving a cycle with 11 Booster batches producing five double and one single batch accelerated to 120~GeV. In normal operation, one double batch would be used for antiproton production and the remaining four large ones and one small sent to NuMI; when the antiproton source was not operating, all the proton batches were sent to NuMI. The slip-stacking technique is demonstrated visually in Fig.~\ref{fig:slipstack}.

The cycle time is the sum of injection, acceleration, and magnet ramp down-time. The injection time is proportional to the number of batches injected, each batch requiring 67~ms. The acceleration and ramp down each take about 0.7~s. Thus the minimum total cycle time for an 11 batch cycle in this mode of operation is about 2.2~s, and for a 6 batch cycle (no slip-stacking) about 1.8~s. 

\subsection{Day-to-Day Operations of the NuMI Beam}
\label{sec:day2day}
From an operational point of view there were four semi-separate systems which formed an integral part of the operation of the MINOS experiment: the accelerator complex, the NuMI beam line, the Near Detector, and the Far Detector. The operation, monitoring, adjustments and repairs of the first two are responsibilities of the Fermilab Accelerator Division, and are carried out from the Fermilab Main Control Room (MCR). The monitoring of the last two were the responsibilities of the MINOS Collaboration and were conducted from a separate MINOS Control Room.

In the MCR Fermilab beam operators monitor the performance of the accelerator complex as well as specific NuMI beam related variables (e.g. beam spot location, beam width, the number of protons on target per spill). The Fermilab control system known as ACNET \cite{ACNET}, using various plotting programs, was used to obtain and record the readings of various monitors described in the previous section. These graphs become a permanent record of the experiment. 

\begin{sidewaysfigure}
\begin{centering}
       \includegraphics[width=0.92\textwidth]{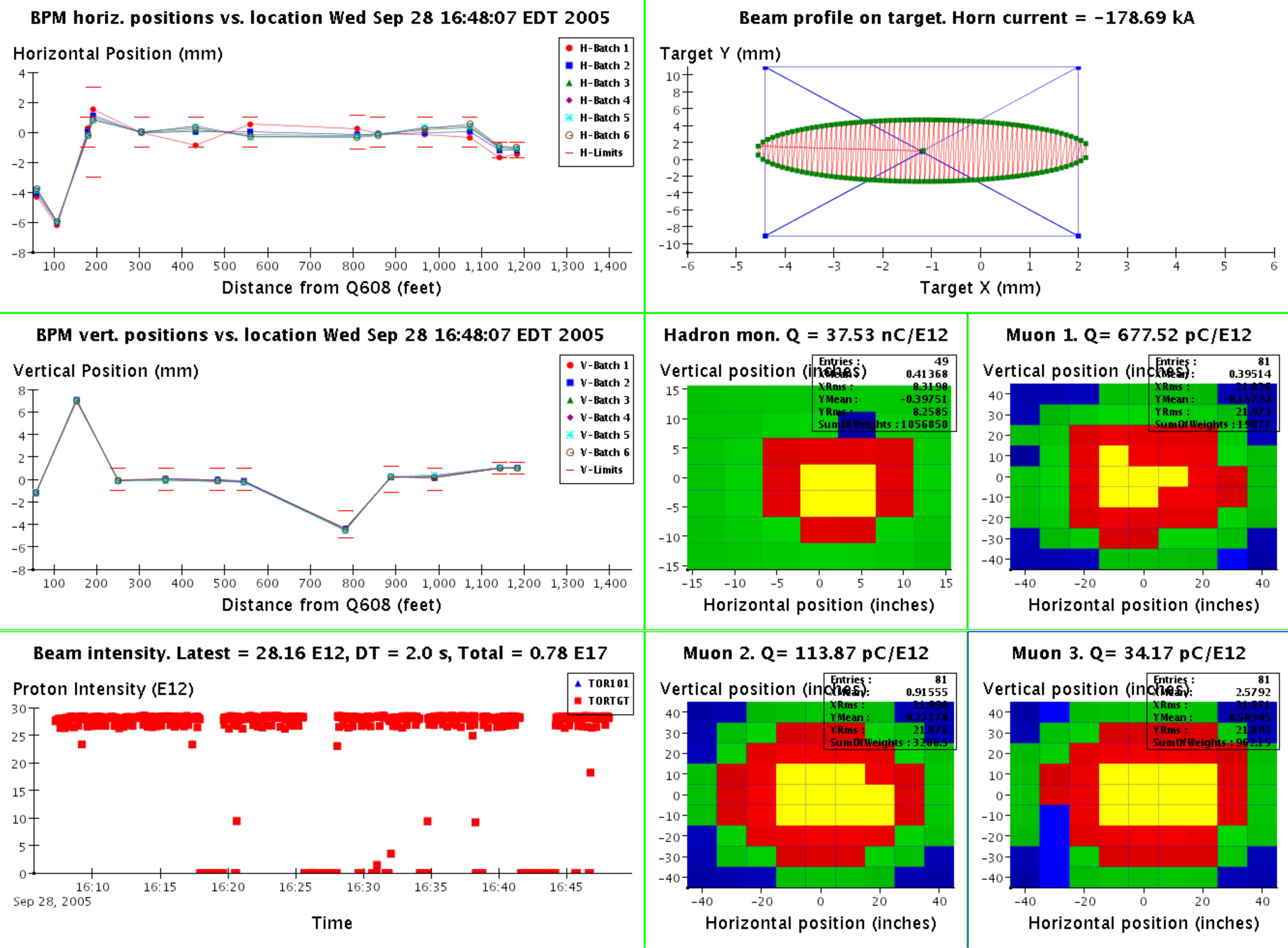}
      \caption{A Screenshot of the JAS3 Monitoring System in the MINOS Control Room. The screenshot was taken during the MINOS running of the NuMI beam on the 28th of September 2005. Each panel in the display shows a different piece of information. The top left panel shows the horizontal position of the beam at several points along the beam line, and similarly, the middle left panel shows the vertical position of the beam at a number of points along the beam line. The bottom left panel shows the intensity of the proton beam delivered to the NuMI target over the preceding hours and the top right panel shows the profile of this beam. The 4 panels in the bottom right show the readings from the hadron monitor and the 3 muon monitors. }
\label{fig:jas3}
 \end{centering}
\end{sidewaysfigure}

The main function of the work in the MINOS Control Room was to monitor and keep record of the performances of the two detectors. Single events from the two detectors were displayed in real time to give visual confirmation that systems were “live”. In addition, the JAVA Analysis Studio 3 (JAS3), an object based data analysis package developed at Brookhaven National Laboratory \cite{jas3} for the analysis of particle physics data, was used to show the most important beam information such as the beam intensity (number of protons per spill), the beam spot size, and beam timing information. Fig.~\ref{fig:jas3} shows a screenshot of the JAS3 monitoring system displayed in the MINOS control room. Furthermore, online monitors for the two MINOS detectors were continuously running and being updated on every spill.

%\subsection{Autotune and Beam Permit System}
%\label{sec:autotune}
In normal circumstances, very little human intervention is needed in the operation of the NuMI beam line. Computer control is essential due to the complexity of monitoring a multitude of systems on a constant basis. Previously two systems which play an integral part in this computerised monitoring and control were mentioned: Autotune and the Beam Permit system. 

The function of the Beam Permit System (BPS)\cite{HB2008} is to prevent extraction of pulses which, because of some malfunction in the accelerator or the beam line, would result in an errant beam. It uses dedicated hardware and is modeled on the Tevatron fast abort system. The Beam Permit system uses as input data the readings of 250 instruments and if any of them fail to meet set tolerances then the beam is either aborted before extraction (if still possible) or the succeeding pulses are suppressed until the problem is solved. 

On a pulse-by-pulse basis, the Autotune program monitors and, as necessary, adjusts the currents in all the NuMI beam trim magnets so that the beam traces the path through the magnets and other apertures without any significant beam loss. The program uses a matrix to relate the measurements of position detectors (in each plane) to currents in all the trim magnets. The inverse of the matrix provides trim-current values required to change the beam positions obtained in the measurements to their desired values. The vertical and horizontal profiles are continuously displayed for the operators so that they can see any major changes that may occur, but most of the time they are just passive observers. 

Initial operation of the NuMI beam has shown that there is a small trajectory difference in the extracted beam between the mixed mode with pbar targeting and the NuMI only mode. Thus slightly different settings of the corrector magnets are required for these two modes. To allow for that, separate corrector files for these two modes have been written for Autotune, which allows switching between the modes and the application of optimum settings in a seamless manner.

The combination of the permit system and Autotune beam position control have also been very important in enabling beam restart after down-times for maintenance or repairs. Typically, after power supply capability is re-enabled and beam permit system status confirmed, operations crew chiefs will load in the desired operating parameters, enable the NuMI beam switch, and normal operation will resume. No manual NuMI beam system tuning is needed.

\subsection{Long-term Beam Performance}
\label{longtermhistory}
The NuMI beam at the time of its design pushed the requirements imposed on its components beyond what had been achieved previously. As a consequence, its operating experience is of great interest to the designers of future beams with similar or even more severe requirements. In this subsection the performance of the NuMI beam is discussed, with emphasis on its most delicate components: the target and the two horns. They were subjected during their lifetimes to very harsh environments arising from very high integrated radiation doses, large mechanical stresses due to electromagnetic pulsing, and/or large thermal stresses from the energy deposited by the beam. 

\begin{figure}
\begin{centering}
      \includegraphics[width=.92\textwidth]{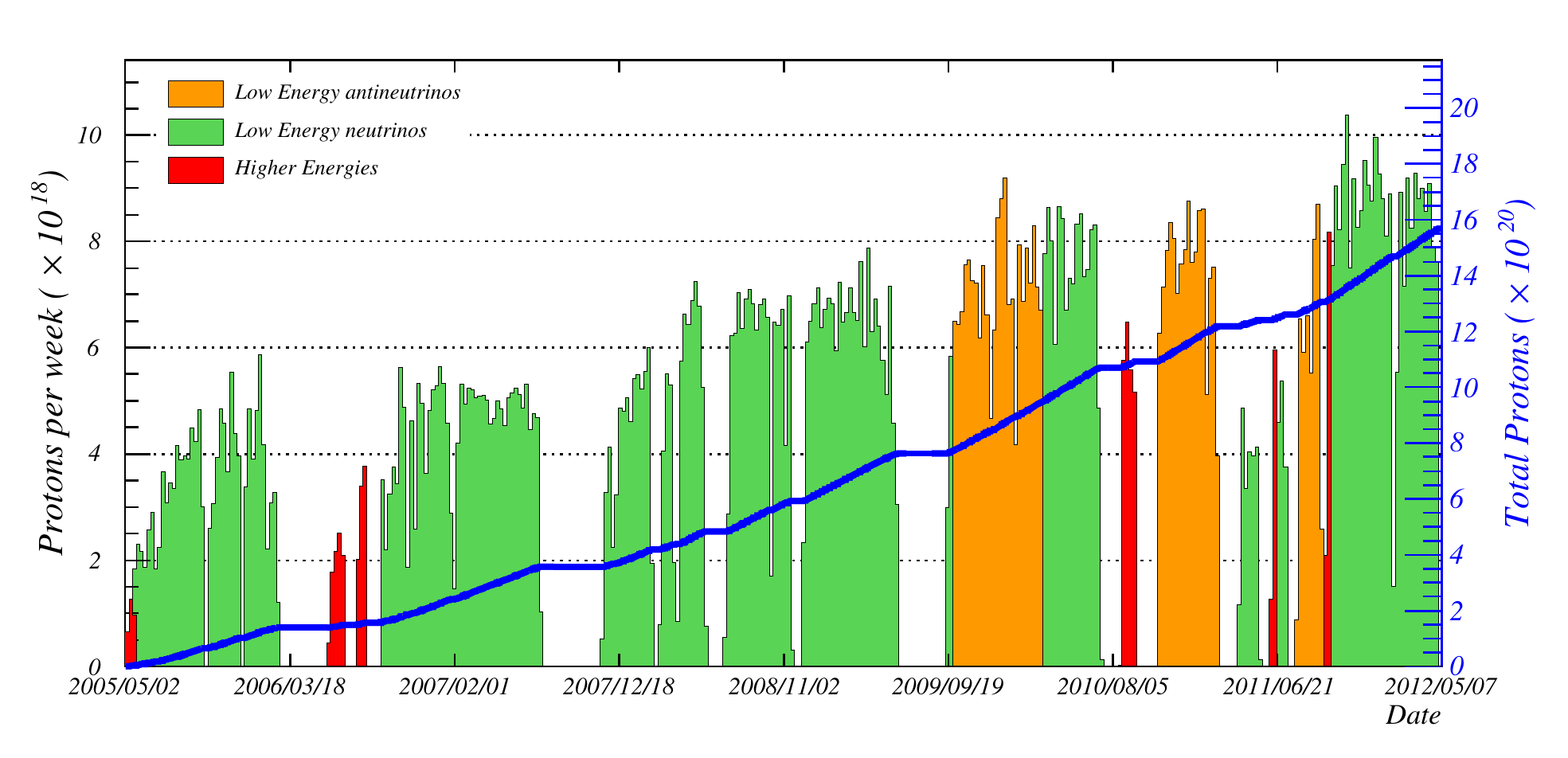}
      \caption{The Cumulative POT on the NuMI Target over the MINOS Run. The whole data taking period from 2005 to 2012 is shown. The experiment took a total of 15.6$\times10^{20}$~POT in various beam configurations; 10.71$\times10^{20}$~POT in $\nu$-beam mode - shown in green, and 3.36~$\times10^{20}$POT in $\bar{\nu}$-beam mode - shown in orange. Special runs like short runs at higher energy or horn off runs are shown in red.}
\label{fig:philpotplot}
 \end{centering}
\end{figure}

The MINOS beam exposure was almost exactly 7 years, starting May 1, 2005 and ending April 29, 2012. Most of it was in the LE (low energy) configuration with a few short interspersed runs with other target positions and/or non-nominal horn currents which were used for beam spectrum measurements and background determination analyses. In total, there were 61 million horn pulses during the experiment. The standard LE configuration data sets corresponded to an exposure of $10.71\times10^{20}$ POT obtained in $\nu$ beam mode, and an exposure of $3.36\times10^{20}$ POT obtained in the $\bar{\nu}$ beam mode. Including the short special runs such as medium or high energy, horn off runs or calibration runs, the total exposure in POT in the experiment was $15.6\times10^{20}$ POT. The time distribution of different runs and the intensity as a function of time are illustrated in Fig.~\ref{fig:philpotplot}. As can be seen, the intensity has gradually increased over the course of the experiment due partially to incremental improvements in the operation of the Booster and the Main Injector, as well as the initiation of the slip-stacking in the latter part of the experiment. 

The primary proton beam line functioned without any serious problem during the whole run. The beam was constantly monitored and adjusted if necessary by Autotune which used the BPM information to steer the beam. The Autotune plus Beam Permit system has been very effective in preventing proton beam loss. In about 61~million total pulses during the 7 years of NuMI/MINOS beam operation, there was never a pulse with large loss\footnote{There was one instance of larger proton beam loss along the primary beam transport during the first 10 million pulses delivered to the subsequent NO$\nu$A experiment. This necessitated the repair of a damaged vacuum pipe flange.}. Over its lifetime, the NuMI beam has averaged about 10 NuMI beam permit system trips per 24 hour day, with monthly averages ranging from three trips per day to as high as twenty one. The great majority of these are reset in 30-45 seconds. Typically, one or two trips per day require a few minutes to clear, and trips associated with a NuMI component problem needing intervention have averaged about one per week. The total downtime from the beam permit system trips is normally less than 10~min per day.

%\begin{table*}
\begin{table}
\centering
\scriptsize
%\begin{tabular} {|c| c| c| c| c| c| p{2.0cm}| c| c|}
\begin{tabular} {| c | c |  c | c |  c | c | c |}
\hline
Target & 1st POT & Last POT & No of weeks &Total POT& Beam Power & max POT/spill \\
\hline
NT01 & 5/1/2005 & 8/13/2006 & 67 & $1.6\times10^{20}$ & 270kW & $3.0\times10^{13}$ \\
\hline
NT02 & 9/11/2006 & 6/12/2009 & 144 & $6.1\times10^{20}$ & 340kW & $4.0\times10^{13}$ \\
\hline
NT03 & 9/11/2009 & 7/12/2010 & 44 & $3.1\times10^{20}$ & 375kW & $4.4\times10^{13}$ \\
\hline
NT04 & 8/22/2010 & 9/17/2010 & 4 & $2.0\times10^{19}$ & 375kW & $4.3\times10^{13}$ \\
\hline
NT05 & 10/29/2010 & 2/24/2011 & 17 & $1.3\times10^{20}$ & 337kW & $4.0\times10^{13}$ \\
\hline
NT06 & 4/7/2011 & 5/16/2011 & 6 & $2.0\times10^{19}$ & 305kW & $3.5\times10^{13}$ \\
\hline
NT01 & 6/11/2011 & 7/8/2011 & 4 & $4.5\times10^{19}$ & 228kW & $2.6\times10^{13}$ \\
\hline
NT02 & 7/29/2011 & 9/15/2011 & 7 & $2.0\times10^{19}$ & 330kW & $3.8\times10^{13}$ \\
\hline
NT07 & 9/24/2011 & 4/29/2012 & 31 & $2.6\times10^{20}$ & 345kW & $4.0\times10^{13}$ \\
\hline
\end{tabular}
\normalsize
\caption { Targets used in the NuMI beam line during the run of the MINOS experiment \cite{targetpos}. Each target is listed with the dates it was in operation, the number of weeks this corresponds to, its total exposure in POT, the maximum beam power it was exposed to, and the maximum POT/spill impinged upon it. Target NT01 was removed after the drive stuck in a high energy position and could not be moved back to the desired low energy position. Target NT02 was changed due to graphite deteriorating resulting in reduced beam flux. Target NT03 was retired due to a break at a ceramic tube holder. Targets NT04-NT06 were retired due to various water leaks. Repaired NT01 was then reinstalled; it also failed with a water leak. Target NT02 was then reused but swapped out when target NT07 became available. Both survived the MINOS experiment run and are remain available as spares. }
\label{tab:targethistory}
\end{table}

There were seven different targets used during the MINOS run \cite{NuMIOperation}. It was expected from the beginning that any single target would not last the whole run; the design called for reliable operation for one year ($10^{7}$ pulses). Thus it was deemed prudent to always try to  have one or even two spare targets available if needed for replacement. The target history is summarised in Table \ref{tab:targethistory}. The first target was removed because it froze in the high energy (HE) position after a special run in that configuration. The motion mechanism was subsequently fixed and the target was briefly used again in 2011 until it sprang a leak after a short operation of about a month. The most problematic failures encountered were in targets 4-6 which sprung water leaks shortly after their installation. The investigation of leaks in targets NT04 and NT05 determined them to be in the laser welds in the downstream water turnaround. That part was redesigned starting with NT06 with those welds replaced with tig welds (tungsten arc gas welds). The NT06 target sprang a leak after 6 weeks in the beam; it was determined to be in the upstream water connection but the precise location was not identified. Modifications were made in that area for NT07 which involved removing ceramic transitions and redoing the tig welds \cite{Hylen8056}. The NT07 target was still operating when the MINOS experiment finished running.

\begin{figure}
\begin{centering}
      \includegraphics[width=.92\textwidth]{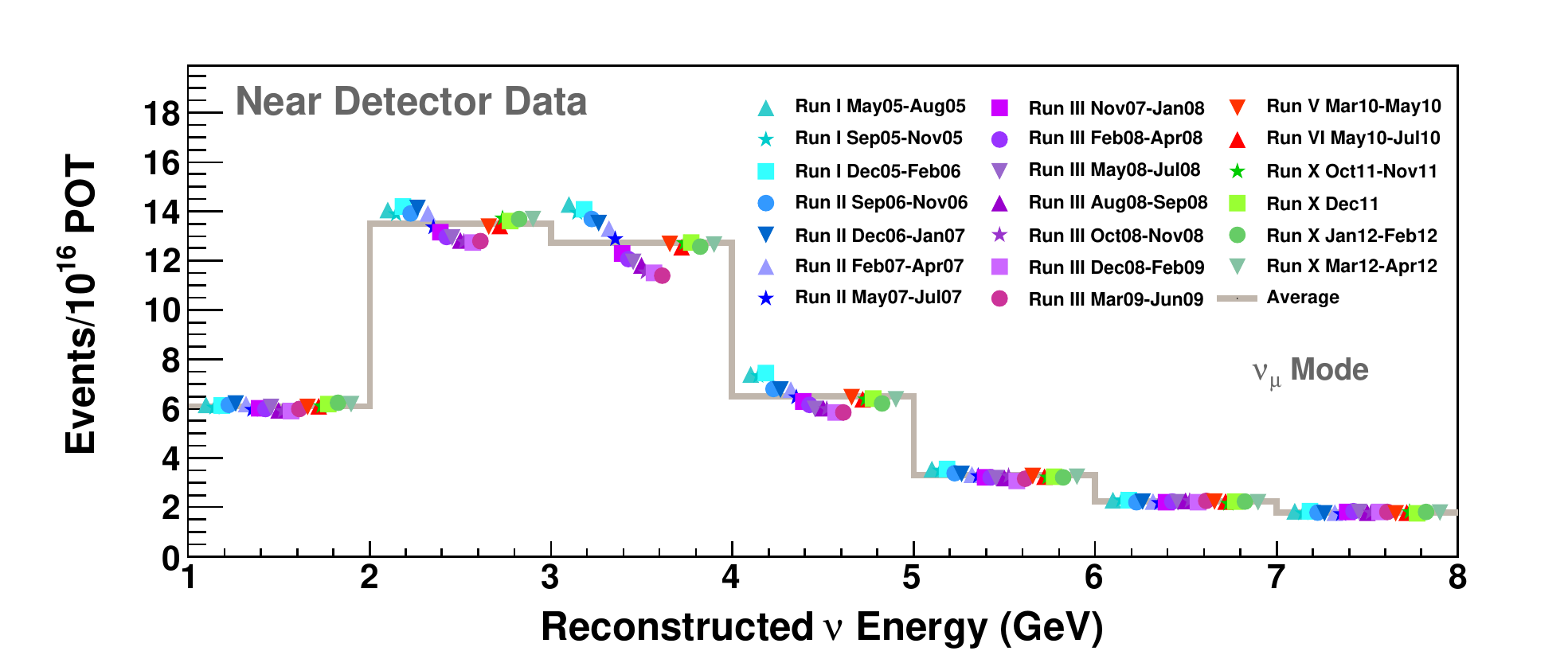}
      \caption{All-Time MINOS LE $\nu$ Beam Mode Reconstructed Neutrino Spectrum. The spectrum is zoomed into the peak region and is broken down by MINOS official runs and time within those runs. The solid line is the POT-weighted average data spectrum over the whole data taking period, while the points represent the data for specific MINOS runs or shorter time periods. The significant drop in runs II and III is due to NT02 target degradation. Furthermore, from run III onwards, helium was added to the NuMI decay pipe, which also leads to a few-percent level loss in neutrino events in the peak of the spectrum. Runs V and VI had new targets, however, these were slightly skewed with respect to the z-axis. Run X used a target which did not show any degradation through to the end of the run, but is nevertheless lower in the peak than run I due to the helium in the decay pipe.}
\label{fig:alldatafhc}
 \end{centering}
\end{figure}

\begin{figure}
\begin{centering}
      \includegraphics[width=.92\textwidth]{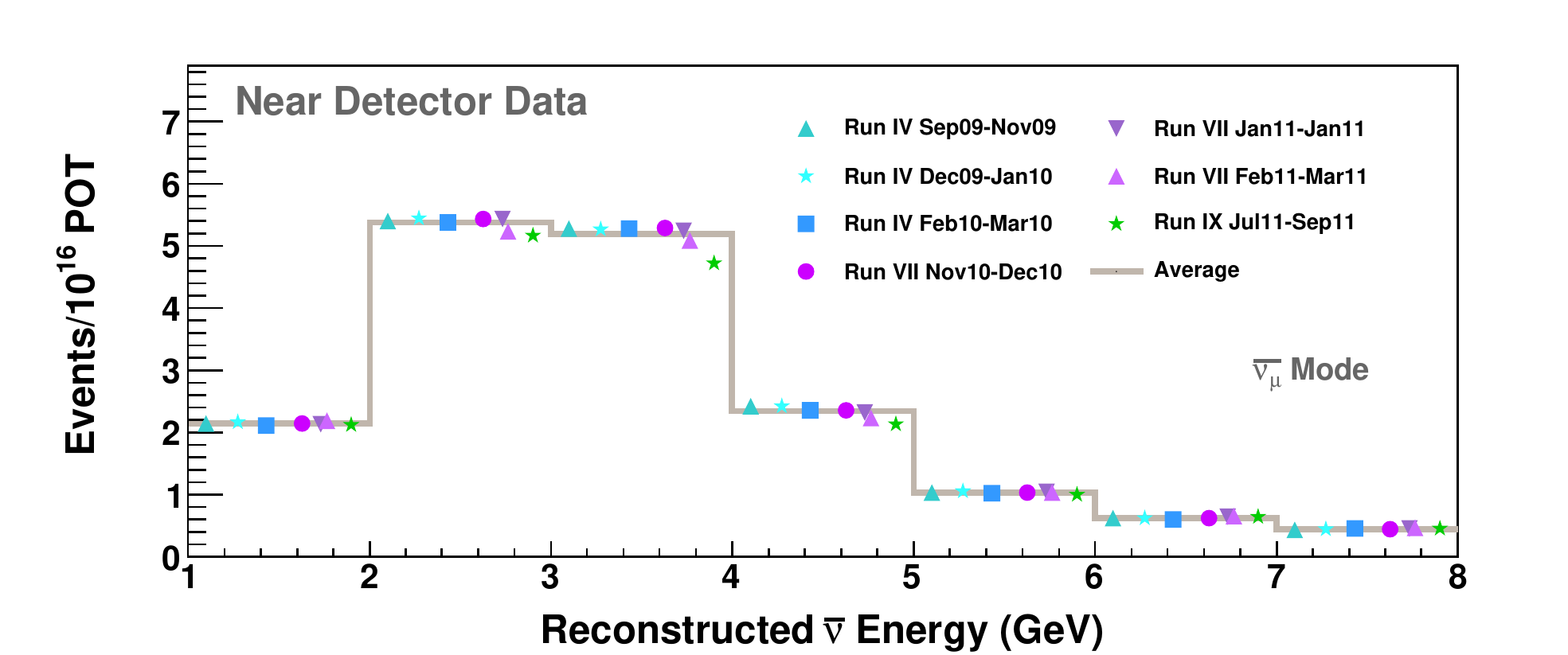}
      \caption{All-Time MINOS LE $\bar{\nu}$ Beam Mode Reconstructed Anti-Neutrino Spectrum. The spectrum is zoomed into the peak region and is broken down by MINOS official runs and time within those runs. The solid line is the POT-weighted average data spectrum over the whole data taking period, while the points represent the data for specific MINOS runs or shorter time periods. The drop at the end of run VII was caused by a water leak in the target, and the drop in event rate was visible in both the MINOS Near Detector and the muon monitors further upstream. The lower rate in run IX was caused by the necessity to reuse the older NT02 target, which showed target degradation.}
 \label{fig:alldatarhc}
 \end{centering}
\end{figure}

The target with the longest life span was NT02. Interestingly it never broke but after a few months of operation, the neutrino spectrum seen by the MINOS Near Detector began showing a significant decrease of flux in the peak of the spectrum. The event rates for the MINOS Near Detector are shown as a function of reconstructed energy in Fig.~\ref{fig:alldatafhc} for the LE $\nu$ beam mode, and Fig.~\ref{fig:alldatarhc} for the LE $\bar{\nu}$ beam mode. In Fig.~\ref{fig:alldatafhc}, a decrease with time can clearly be seen in the interval of 3-4~GeV. Monte Carlo simulations using FLUKA with GNuMI and MARS \cite{bishai7029, FLUKA} showed that such a decrease is consistent with the gradual disappearance of at least two target fins at a location corresponding with the peak of the energy deposition in the target. Target NT02 continued degrading with time throughout the course of its use in a manner consistent with progressive damage to the target. Because this target has been heavily irradiated during its use in the beam, as of this writing it has still not been possible to perform an autopsy on it to conclusively determine the cause of the flux decrease. The irradiation of the targets is high enough that one does expect to see changes in graphite properties, and the high stress induced in the target with each beam pulse could then induce cracking and fin failure.  However, the other targets have not shown deterioration even though some were irradiated beyond the level where the onset of deterioration was seen in NT02. So one may speculate that some other factor such as gas contamination may have contributed in that case.

Both beam horns were replaced once. The first version of Horn~1 had 24~million pulses from March 2005 to June 2008. The failure mode was a breaking of the water suction line near the electrical insulator. Its replacement had thicker steel installed in that region. The replacement went through 36~million pulses and was still operating at the end of the MINOS run. The first version of Horn~2 lasted until December 2008 and survived 28 million pulses. It failed when the power strip line fractured after a high strength steel washer broke due to hydrogen embrittlement. The Horn~2 replacement used lower strength steel washers; it successfully went through 32~million pulses through the end of the MINOS experiment. 

The hadron monitor is another beam element that is subjected to a high radiation environment and the build up of damage due to radiation was such that the hadron monitor had to be replaced in August 2009 after 7.64$\times10^{20}$~POT. The muon monitors are in a relatively shielded environment and they remained operational during the whole MINOS run. 

\section{Post-MINOS Plans and Developments}
\label{sec:future}
After a long accelerator shutdown following the completion of the MINOS experiment, the NuMI beam was restarted in September 2013 for two new experiments, NO$\nu$A and MINOS+, and for the continuation of MINER$\nu$A data taking. NO$\nu$A is an off-axis long-baseline neutrino experiment. The off-axis location means that the NO$\nu$A experiment is able to utilize a medium energy NuMI beam in order to observe a narrow band beam centered around approximately 2~GeV, at the neutrino oscillation maximum, and thus maximize the experiment's neutrino oscillation sensitivity. To generate the required medium energy beam, the Horn~2 location was optimized and is further away from Horn~1 than it was for the MINOS experiment run.  Horn~2 was moved to its present location of 19.2~m downstream of Horn~1 during the long shutdown. The MINOS+ experiment is effectively a continuation of MINOS but with a ``true'' medium energy spectrum where both horns are in their optimal positions for this energy configuration. The MINER$\nu$A experiment will continue as previously but also with the same medium energy spectrum that MINOS+ sees. The long decay pipe is more of an advantage for NO$\nu$A than it was for MINOS because the former is off-axis and therefore more of its neutrino flux originates from higher energy mesons.

The eventual design for NO$\nu$A operation is 700~kW with 4.9$\times 10^{13}$ protons per pulse. Several technical modifications in the accelerator complex and its operation are necessary to achieve these parameters. Advantage is taken of the availability of the Recycler since the Tevatron is no longer operational. The Recycler is a fixed energy machine of 8~GeV located on the top of the MI with the same circumference of 3.3~km. The adopted eventual mode of operation calls for injecting the Booster batches into the Recycler during the MI acceleration cycle. When all the batches are stored in the Recycler and the beam from the MI is extracted and its magnets are ramped down to 8~GeV, the 8~GeV protons in the Recycler are transferred to the MI for acceleration; new Recycler loading and Main Injector acceleration cycles then begin. Thus the two most time-consuming operations, the transfer from the Booster to the next accelerator, and the Main Injector acceleration, are done in parallel rather than in series. In this way it is possible to gradually reduce the overall cycle time to 1.3~s. To achieve the design proton intensity, eleven batches will have to be accelerated in the Booster and injected into the Recycler. This increased Booster proton yield requires some upgrades to the Booster RF power hardware, scheduled for completion in 2016. The 1.3~s cycle time has already been achieved for the six-batch operation.

The increase in the average beam power required some modifications in the NuMI beam operation and some redesign of sensitive beam components. The beam spot size at the target will be increased to 1.3~mm RMS in both directions to keep stress on the target acceptable\footnote{The beam spot size is currently still 1.1~mm as of writing this paper, however, it will be increased to 1.3~mm when the intensity upgrade is finished.}. A new baffle with a 13~mm diameter hole has been constructed to acccommodate this larger beam. A new more robust target was designed and constructed; it abandons the ability to move the target, an option not deemed necessary for the NO$\nu$A operation.  It does not extend into Horn~1 but is located further upstream with its vertical fins extending to 194~mm in front of Horn~1. While Horn~2 was unchanged except for its location, a new Horn~1 was constructed with the same inner conductor shape and current as for MINOS, but with thinner outer conductor and increased cooling. A new hadron monitor (of unchanged design) was constructed and installed. Finally there were a number of smaller changes focused on improving the proton beam monitoring, air circulation in the Target Hall and the cooling of systems in general.

\section{Conclusion}

The design and success of the NuMI beam profited from the experience gained over the preceding 50 years in building and operating accelerator-based neutrino beams in Western Europe, the Soviet Union, Japan and the United States. However, because of NuMI's unprecedented intensity, its long envisaged running time, and the high level of accuracy required in pointing and locating the beam at the target, a number of different challenges had to be met and solved. Over the years it has been running, the NuMI beam has been delivering neutrinos to the experiments located in its path reliably day after day. The NuMI operating experience during the MINOS experiment has been invaluable in guiding the design of the modifications of the most sensitive components of the beam for the 700~kW NO$\nu$A era operation and in planning the LBNF beam for the proposed DUNE experiment \cite{DUNE}.

\section*{Acknowledgments}
In a very real sense, it takes a whole community of thought, planning, and effort to carry out an enterprise of the scope of the NuMI beam line. Our acknowledgement and thanks go out first of all to the support groups of the Fermilab Accelerator Division, including the Mechanical Support Department - and in particular the Target Hall Support Group,  the Electrical Engineering Support Department, the Controls Department and the Instrumentation Department.

We recognize and gratefully acknowledge the support of the Fermilab Particle Physics Division, who provided crucial design and support via its Mechanical Engineering and Alignment and Metrology Departments.

The beamline could never have been completed or operated without the help of the Fermilab Facilities Engineering Services Section (FESS); we recognize their contributions with thanks.

We extend our grateful acknowledgement to IHEP, Protvino, for its vital work in design and fabrication of the NuMI targets.

The dedication, enthusiasm, and experience of the operations groups of the Fermilab Accelerator Division were essential in commissioning and running the NuMI beam, and receive our special appreciation.

This work was supported by the US DOE, the United Kingdom STFC, the US NSF, the state and University of Minnesota, and Brazil's FAPESP, CNPq and CAPES. Fermilab is operated by Fermi Research Alliance, LLC under Contract No. De-AC02-07CH11359 with the United States Department of Energy.

%\section*{References}

%\newpage


\addcontentsline{toc}{chapter}{Bibliography}
\begin{thebibliography}{0}


\bibitem{numitdr} K. Anderson et al., The NuMI Facility, Technical Design Report, Fermilab-TM-2018 (1998).

\bibitem{numitdh} NUMI Beam Technical Design Handbook (2002), http://www-numi.fnal.gov/numwork/tdh/tdh\_index.html

\bibitem{PRD77} P. Adamson et al. (MINOS Collaboration), Phys. Rev. D \textbf{77}, 072002 (2008).

\bibitem{cosmosref} N.W. Reay et al. (COSMOS Collaboration), Proposal for a short baseline experiment at Fermilab, Fermilab Proposal P803 (1993).

\bibitem{MINERvA} L. Fields et al. (MINER$\nu$A Collaboration), Phys. Rev. Lett. \textbf{111}, 022501 (2013).

\bibitem{Argoneut} C. Anderson et al. (ArgoNeuT Collaboration), Phys. Rev. Lett. \textbf{108}, 161802 (2012).

\bibitem{novaref} NO$\nu$A Collaboration, Fermilab Proposal P929 (2004), hep-ex/0503053.

\bibitem{minosplusref} A Proposal to FNAL to run MINOS with the medium energy NuMI beam, Internal MINOS-doc-7923 (2011), Fermilab Proposal P1016 (2011).

\bibitem{minibooneref} P. Adamson et al. (MiniBooNE Collaboration), Phys. Rev. Lett. \textbf{102} (2009) 211801.

\bibitem{fluxpaper} P. Adamson et al., The NuMI Beam Neutrino Flux, in preparation, to be published.

\bibitem{Davis} R. Davis, Jr, D. S. Harmer and K. C. Hoffman, Phys. Rev. Lett. \textbf{20}, 1205 (1968).

\bibitem{Haines} T. J. Haines et al., Phys. Rev. Lett. \textbf{57}, 1986 (1986).

\bibitem{Kam0} K. S. Hirata et al. (Kamiokande Collaboration), Phys. Lett. B \textbf{205}, 416 (1988).

\bibitem{Kam} K. S. Hirata et al. (Kamiokande Collaboration), Phys. Lett. B \textbf{280}, 146 (1992).

\bibitem{Beier} E. W. Beier et al. (Kamiokande Collaboration), Phys. Lett. B \textbf{283}, 446 (1992).

\bibitem{Fukuda} Y. Fukuda et al. (Kamiokande Collaboration), Phys. Lett. B \textbf{335}, 237 (1994). 

\bibitem{IMB3} D. Casper et al. (IMB-3 Collaboration), Phys. Rev. Lett. \textbf{66}, 2561 (1991).

\bibitem{IMB} R. Becker-Szendy et al. (IMB Collaboration), Phys. Rev. D \textbf{46}, 3720 (1992).

\bibitem{LSND} C. Athanassopoulos et al. (LSND Collaboration), Phys. Rev. Lett. \textbf{75}, 2650 (1995).

\bibitem{macro} S. Ahlen et al., (MACRO Collaboration), Phys. Lett. B \textbf{357}, 481 (1995).
 
\bibitem{Ambrosio} M. Ambrosio et al. (MACRO Collaboration), Phys. Lett. B \textbf{434}, 451 (1998).

\bibitem{Soudan2a} W. W. M. Allison et al. (Soudan 2 Collaboration), Phys. Lett. B \textbf{391}, 491 (1997).

%\bibitem{Soudan2} Sanchez M et al., Phys. Rev. D \textbf{68}, 112005 (2003).

\bibitem{fnalnumitdh} The Fermilab Main Injector Technical Design Handbook, FERMILAB-DESIGN-1994-01 (1994).

\bibitem{P822ref} Proposal for a long baseline neutrino oscillation experiment using the Soudan 2 neutrino detector, Fermilab Proposal P822 (1991).

\bibitem{workshop1991} Proceedings of the Workshop on Long-Baseline Neutrinos at Fermilab 17-20 November 1991.

\bibitem{MINOS} A Long-baseline Neutrino Oscillation Experiment at Fermilab, Fermilab Proposal P875 (1995).

\bibitem{Abramov:2001nr}  A.~G.~Abramov et al., Beam optics and target conceptual designs for the NuMI project,  Nucl. Instrum. Meth. A {\bf 485}, 209 (2002).

\bibitem{Brown:2013idd}  B.~C.~Brown et al., The Fermilab Main Injector: high intensity operation and beam loss control,  Phys. Rev. ST Accel. Beams {\bf 16}, no. 7, 071001 (2013)  [arXiv:1307.2934 [physics.acc-ph]].

\bibitem{NuMIBeam} P. Adamson et al., Primary Proton Beamline for the Fermilab Main Injector Neutrino Program, Internal MINOS-Doc-1669 (2007).

\bibitem{MARSref} N.V. Mokhov, The MARS Code System Users' Guide (1995), Fermilab-FN-628.

\bibitem{Budal} K. Budal, IEEE Trans. Nucl. Sci. \textbf{14}, 1132 (1967).

\bibitem{VariableEnergy} M. Kostin, Proposal for Continuously Variable Beam Energy, Internal NuMI-B-783 (2001).

\bibitem{MarinoMorfin} A. Marino and J. Morfin, Study of Neutrino Event Rates for Various Target Positions in the Medium-Energy Horn Configuration, Internal MINOS-Doc-1520 (2006).

\bibitem{hornsref} S. van der Meer, CERN Report CERN-61-07 (1961).

\bibitem{hornmap} J. Hylen, Magnetic Field Mapping of NuMI Horn PH1-03, Internal MINOS-Doc-6122 (2009).

\bibitem{horn710} A. Abramov et al., IHEP Group, Calculations and Mapping of the Magnetic Field in the Prototype Horn~1, MINOS-Doc-710 (2000)

\bibitem{hadronichose} J. Hylen et al., Nucl. Instrum. Meth. A \textbf{498}, p.29-51 (2003).

\bibitem{absorberconcept} A.Abramov et al., Advanced Conceptual Design of the NuMI Hadron Beam Absorber Core, Internal NuMI-B-652 (2000).

\bibitem{docdb2769}  Hadron Absorber and Muon Alcoves composition and G4NuMI representation, Internal MINOS-Doc-2769 (2007).

\bibitem{BeamMonitors2} S.E. Kopp et al., Beam test of a segmented foil SEM grid, Nucl. Instrum. Meth. A 554 (2005) 138-146; also Fermilab-Pub-05-045-AD, July 2005.

\bibitem{Ref14in1669} R. Joshel et al., Automated beam position and split control for the Fermilab switchyard, Presented at PAC1987, Washington, Vol. \textbf{1}, p.515.

\bibitem{BeamMonitors} S. Kopp et al., Secondary beam monitors for the NuMI facility at FNAL, Nucl. Instrum. Meth. A \textbf{568}, p. 503-519, (2006), Fermilab-Pub-06-007-AD (2006), arXiv:physics/0607229.

\bibitem{HadronMuonMonitors} J. McDonald et al., Ionization chambers for monitoring in high-intensity charged particle beams, Nucl. Instrum. Meth. A \textbf{496}, 293-304, (2003).

\bibitem{HadronMuonMonitors2} R.M. Zwaska et al., Beam tests of ionization chambers for the NuMI neutrino beam, IEEE Trans. Nucl. Sci. \textbf{50}, 1129-1135 (2003), IEEE 2002 Nuclear Science Symposium, Norfolk, VA, 10-16 Nov. 2002; also hep-ex/0212011.

\bibitem{LauraThesis}  L. Loiacono, Ph.D thesis, University of Texas, Austin (2010).

\bibitem{ZarkoThesis} Z. Pavlovich, Ph.D thesis, University of Texas, Austin, (2008).

\bibitem{NIMdetectorpaper} D.H. Michael et al., Nucl. Instr. Meth. A \textbf{596}, 190 (2008).

\bibitem{BeamAlignment} R. Zwaska et al., Beam-based alignment of the NuMI target station components at FNAL, Nucl.Instrum.Meth.A \textbf{568}, p.548-560 (2006), Fermilab-Pub-06-171-AD, arXiv:physics/0609106.

\bibitem{bobthesis} R. M. Zwaska PhD Thesis, 2005, Accelerator systems and instrumentation for the NuMI neutrino beam, Fermilab-Thesis-2005-73, 2005

%\bibitem{SlipStacking} K. Seiya et al., Multi-batch slip stacking in the Main Injector at Fermilab, Proceedings of the 2007 PARTICLE ACCELERATOR CONFERENCE, pp. 742-744.

\bibitem{SlipStacking} K. Seiya et al., Progress in Multi-Batch Slip Stacking in the Fermilab Main Injector and Future Plans, FERMILAB-CONF-09-155-AD.

\bibitem{ACNET} J. Patrick, Fermilab Accelerator Control System, Proceedings of ICP 2006, Chamonix, France.

\bibitem{jas3} A.S. Johnson, Java Analysis Studio, paper presented at the 1998 conference on Computing in High Energy Physics (CHEP1998), Chicago, IL (1998).

\bibitem{HB2008} S. Childress, NuMI Proton Beam Diagnostics and Control, paper presented at the Workshop on High Intensity, High Brightness Hadron Beams (HB2008), August 2008.

\bibitem{targetpos} J. Hylen, Target location along beam-line; summary of entire MINOS run, Internal MINOS-Doc-9314 (2012).

\bibitem{NuMIOperation} S. Childress, The NuMI Proton Beam at Fermilab, Successes and Challenges, Fermilab-CONF-08-380-AD.

\bibitem{Hylen8056} J. Hylen, Target Saga, Internal MINOS-Doc-8056; NUMI target story, Internal MINOS-Doc-8517 (2011)

\bibitem{bishai7029} M. Bishai, Towards a better target decay model, Internal MINOS-Doc-7029 (2010)

\bibitem{FLUKA} A. Fasso‘, A. Ferrari, J. Ranft, and P.R. Sala, FLUKA: a multi-particle transport code, CERN-2005-10 (2005), INFN/TC 05/11, SLAC-R-773

%\bibitem{GNuMI} GNuMI

\bibitem{DUNE} DUNE Letter of Intent, Fermilab Proposal P1062 (2015).


\end{thebibliography}
\end{document}